\documentclass[superscriptaddress,amsmath,amssymb, aps, prl,longbibliography,twocolumn, footinbib]{revtex4-2}

\usepackage{upgreek}
\usepackage{color}
\usepackage{graphicx}
\usepackage{nccmath}
\usepackage{bm}
\usepackage{hyperref}
\hypersetup{colorlinks,breaklinks,
            urlcolor=[rgb]{0,0,0.64},
            linkcolor=[rgb]{0,0,0.64},
            citecolor=[rgb]{0,0,0.64}}
\usepackage[mathlines]{lineno}
\usepackage{dcolumn}
\usepackage[mathscr]{euscript}

\begin{document}

\title{Unconventional light-induced states visualized by ultrafast electron diffraction and microscopy}

\author{Alfred~Zong}
\affiliation{University of California at Berkeley, Department of Chemistry, Berkeley, California 94720, USA.}
\author{Anshul~Kogar}
\affiliation{University of California at Los Angeles, Department of Physics and Astronomy, Los Angeles, California 90095, USA.}
\author{Nuh~Gedik}
\email[Correspondence to: ]{gedik@mit.edu}
\affiliation{Massachusetts Institute of Technology, Department of Physics, Cambridge, Massachusetts 02139, USA.}

\date{\today}

\begin{abstract}

Exciting electrons in solids with intense light pulses offers the possibility of generating new states of matter through nonthermal means and controlling their macroscopic properties on femto- to picosecond timescales. One way to manipulate a solid is by altering its lattice structure, which often underlies the electronic, magnetic and other phases. Here, we review how structures of solids are affected by photoexcitation and how their ultrafast dynamics are captured with time-resolved electron diffraction and microscopy. Specifically, we survey how a strong light pulse has been used to tailor the nonequilibrium characteristics to yield on-demand properties in various material classes. In the existing literature, four main routes have been exploited to control material structures: (i)~phase competition, (ii)~electronic correlations, (iii)~excitation of coherent modes, and (iv)~defect generation. In this review, we discuss experiments relevant to all four schemes and finish by speculating about future directions.
\end{abstract}

\maketitle


This review details methods to instigate crystallographic phase transitions and to control material structures with femtosecond light pulses. We focus on experimental studies using ultrafast electron diffraction (UED) and microscopy (UEM), where, as detailed below, recent progress has been substantial. For a more general survey of experimental and theoretical advances in the ultrafast control of quantum materials, we refer readers to a recent review by de~la~Torre \textit{et al.} \cite{delaTorre2021}. Here, we concentrate on four main themes of material control that have arisen in the past few years: (i)~phase competition, (ii)~electronic correlations, (iv)~excitation of coherent modes, and (iv)~defect generation. 

Many of the studies highlighted here involve charge density wave (CDW) systems because they serve as excellent model platforms to investigate light-induced phenomena using UED and UEM. Charge density waves exhibit rich phase diagrams, diverse experimental phenomenology, and are ideally suited to UED and UEM experiments, which can quantitatively track the evolution of order parameter amplitudes, correlation lengths and fluctuations. Early work dating back to 2010 \cite{Eichberger2010} demonstrated that the CDW order parameter, its interplay with the underlying lattice and its collective excitations can all be measured by existing time-resolved probes. Thus, we focus here on lattice order parameters, many of those arising from CDWs, where photoexcitation drives structural transitions that can affect the electronic and other properties.

\section{Phase competition}

Many classes of solids exhibit a rich phase diagram when one tunes temperature, pressure, chemical composition, or magnetic field. In these systems, phase competition as a result of intertwined degrees of freedom often leads to exotic phenomena from high-$T_c$ superconductivity and colossal magnetoresistance to hidden orders. Inspired by these equilibrium phase diagrams, we can envisage a dynamical control of the competition, where one ground state is transiently suppressed by photoexcitation while a neighboring state emerges out of equilibrium. This idea is used to account for certain light-induced superconductor-like behavior in cuprates, where charge order and $d$-wave superconductivity are thought to compete near a hole doping at $p=1/8$ \cite{Kaiser2017}. As photoexcitation weakens the charge order, superconductivity can be transiently enhanced; their distinct photo-responses can be rationalized by the different relaxation rates after the excitation event \cite{Sun2020}.

\begin{figure*}[htb!]
	\centering
	\includegraphics[width=0.8\textwidth]{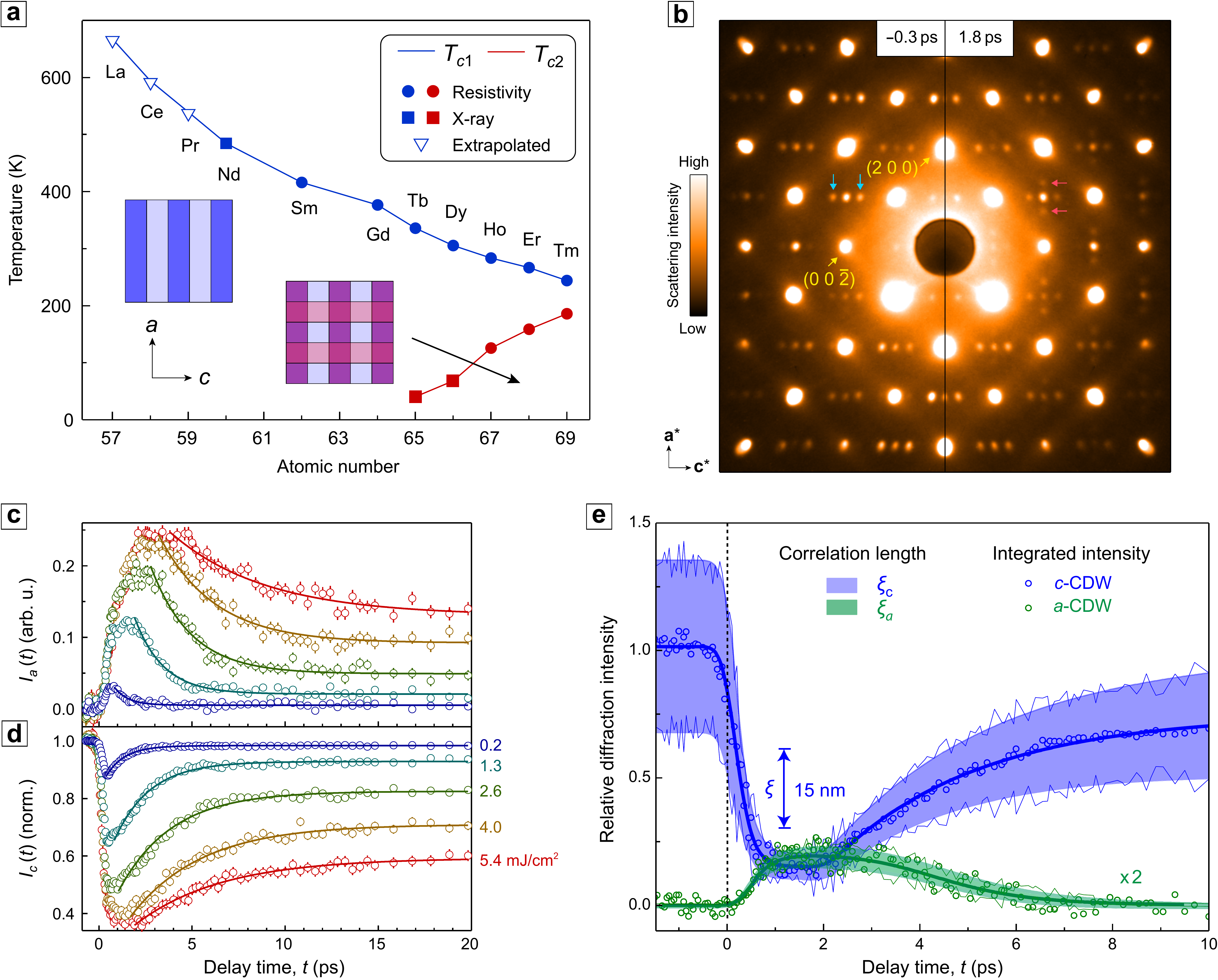}
	\caption{\textbf{Light-induced states in systems with competing orders.} (a)~Summary of transition temperatures of two competing charge density wave (CDW) orders in rare-earth tritellurides ($R$Te$_3$). Insets show the schematic of the CDW states below $T_{c1}$ and $T_{c2}$. (b)~Electron diffraction pattern of LaTe$_3$ at 307~K, taken at 0.3~ps before (left) and 1.8~ps after (right) the arrival of an 80-fs, 800-nm laser pulse. It shows the suppression of the equilibrium $c$-axis CDW peaks (blue arrows) and the emergence of light-induced $a$-axis CDW peaks (red arrows). (c,d)~Intensity evolutions of transient $a$-axis and equilibrium $c$-axis CDW peaks after photoexcitation at different incident fluences, showing similar relaxation time to the quasi-equilibrium state. (e)~Similar photoinduced evolution of $c$-axis and $a$-axis CDW peaks in CeTe$_3$. The correlation length ($\xi$) of each CDW extracted from the diffraction intensity profile is plotted as the shaded envelope. Adapted from ref.~\cite{Kogar2020} (a--d) and ref.~\cite{Zhou2021} (e).}
\label{fig:competition}
\end{figure*}

From an experimentalist's point of view, the study of nonequilibrium phase competition necessitates a time-resolved probe that can capture all order parameters at the same time. This would eliminate the uncertainty of varying photoexcitation conditions or sample dimensions in different ultrafast setups. In this regard, recent UED investigations into the phase competition in rare-earth tritellurides ($R$Te$_3$) serve as a representative example that gives a direct comparison of two competing orders, showing that a hidden CDW state suppressed in equilibrium can be ``unleashed'' by a femtosecond light pulse \cite{Kogar2020,Zhou2021}. 

$R$Te$_3$ are layered van der Waals materials that crystallize in nearly tetragonal structures with only a slight anisotropy between the two in-plane axes, $a$ and $c$ \cite{RuThesis}. All members share similar structural and electronic properties except for the different chemical pressures exerted by the rare-earth elements with varying ionic radii. As one moves from La to Tm, shown in Fig.~\ref{fig:competition}(a), the material can host either a unidirectional CDW along the $c$-axis or a bidirectional CDW along both in-plane axes. The critical temperatures of the respective transitions display opposite trends, suggesting a phase competition between the $c$- and $a$-axis CDWs. In equilibrium, the slight lattice anisotropy dictates that the $c$-axis CDW always forms first. If this order is sufficiently strong, the $a$-axis CDW cannot be stabilized under ambient pressure.

The $R$Te$_3$ family has been extensively studied by time-resolved optical spectroscopy \cite{Yusupov2008,Yusupov2010,Zong2019a,Zong2019b}, photoemission \cite{Schmitt2008,Schmitt2011,Leuenberger2015,Rettig2016,Zong2019a,Maklar2021}, X-ray scattering \cite{Moore2016,Trigo2019,Trigo2021,Maklar2021}, and electron diffraction \cite{Han2012,Zong2019a}, providing a prototypical platform for investigating uniquely nonequilibrium phenomena associated with a photoinduced transition. However, most studies only concentrate on the dominant $c$-axis CDW while neglecting the competing $a$-axis order. Starting from the unidirectional CDW state in LaTe$_3$, Kogar \textit{et al.} found that while a femtosecond laser pulse suppressed the equilibrium $c$-axis CDW, it also seeded the growth of the competing $a$-axis order \cite{Kogar2020}. This light-induced CDW is evident from electron diffraction images taken before and after the incidence of the pump laser pulse, shown in Fig.~\ref{fig:competition}(b), where blue and red arrows indicate pairs of CDW peaks along the $c^*$ and $a^*$ axes, respectively. The same observation was reported by Zhou \textit{et al.} in CeTe$_3$ [Fig.~\ref{fig:competition}(e)] \cite{Zhou2021}, and we expect other members of the $R$Te$_3$ family to exhibit similar phenomenology in their unidirectional CDW phase. The competing nature of the equilibrium and the light-induced CDWs is best illustrated by the evolution of their superlattice peak intensities, shown in Fig.~\ref{fig:competition}(c,d). As photoexcitation quenches the $c$-axis CDW peak within 400~fs, the $a$-axis peak emerges at the same time. The slightly slower rise along the $a$ direction compared to the faster intensity drop in the $c$ direction hints at the incoherent development of the light-induced CDW in different spatial regions of the sample, in contrast to a coherent displacive motion that drives the melting of the equilibrium order. On the other hand, during the relaxation to a quasi-equilibrium plateau, the intensity curves are perfectly anti-correlated between the two CDWs for all pump fluences measured, suggesting a strong competition scenario.

There are a number of peculiarities observed in the light-induced CDW that warrant further research. For example, the modulation wavevector of the nonequilibrium $a$-axis CDW appears to be distinct from the value of the equilibrium order in the checkerboard-like state \cite{Kogar2020}, pointing towards a truly nonequilibrium state with no equilibrium counterpart. Furthermore, the dominant $c$-axis modulation acquires a persistent wavevector shift that cannot be explained by laser-induced heating \cite{Zhou2021}, further underscoring the nonthermal nature of the light-induced state. With improved momentum resolution, the correlation length of the photoinduced CDW can be more accurately quantified along different directions in future experiments, offering insights into whether the CDW can transiently acquire three-dimensional phase coherence, an important question with broad implications in other light-induced states. By tailoring the photoexcitation conditions, for instance by adjusting the pulse width, fluence, or even applying a particular pulse sequence, one could also engineer ways to prolong the lifetime of the light-induced order beyond just a few picoseconds \cite{Budden2021}.

\section{Electronic correlations}

\begin{figure*}[htb!]
	\centering
	\includegraphics[width=0.8\textwidth]{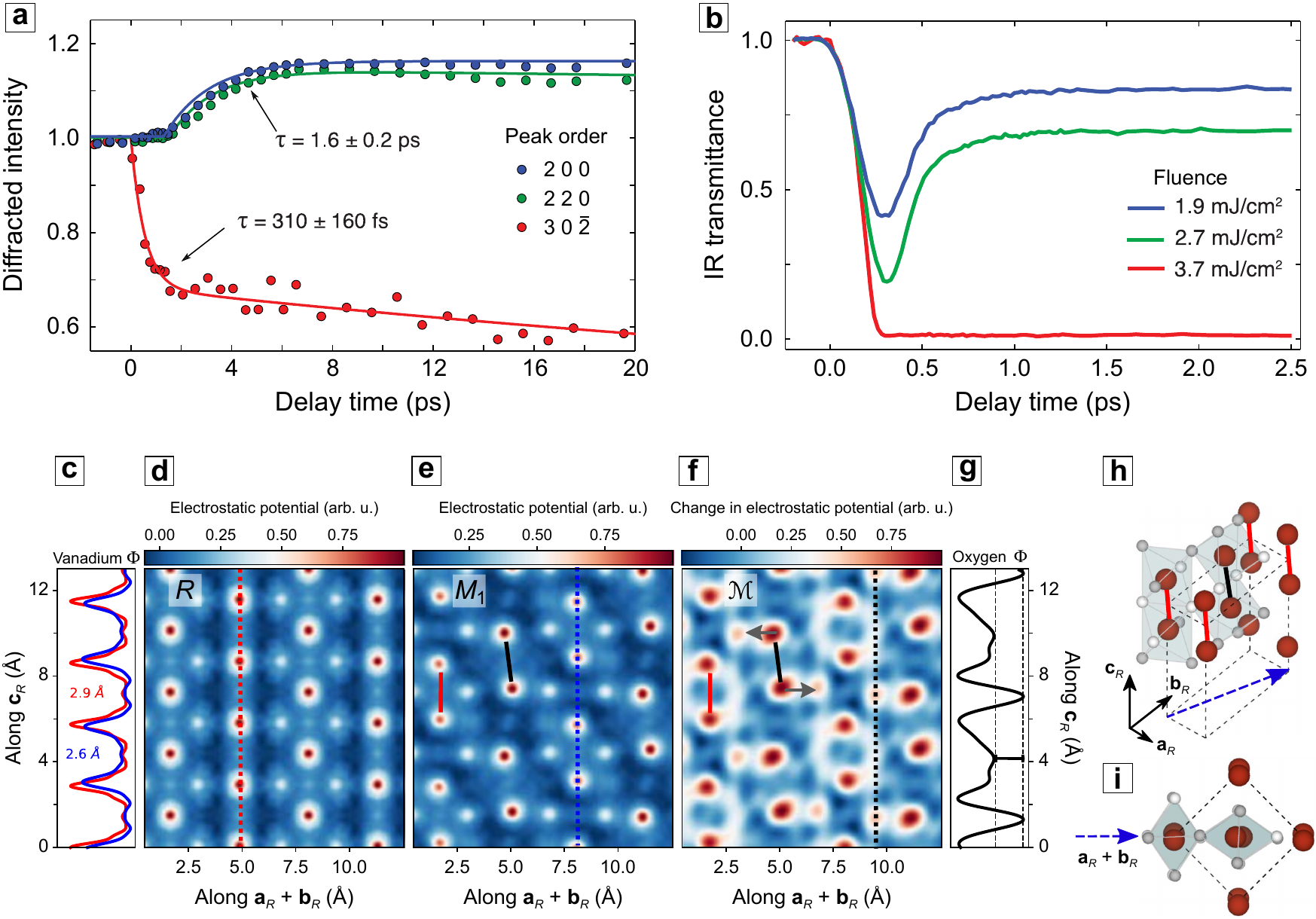}
	\caption{\textbf{Metastable states in strongly correlated materials.} (a)~Intensity evolution of three Debye-Scherrer rings of a polycrystalline VO$_2$ film after photoexcitation by 35-fs, 800-nm pulses at 310~K with a 20~mJ/cm$^2$ pump laser fluence. There is a fast decay in intensity of diffraction peaks associated with the periodic lattice distortion ($30\bar{2})$ and a slow increase for peaks that are particularly sensitive to the valence charge distribution (200 and 220). (b)~Fluence dependence of transient infrared transmissivity at 5~$\upmu$m, showing a nearly complete decrease at 3.7~mJ/cm$^2$. (c--g)~Reconstruction of the electrostatic crystal potential $\Phi(\mathbf{r})$ based on UED intensities for the equilibrium high-temperature rutile ($R$) and low-temperature monoclinic ($M_1$) phases, as well as the photoinduced metastable monoclinic metallic phase $(\mathscr{M})$ at 10~ps after photoexcitation at a fluence of 6~mJ/cm$^2$. The $\mathscr{M}$ phase shows an antiferroelectric order (arrows) in addition to V-V dimerization both in-plane (black line) and out-of-plane (red line). Panels (c) and (g) are line cuts in (d--f) with respective colors. (h)~3D schematic of the VO$_2$ structure, where V/O atoms are red/gray. Lines indicate dimerization. (i)~The structure in (h) projected along the $\mathbf{c}_R$ axis. Adapted from ref.~\cite{Morrison2014} (a,b) and ref.~\cite{Otto2019} (c--i).}
\label{fig:correlation}
\end{figure*}

In materials with strong electronic correlations, charge is often coupled to the spin, orbital, or structural degrees of freedom. As many of these systems possess $d$-electrons that sit close to the boundary between electron itinerancy and localization, there often exists a wide variety of phase transitions that can be triggered with small external perturbations \cite{Zhang2014}. Vanadium dioxide (VO$_2$) is a prototypical example of such a material where, in thermal equilibrium, a high-temperature metallic phase gives way to a low-temperature insulating phase below $\sim 341$~K. Accompanying the orders-of-magnitude change in conductivity is a crystallographic phase transition from a high-temperature rutile to a low-temperature monoclinic structure. The transition has remained controversial for decades with researchers arguing about whether it is primarily driven by electron correlations or electron-phonon interactions \cite{Shao2018}. Roughly two decades ago, it was found that a femtosecond laser pulse could instigate the transition between the two phases, evidenced by photoinduced changes in the optical response and diffraction peaks \cite{Becker1996,Cavalleri2001,Kim2006,Baum2007}, yielding new experimental insight into the transition in VO$_2$.

Recently, ultrafast electron diffraction has provided a fresh perspective concerning this particular transition through strong evidence of a metastable metallic phase with a monoclinic structure \cite{Morrison2014, Otto2019, Sood2021}. Of utmost importance in these studies is the sensitivity of low-angle electron diffraction peaks to the valence electron distribution. In contrast to scattering probes such as X-ray and neutron diffraction, electron diffraction is sensitive to the valence electron density in addition to the lattice structure \cite{Zheng2005}. This key detail allowed Otto \textit{et al.} to determine the change in the electric potential and electronic distribution after photoexcitation with an 800~nm wavelength laser pulse \cite{Otto2019}. 

Figure~\ref{fig:correlation}(a) shows that VO$_2$ exhibits two diffraction timescales after photoexcitation at room temperature, depending on which peaks are examined. Faster timescales below 1~ps are associated with the crystallographic transformation to the rutile structure while the slow timescales over several picoseconds are assigned to the electronic redistribution. Remarkably, below a certain fluence threshold ($\sim9$~mJ/cm$^2$), only the slow timescales are observed, indicating that the system retains its monoclinic structure. By also measuring the infrared transmissivity at a wavelength of 5~$\upmu$m [Fig.~\ref{fig:correlation}(b)], Siwick and coworkers were able to determine that this monoclinic phase is metallic by $\sim 3.7$~mJ/cm$^2$. The electronic redistribution was estimated using the Patterson method based on the diffraction pattern, and is plotted in Fig.~\ref{fig:correlation}(d--f). In Fig.~\ref{fig:correlation}(d), the electrostatic potential of the equilibrium rutile phase ($R$) is shown, demonstrating that the high-symmetry phase consists of equally spaced vanadium atoms along the \textbf{c}$_R$ axis [red curve in Fig.~\ref{fig:correlation}(c)]; see Fig.~\ref{fig:correlation}(h,i) for the definition of crystalline axes. A similar plot for the equilibrium monoclinic phase ($M_1$) is shown in Fig.~\ref{fig:correlation}(e), where a clear dimerization of the vanadium atoms takes place [blue curve in Fig.~\ref{fig:correlation}(c)]. Figure~\ref{fig:correlation}(f) shows the \emph{change} in the electric potential 10~ps after photoexcitation with a fluence of 6~mJ/cm$^2$. Despite the reorganization of electrons, the vanadium atoms retain their dimerized structure. Interestingly, the electronic redistribution in the nonequilibrium metastable phase ($\mathscr{M}$) discriminates between two oxygen atoms that are symmetry equivalent in the rutile phase, which is highlighted by the line cut through the oxygen atoms in Fig.~\ref{fig:correlation}(g). 

There are still many open questions about this metallic monoclinic phase, but the data are suggestive that the rutile structure is not a necessary requirement for metallicity in this system. One question apparent from a comparison between Fig.~\ref{fig:correlation}(a) and (b) is that the infrared transmissivity has a timescale corresponding to the ``fast'' timescale in the diffraction peaks, although only the relatively slow transition into a monoclinic metallic phase is expected in transmissivity at fluences below $\sim9$~mJ/cm$^2$. While this and other issues abound, the metastable metallic phase with monoclinic structure represents a case study where ultrafast electron diffraction yielded profound insight into a phase transition that is commonly associated with electronic correlations. Importantly, the sensitivity of the probing electrons to the valence electron distribution is exclusively accessed with ultrafast electron diffraction.

\section{Excitation of coherent modes}

In a broken-symmetry phase, coherent photoexcitation of a collective mode -- be it a phonon, magnon, plasmon, or a hybrid -- offers important clues about the ground state properties and associated phase transitions. Here, the word \emph{coherent} means that microscopic entities such as lattice ions, charge or spin undergo a phase-locked motion over a macroscopic length scale. The synchronized motion typically manifests as an oscillatory signal when plotted against pump-probe time delay, and its amplitude and frequency encode information about the local environment in a highly excited medium. When several intertwined degrees of freedom are present, these coherent modes are particularly useful in isolating the driving force behind the formation of the ground state. If oscillations with the same frequency are detected in different observables, such as ionic position and electronic energy, these modes also allow a precise measurement of the coupling strength between various microscopic entities, free from interference of incoherent dynamics that are often too complex to model \cite{Gerber2017}.

\begin{figure*}[htb!]
	\centering
	\includegraphics[width=1\textwidth]{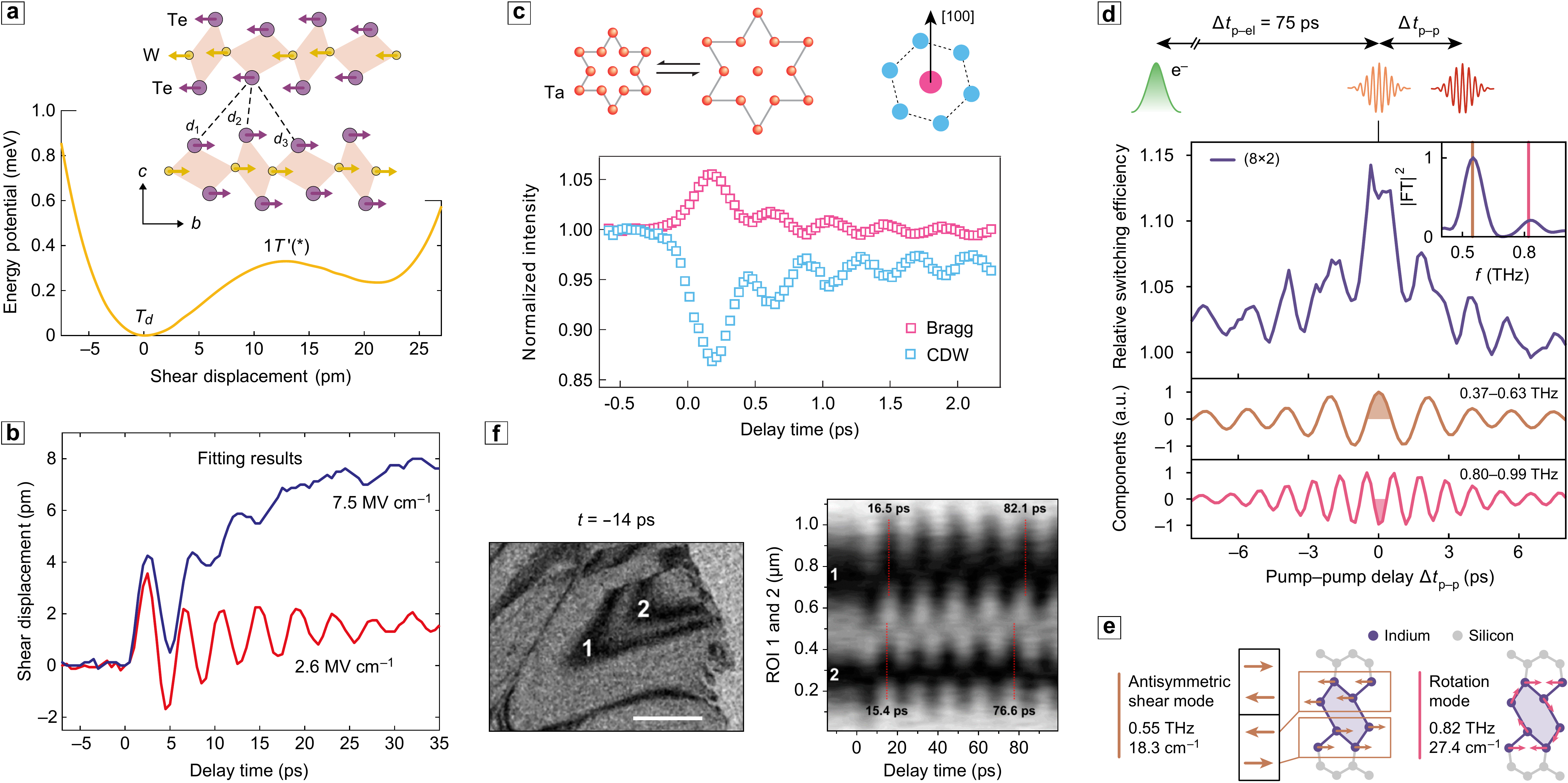}
	\caption{\textbf{Coherent manipulation via ultrafast excitation.} (a)~Calculated energy potential for different interlayer shear displacement in WTe$_2$, ranging from its $T_d$ to 1$T'(^*)$ polytypes. Inset shows a schematic of the shear motion, where positive displacement is defined as towards $d_1 < d_3$. (b)~Measured shear displacement from intensity change in electron diffraction after excitation by a 23~THz pulse at two different field strengths. The blue curve suggests a transient trapping into the metastable 1$T'(^*)$ state.	(c)~Photoinduced amplitude mode oscillation of the commensurate CDW in 1$T$-TaS$_2$, showing a perfect $\pi$ phase shift between the intensity modulation of Bragg and CDW peaks. The amplitude mode is characterized by a breathing motion of 12 Ta atoms relative to a central Ta atom (top left). A schematic of the Bragg and first-order CDW peaks projected to the (001) plane is shown on the top right. (d)~Double-pulse coherent control of the $(8\times2)$ to $(4\times1)$ transition in atomic indium wires grown on the (111) surface of silicon. Two photoexcitation pulses (1030~nm and 800~nm) are temporarily varied by $\Delta t_{\text{p-p}}$ while the probing electron pulse is fixed at 75~ps after the second pump pulse, when the fast transients have already relaxed into a metastable state. Top graph shows the $(8\times2)\rightarrow(4\times1)$ switching efficiency while the bottom two plots are the Fourier-filtered components. Inset shows the spectral density of the switching efficiency. (e)~Schematic of the two phonon modes that correspond to the two spectral contents in (d). (f)~Few-layer dephasing of photoinduced strain waves in MoS$_2$. A bright-field electron micrograph featuring two bend contours that are separated by a crystal step edge, imaged at 14~ps prior to photoexcitation. Scale bar: 500~nm. A space-time contour plot of the intensity modulation is shown for the two bend contours, where the first and the fifth periods are marked by vertical lines. Adapted from ref.~\cite{Sie2019} (a,b), ref.~\cite{Zong2018} (c), ref.~\cite{Horstmann2020} (d,e) and ref.~\cite{Zhang2019} (f).
	}
\label{fig:coherent}
\end{figure*}

In the context of UED and UEM, the most readily detected coherent modes are phonons. Besides resonant driving through a terahertz or mid-infrared pulse \cite{Subedi2014}, coherent \emph{optical} phonons can be generated by impulsive stimulated Raman scattering \cite{Stevens2002}, displacive excitation \cite{Zeiger1992}, or transient depletion field screening \cite{Forst2007}. For coherent \emph{acoustic} phonons, they may instead result from electron-phonon deformation potential, thermoelasticity, the inverse piezoelectric effect, electrostriction, and magnetostriction \cite{Ruello2015,KorffSchmising2008}. Early studies of coherent phonons in UED were focused on elemental thin films such as aluminum \cite{Nie2006}, silicon \cite{Harb2009}, bismuth \cite{Bugayev2011,Moriena2012}, and graphite \cite{Park2009,Chatelain2014}, revealing important information such as the Gr\"{u}neisen parameter of a material. Here, we review more recent studies in which coherent phonons are driven to large amplitude so that the initial harmonic motion can transform into a plastic distortion, giving rise to a long-lasting metastable state. Via a multi-pulse sequence, the coherent motion can also be amplified or suppressed with femtosecond precision, allowing one to manipulate the nonequilibrium pathway of a photoinduced transition.

A notable example is the light-induced topological transition from a type-II Weyl semimetal to a normal semimetal in WTe$_2$ \cite{Sie2019}. As illustrated in Fig.~\ref{fig:coherent}(a), the transition originates from a shear motion between van der Waals-bonded layers, leading to a structural change from the noncentrosymmetric $T_d$ ground state towards a centrosymmetric orthorhombic 1$T'(^*)$ phase. In a UED experiment, the coherent shear mode was evidenced by a periodic modulation of the Bragg peak intensity. Through a structure factor calculation, Sie \textit{et al.} were able to extract the exact atomic displacement down to picometer precision [Fig.~\ref{fig:coherent}(b)]. In the UED measurement, intense terahertz pulses were used (1.5~THz or 23~THz) to instigate the coherent shear mode, though excitation at higher photon energy up to 1.55~eV was shown to exhibit similar phenomenology \cite{Zhang2019b}. For very high excitation field (7.5~MV/cm), the oscillatory motion starts to deviate from its equilibrium position and slowly morphs into a metastable state over 25~ps [blue curve in Fig.~\ref{fig:coherent}(b)]. This trend signifies a large-amplitude, anharmonic lattice motion that drives WTe$_2$ into a new polytype, 1$T'(^*)$, which persists for more than 70~ps. As an independent check for the photoinduced symmetry switch, rotation anisotropy-second harmonic generation (RA-SHG) measurement was carried out before and after $T_d$-WTe$_2$ was photoexcited. The large RA-SHG signal from an inversion symmetry-broken $T_d$ state almost completely vanishes in all polarization channels upon photoexcitation, lending further support to a scenario that WTe$_2$ enters a topologically trivial $1T'(^*)$ phase with restored inversion symmetry.

With significant progress in temporal resolution, coherent optical phonons with frequencies up to several terahertz can be resolved by UED at present. A special type of optical phonon is the amplitude mode of a CDW phase, which describes the atomic motion along the trajectory that connects the CDW and the high-temperature states. A hallmark of the amplitude mode is the anti-phase relation of intensity modulation between the CDW superlattice peak and the crystal Bragg peak, which was first visualized by UED in a paradigmatic CDW material, 1$T$-TaS$_2$ [Fig.~\ref{fig:coherent}(c)] \cite{Zong2018}. In this case, the amplitude mode is a breathing motion of twelve Ta atoms surrounding a central Ta atom, which carries a characteristic frequency of 2.4~THz at 40~K. The use of UED is instrumental in this measurement because both Bragg and superlattice peaks need to be captured within the same detector frame to directly compare their phase relation in the oscillatory dynamics. At sufficiently high pumping fluence, it is possible that atoms traveling along their amplitude mode coordinate eventually ``overshoot'' across the high-symmetry position, entering an inverted CDW state. This scenario was first postulated in K$_{0.3}$MoO$_3$ \cite{Huber2014} and subsequently proposed for SmTe$_3$ \cite{Trigo2019} and 1$T$-TaSe$_2$ \cite{Zhang2020}. As the inversion only takes place above a threshold fluence, an intriguing possibility arises due to different absorbed fluences at different depths of the sample, leading to a domain wall between the normal and the inverted CDW states \cite{Yusupov2010,Wang2019,Trigo2021,Duan2021}. This domain wall is shown to exist in classic CDW systems such as rare-earth tritellurides \cite{Yusupov2010,Trigo2021} and 1$T$-TiSe$_2$ \cite{Duan2021}, and its generation is expected to be generic in other broken-symmetry states possessing an amplitude mode. As the CDW locally collapses at the domain wall, the persistent nature of the domain wall would allow the proliferation of other phases that compete with the equilibrium CDW, possibly leading to novel photoinduced orders.

Leveraging a well-defined frequency of a coherently excited phonon mode, one can further design a pulse sequence to achieve a surgical control over the photoinduced phase transition and the resulting metastable state. Horstmann \textit{et al.} demonstrated such an example in atomic indium wires grown on the (111) surface of silicon \cite{Horstmann2020}. At room temperature, the atomic wire is metallic and forms a $(4 \times 1)$ superstructure. It undergoes a first-order Peierls-like transition into an insulating $(8 \times 2)$ state upon cooling \cite{Snijders2010}. The reverse $(8 \times 2)\rightarrow (4 \times 1)$ transition can be instigated by an ultrafast light pulse \cite{Wall2012,Frigge2017,Horstmann2020}; owing to the first-order character, the system can be trapped in a $(4 \times 1)$ metastable state that persists over nanoseconds. Horstmann \textit{et al.} studied this photoinduced transition with two pump pulses that are delayed by $\Delta t_{\text{p-p}}$ and recorded the excited structure with an electron pulse at 75~ps after the second pump pulse, when the system has already relaxed into the metastable $(4 \times 1)$ phase [see schematic in Fig.~\ref{fig:coherent}(d)] \cite{Horstmann2020}. It was discovered that the efficiency of switching from the $(8 \times 2)$ ground state into the $(4 \times 1)$ metastable phase sensitively depends on the pump-pump temporal separation, exhibiting a pronounced oscillation as $\Delta t_{\text{p-p}}$ is varied [Fig.~\ref{fig:coherent}(d)]. Based on the Fourier transform of the switching efficiency, two spectral components are found to be responsible for the oscillation, indicated by the brown and red curves in Fig.~\ref{fig:coherent}(d). They correspond to an antisymmetric shear mode and a rotational mode of the indium hexagons -- shown in Fig.~\ref{fig:coherent}(e) -- which constitute the principal atomic motions for the $(8 \times 2)\rightarrow (4 \times 1)$ transition. The identification of these driving phonons allowed the authors to harness their long-lasting vibrational coherence for phase engineering, opening the possibility of targeting specific modes to steer the phase transition pathway.

The coherent atomic motion triggered by an ultrafast light pulse can also be directly visualized in real space through UEM. With sub-100-nm spatial resolution and sub-picosecond temporal resolution, UEM has played an instrumental role in studying the photo-generation and propagation of acoustic waves, which locally modify the strain and hence the electronic properties in a wide class of materials \cite{Park2009,Zewail2010,Zhang2019,Cremons2016,McKenna2017,Feist2018,Kim2019,Nakamura2020,Kim2020}. While UED obtains spatially averaged information that could contain multiple frequencies from several coherent modes, UEM is capable of identifying the spatial origins for individual components. For example, Zhang and Flannigan examined a mechanically exfoliated van der Waals material, 2$H$-MoS$_2$, where two bend contours [1 and 2 in Fig.~\ref{fig:coherent}(f)] are separated by a crystal step edge \cite{Zhang2019}. Due to the different layer numbers at positions 1 and 2, photoinduced strain waves traveling back and forth across the layers acquire different round-trip times, leading to a gradual dephasing of the oscillatory signal between the two regions [Fig.~\ref{fig:coherent}(f)]. This experiment also reminds us that realistic materials are often characterized by defects, terraces, step edges, layer twists, or wrinkles, so the spatial resolution afforded by UEM is critical for studying their effects on the light-induced states.

\section{Defect generation}

One of the central questions pertaining to phase transitions triggered with light concerns how these transitions differ from those in thermal equilibrium. In the past few years, it has become clear that light-triggered phase transitions can give rise to topological defects, even in materials where the density of defects is negligible in thermal equilibrium. Topological defects are characterized by a region in space where the amplitude of the order parameter vanishes at a ``core''. Depending on the broken symmetry, these defects can be points, lines, domain walls or consist of more exotic order parameter textures \cite{Srivastava2001}. Engineering these defects with light gives rise to the possibility of controlling material properties in a metastable fashion and realizing unconventional phase transitions that are not present in thermal equilibrium \cite{Stojchevska2014,Zong2018,Gerasimenko2019,Stoica2019,Buttner2021}.

The importance of topological defects to nonequilibrium phase transitions was originally recognized in a cosmological context. Kibble predicted that cosmic strings would be left over from the rapid expansion of the early universe. Soon after, the idea was adapted by Zurek to look at phase transitions in a condensed matter setting where a rapid thermal quench from a high-temperature disordered state to a low-temperature ordered one would similarly give rise to topological defects. Quantitatively, Zurek suggested that the density of topological defects in the ordered phase would scale with the quench time, exhibiting characteristic critical exponents~\cite{Kibble2007}. Although light-induced phase transitions are qualitatively different from thermal quenching transitions (in that photons are usually used to excite electrons to high energy scattering states), a variety of experiments suggest that topological defects may play a significant role in light-triggered phase transitions as well.

\begin{figure*}[htb!]
	\centering
	\includegraphics[width=0.85\textwidth]{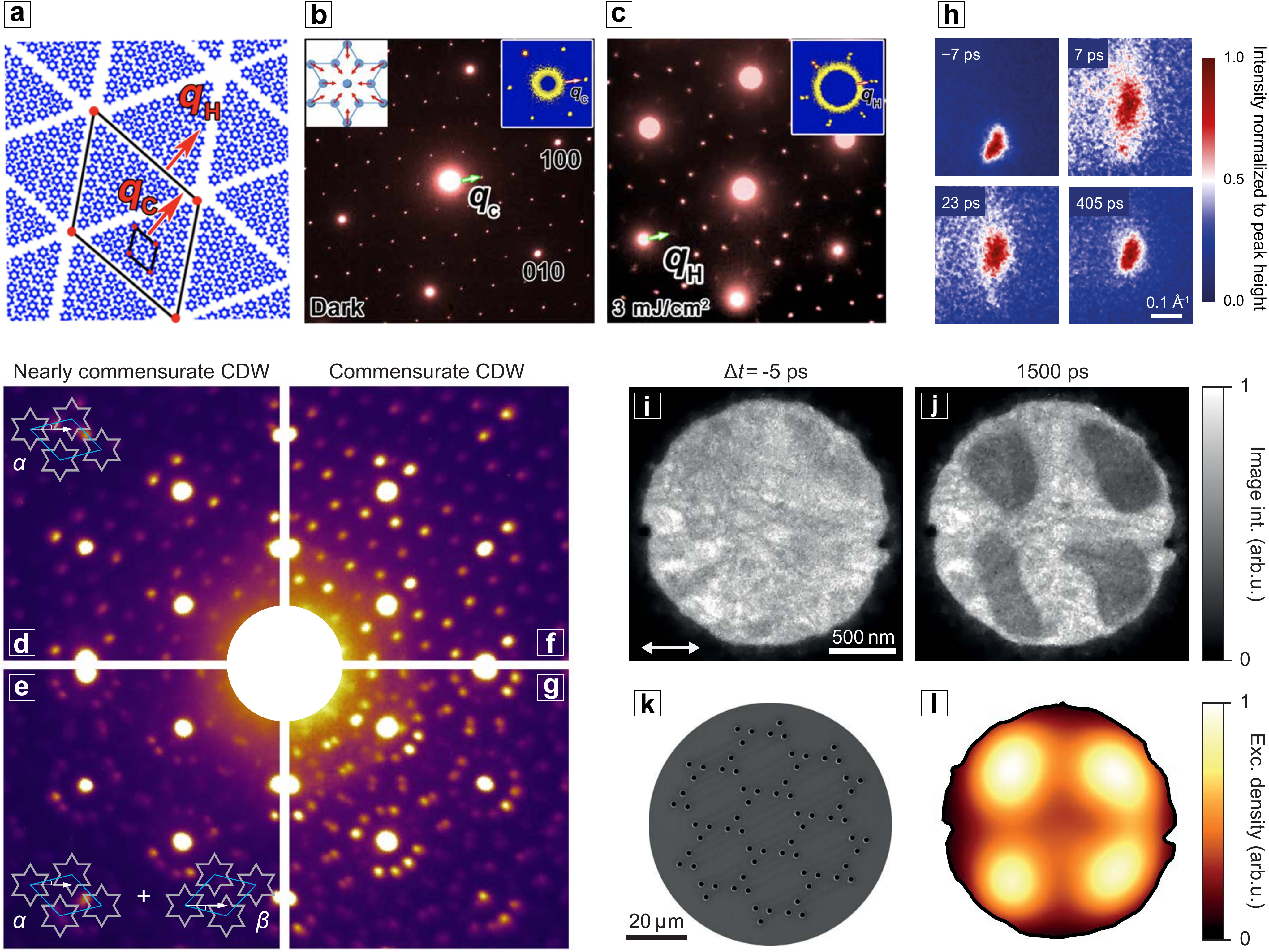}
	\caption{\textbf{Material control via photoinduced defects.} (a)~Simulated real-space pattern of the light-induced ``hidden'' state in the commensurate CDW phase of 1$T$-TaS$_2$, which corresponds to the electron diffraction pattern in (c). Each star corresponds to one Star-of-David polaron unit. The photoinduced incommensurability is $\delta=1/9$, with $\mathbf{q}_H=(1-\delta)\mathbf{q}_c$. (b)~Equilibrium electron diffraction pattern of 1$T$-TaS$_2$ at 10~K in the commensurate CDW phase, where top left inset shows a real-space Star-of-David polaron unit for the superlattice with modulation wavevector $\mathbf{q}_c$. (c)~Diffraction pattern of the same sample after femtosecond pulse excitation with 3~mJ/cm$^2$ fluence. Inset shows an enlarged view surrounding a Bragg peak, illustrating an additional superlattice modulation at vector $\mathbf{q}_H$. (d,e)~Switchable planar chiral domains of nearly-commensurate CDW in 1$T$-TaS$_2$ at room temperature. The electron diffraction pattern of a single planar chirality ($\alpha$) is shown in (d), while that of mixed chiralities ($\alpha+\beta$) is in (e). The two states can be reversibly switched with a single 80-fs, 800-nm pulse at incident fluence of 7~mJ/cm$^2$. Thermal annealing returns the sample to state (d). Insets show the real space arrangement of Stars-of-David that constitute the CDW state at room temperature. (f,g)~Corresponding diffraction patterns for (d) and (e), respectively, upon cooling the sample to 40~K in the commensurate CDW state. These two states can no longer be switched into the other by a single pulse up to the highest fluence attempted (11~mJ/cm$^2$). (h)~Evolution of CDW diffraction peak in 1$T$-TaS$_2$ during the photoinduced nearly commensurate to incommensurate transition, showing the narrowing of the incommensurate peak after photoexcitation. (i,j)~Dark field electron micrograph of the nearly commensurate CDW in 1$T$-TaS$_2$ before photoexcitation (i) and domains of incommensurate CDW [dark patches in (j)] at 1,500~ps after the incidence of an 80-fs, 800-nm, 2.6-mJ/cm$^2$ pulse with linear polarization indicated by the double-headed arrow. (k)~Mask for ultrafast dark field imaging, allowing the passage of second-order nearly-commensurate CDW peaks. (l)~Photoexcitation profile for producing the domain structure in (j). Adapted from ref.~\cite{Sun2018} (a--c), ref.~\cite{Zong2018} (d--g), ref.~\cite{Vogelgesang2018}~(h) and ref.~\cite{Danz2021}~(i--l).}
\label{fig:defect}
\end{figure*}

One of the best examples of light-induced defects is seen in the charge density wave state of 1$T$-TaS$_2$. This compound and its isovalent variants have been extensively studied by UED and UEM in recent years \cite{Eichberger2010,Erasmus2012,Sun2015,Han2015,Haupt2016,Wei2017,LeGuyader2017,Zong2018,Vogelgesang2018,Li2019,Ji2020,Danz2021}. Here, we restrict our discussion to cases where light-induced defects lead to newfound states of matter. In 2014, Stojchevska \textit{et al.} suggested that defects were key to the stable and reversible control over an insulator-to-metal transition \cite{Stojchevska2014}. In this experiment, the authors used a single 1--3~mJ/cm$^2$, 800~nm femtosecond light pulse to convert the low-temperature insulating state of 1$T$-TaS$_2$ into a ``hidden'' metastable metal. At temperatures below 4~K, the metallic state was present for over a week. A train of 10,000 subsequent pulses was then able to drive this metastable metal back into an insulator. Below 180~K, 1$T$-TaS$_2$ possesses a commensurate charge density wave, whose real space structure consists of tiled Star-of-David hexagrams with a $\sqrt{13} \times \sqrt{13}$ superstructure [Fig.~\ref{fig:coherent}(c)]. The authors suggested that the metallicity arose out of ordered domain walls within this CDW texture, which has a configuration similar to that shown in Fig.~\ref{fig:defect}(a). Subsequent electron diffraction studies, shown in Fig.~\ref{fig:defect}(b,c), confirmed their suspicions. Figure~\ref{fig:defect}(b) shows the equilibrium diffraction pattern in 1$T$-TaS$_2$ at 10~K, where superlattice peaks arising from commensurate charge density wave are easily visible. Once the system is photoexcited with a single pulse, the CDW peaks split into pairs of peaks along the CDW wavevector, characteristic of their predicted pattern of ordered domain walls. Although electron diffraction used in this way is not ``ultrafast'', the physics demonstrated in this set of experiments is rather remarkable in the degree of control over the material properties. It should be emphasized that the ``hidden'' state is nowhere to be found on the equilibrium phase diagram and is thus a true nonequilibrium ordered state with no equilibrium counterpart \cite{Vaskivskyi2015,Cho2016,Ma2016}.

These works on the ``hidden'' low-temperature state in 1$T$-TaS$_2$ have also spawned other experiments looking at the same material in different temperature regimes where 1$T$-TaS$_2$ possesses different phases. Notably, another qualitatively different switching phenomenon in 1$T$-TaS$_2$ was found in its room temperature nearly-commensurate (NC) CDW state. In equilibrium, the nearly-commensurate state consists again of commensurate patches separated by ordered domain walls. However, the ordering of the domain walls is completely different from that in the ``hidden'' state. In the NC-CDW state, CDW domains are roughly hexagonal, instead of the triangular ones observed in the ``hidden'' state, with larger regions of zero order parameter amplitude \cite{Spijkerman1997}. Of particular importance is that the NC-CDW state possesses broken mirror symmetry stemming from how the hexagrams are tiled. The inset of Fig.~\ref{fig:defect}(e) shows the two possible domain orientations, which are labeled $\alpha$ and $\beta$. Zong \textit{et al.} used a single $\sim7$~mJ/cm$^2$, 800~nm femtosecond light pulse to trigger a stable conversion to a qualitatively different diffraction pattern \cite{Zong2018}. A single light pulse converted the diffraction pattern from that of Fig.~\ref{fig:defect}(d) to that in Fig.~\ref{fig:defect}(e). Again, the altered pattern was stable for an indefinite period of time. A subsequent identical pulse sent the diffraction pattern back into its original state [Fig.~\ref{fig:defect}(d)]. At the moment, however, this switching behavior, unlike that in the ``hidden'' state, is not perfectly controllable. It is currently probabilistic in nature and sometimes only switches with the application of two or three pulses. Due to the stability of this new state at room temperature, the sample was examined with a conventional transmission electron microscope to map its real space texture. These latter studies revealed that the laser pulse had instigated the growth of domains of opposite mirror symmetry, which were separated by rigid domain walls, again demonstrating the importance of topological defects in changing the material properties in a metastable fashion. With both $\alpha$ and $\beta$ domains present, the authors cooled the sample down into the commensurate CDW state, showing that the domains persisted and that the diffraction peaks get more intense [Fig.~\ref{fig:defect}(g)]. At low temperatures, the kinetic barrier cannot be surmounted without damaging the sample, and it thus cannot be returned to its pristine state [Fig.~\ref{fig:defect}(f)]. It remains to be seen whether the $\alpha/\beta$-domains can coexist with the metastable metallic ``hidden'' state, though there are no conceptual issues preventing such a ``double metastability''.

Though the description of topological defects up to now have concentrated on their stability, other studies have focused on their dynamics. Using a newly developed technique, ultrafast low-energy electron diffraction (ULEED) with a nanometer-sized emission tip, Vogelgesang \textit{et al.} measured the peak width in 1$T$-TaS$_2$ as a function of time after exciting the material from its nearly-commensurate state into an incommensurate state \cite{Vogelgesang2018}. The low energy of the electrons (50--100~eV) permits a much better momentum resolution ($\Delta k_s = 0.03~\text{\AA}^{-1}$) compared to high-energy electron diffraction setups, making it possible to more accurately extract the correlation lengths of the CDW based on the peak width. In thermal equilibrium, the incommensurate state is present above room temperature at $\sim355$~K. In their work, the authors found that the incommensurate peak width narrows as a function of time after photoexcitation, shown in Fig.~\ref{fig:defect}(h), and they interpreted that the incommensurate phase is stabilized by the annihilation of topological defects. Such domain growth is referred to as domain coarsening, and the authors found a characteristic scaling behavior between the domain size and the time after photoexcitation, suggesting that such coarsening dynamics may be universal \cite{Laulhe2017,Bray1994}. These domain dynamics were further studied using an inhomogeneous excitation pulse profile [Fig.~\ref{fig:defect}(l)] and imaged in real space using time-resolved dark-field microscopy with a tailor-made mask [Fig.~\ref{fig:defect}(k)] \cite{Danz2021}. Figure~\ref{fig:defect}(i,j) shows the imaged sample before and after photoexcitation, where the incommensurate regions appear, interact, shrink, and then disappear. Although such methods currently do not have the spatial resolution to visualize individual defects, it seems like this may be possible in the future. Overall, these case studies present us with the view that topological defects are a key component in stabilizing thermodynamically inaccessible phases and in governing the kinetics of light-induced phase transitions. Ultrafast electron diffraction and microscopy will no doubt continue to play a significant role in the visualization of these defect dynamics.

\section{Conclusion and outlook}

To realize a photoinduced state on demand, the most common tuning parameters in a UED/UEM experiment are pump photon flux, wavelength, polarization, pulse duration, and sample temperature. With advances in thin flake fabrication techniques, a current trend is to integrate more sophisticated perturbing fields that provide additional degrees of freedom to manipulate material properties \textit{in situ}. A recent example is realized in the study of VO$_2$ \cite{Sood2021}, where metal electrodes were patterned and deposited on a 60-nm-thick polycrystalline film via photolithography and electron-beam evaporation. This setup enables concurrent UED and transport measurements to characterize the photoinduced insulator-to-metal transition, and it further allows the application of an electrical pulse to trigger a similar metastable state. Another fruitful avenue is to incorporate a strain device that modifies the equilibrium lattice constants. It was recently shown that the two competing CDWs in $R$Te$_3$ can be selectively enhanced depending on the direction of an applied uniaxial stress \cite{Straquadine2020}, raising the intriguing question of whether the lifetime of the light-induced CDW can be adjusted by strain engineering. Although it is challenging to apply a significant stress on a nanometer-thick  flake without buckling or sample damage, a strain device may be adopted in a ULEED setup for bulk crystals, as long as the stray electric field from the piezoelectric control is properly shielded from the low-energy electron probe.

If the photoinduced transition is irreversible or the metastable state does not relax before the arrival of the next pump pulse, a single-shot UED/UEM experiment is necessary. The leading challenge is to pack sufficient electrons within a single pulse without jeopardizing the spatiotemporal profile due to the space charge effect. Various schemes have been implemented in the past two decades \cite{Siwick2003,Sciaini2009,Tokita2010,Musumeci2010,Li2010,Speirs2015,Mo2018} and most studies focus on the photoinduced melting of elemental films. With improved momentum resolution and signal-to-noise ratio, the single-shot scheme may be employed to answer more complex questions, such as solving the exact atomic trajectory towards the metastable ``hidden'' or ``jammed'' state in 1$T$-TaS$_2$ \cite{Stojchevska2014,Gerasimenko2019}.

To obtain real-space visualization of a photoinduced state, the dark-field imaging setup by Danz \textit{et al.} may be extended to other systems \cite{Danz2021}. However, the mask needs to be tailored for each sample and it requires precision positioning to sieve through diffraction peaks of interest. An area for development is to have an \textit{in situ} programmable mask, in the same spirit as a spatial light modulator but for electrons instead. A crude implementation is to use a material with low melting point -- such as elemental indium or bismuth -- coated on a silicon nitride support film. The excitation pulse with an appropriate fluence may be focused to drill tiny holes on the metallic film to produce a diffraction mask, and subsequently the holes could be erased via a temperature cycle for repeated usage.

We also envision continued research effort on electron pulse compression to push towards sub-10-fs total resolution using, for example, particle accelerator technology \cite{Wang1996,Wang2003,Weathersby2015,Kim2019b,Qi2020}. In addition, improvements in signal-to-noise ratios can be leveraged to visualize diffuse signals away from Bragg peaks as described by D\"{u}rr \textit{et al.} in this MRS Bulletin special issue \cite{Durr2021}. These developments would enable the discovery of novel short-lived states, high-frequency collective modes, as well as unconventional pathways of energy transfer in an excited medium. Together, they would make UED and UEM indispensable tools for discovering and shaping nonequilibrium properties of matter.\\

\noindent{\textbf{Acknowledgments}} We acknowledge the support from Gordon and Betty Moore Foundation's EPiQS Initiative grant GBMF9459 (manuscript writing). A.Z. acknowledges support from the Miller Institute for Basic Research in Science.


\begin{thebibliography}{100}%
\makeatletter
\providecommand \@ifxundefined [1]{%
 \@ifx{#1\undefined}
}%
\providecommand \@ifnum [1]{%
 \ifnum #1\expandafter \@firstoftwo
 \else \expandafter \@secondoftwo
 \fi
}%
\providecommand \@ifx [1]{%
 \ifx #1\expandafter \@firstoftwo
 \else \expandafter \@secondoftwo
 \fi
}%
\providecommand \natexlab [1]{#1}%
\providecommand \enquote  [1]{``#1''}%
\providecommand \bibnamefont  [1]{#1}%
\providecommand \bibfnamefont [1]{#1}%
\providecommand \citenamefont [1]{#1}%
\providecommand \href@noop [0]{\@secondoftwo}%
\providecommand \href [0]{\begingroup \@sanitize@url \@href}%
\providecommand \@href[1]{\@@startlink{#1}\@@href}%
\providecommand \@@href[1]{\endgroup#1\@@endlink}%
\providecommand \@sanitize@url [0]{\catcode `\\12\catcode `\$12\catcode
  `\&12\catcode `\#12\catcode `\^12\catcode `\_12\catcode `\%12\relax}%
\providecommand \@@startlink[1]{}%
\providecommand \@@endlink[0]{}%
\providecommand \url  [0]{\begingroup\@sanitize@url \@url }%
\providecommand \@url [1]{\endgroup\@href {#1}{\urlprefix }}%
\providecommand \urlprefix  [0]{URL }%
\providecommand \Eprint [0]{\href }%
\providecommand \doibase [0]{https://doi.org/}%
\providecommand \selectlanguage [0]{\@gobble}%
\providecommand \bibinfo  [0]{\@secondoftwo}%
\providecommand \bibfield  [0]{\@secondoftwo}%
\providecommand \translation [1]{[#1]}%
\providecommand \BibitemOpen [0]{}%
\providecommand \bibitemStop [0]{}%
\providecommand \bibitemNoStop [0]{.\EOS\space}%
\providecommand \EOS [0]{\spacefactor3000\relax}%
\providecommand \BibitemShut  [1]{\csname bibitem#1\endcsname}%
\let\auto@bib@innerbib\@empty
\bibitem [{\citenamefont {de~la Torre}\ \emph {et~al.}(2021)\citenamefont
  {de~la Torre}, \citenamefont {Kennes}, \citenamefont {Claassen},
  \citenamefont {Gerber}, \citenamefont {McIver},\ and\ \citenamefont
  {Sentef}}]{delaTorre2021}%
  \BibitemOpen
  \bibfield  {author} {\bibinfo {author} {\bibfnamefont {A.}~\bibnamefont
  {de~la Torre}}, \bibinfo {author} {\bibfnamefont {D.~M.}\ \bibnamefont
  {Kennes}}, \bibinfo {author} {\bibfnamefont {M.}~\bibnamefont {Claassen}},
  \bibinfo {author} {\bibfnamefont {S.}~\bibnamefont {Gerber}}, \bibinfo
  {author} {\bibfnamefont {J.~W.}\ \bibnamefont {McIver}},\ and\ \bibinfo
  {author} {\bibfnamefont {M.~A.}\ \bibnamefont {Sentef}},\ }\href@noop {}
  {\bibinfo {title} {{Nonthermal pathways to ultrafast control in quantum
  materials}}} (\bibinfo {year} {2021}),\ \Eprint
  {https://arxiv.org/abs/2103.14888} {arXiv:2103.14888} \BibitemShut {NoStop}%
\bibitem [{\citenamefont {Eichberger}\ \emph {et~al.}(2010)\citenamefont
  {Eichberger}, \citenamefont {Sch{\"a}fer}, \citenamefont {Krumova},
  \citenamefont {Beyer}, \citenamefont {Demsar}, \citenamefont {Berger},
  \citenamefont {Moriena}, \citenamefont {Sciaini},\ and\ \citenamefont
  {Miller}}]{Eichberger2010}%
  \BibitemOpen
  \bibfield  {author} {\bibinfo {author} {\bibfnamefont {M.}~\bibnamefont
  {Eichberger}}, \bibinfo {author} {\bibfnamefont {H.}~\bibnamefont
  {Sch{\"a}fer}}, \bibinfo {author} {\bibfnamefont {M.}~\bibnamefont
  {Krumova}}, \bibinfo {author} {\bibfnamefont {M.}~\bibnamefont {Beyer}},
  \bibinfo {author} {\bibfnamefont {J.}~\bibnamefont {Demsar}}, \bibinfo
  {author} {\bibfnamefont {H.}~\bibnamefont {Berger}}, \bibinfo {author}
  {\bibfnamefont {G.}~\bibnamefont {Moriena}}, \bibinfo {author} {\bibfnamefont
  {G.}~\bibnamefont {Sciaini}},\ and\ \bibinfo {author} {\bibfnamefont
  {R.~J.~D.}\ \bibnamefont {Miller}},\ }\bibfield  {title} {\bibinfo {title}
  {{Snapshots of cooperative atomic motions in the optical suppression of
  charge density waves}},\ }\href {https://doi.org/10.1038/nature09539}
  {\bibfield  {journal} {\bibinfo  {journal} {Nature}\ }\textbf {\bibinfo
  {volume} {468}},\ \bibinfo {pages} {799} (\bibinfo {year}
  {2010})}\BibitemShut {NoStop}%
\bibitem [{\citenamefont {Kaiser}(2017)}]{Kaiser2017}%
  \BibitemOpen
  \bibfield  {author} {\bibinfo {author} {\bibfnamefont {S.}~\bibnamefont
  {Kaiser}},\ }\bibfield  {title} {\bibinfo {title} {{Light-induced
  superconductivity in high-$T_c$ cuprates}},\ }\href
  {https://doi.org/10.1088/1402-4896/aa8201} {\bibfield  {journal} {\bibinfo
  {journal} {Phys. Scr.}\ }\textbf {\bibinfo {volume} {92}},\ \bibinfo {pages}
  {103001} (\bibinfo {year} {2017})}\BibitemShut {NoStop}%
\bibitem [{\citenamefont {Sun}\ and\ \citenamefont {Millis}(2020)}]{Sun2020}%
  \BibitemOpen
  \bibfield  {author} {\bibinfo {author} {\bibfnamefont {Z.}~\bibnamefont
  {Sun}}\ and\ \bibinfo {author} {\bibfnamefont {A.~J.}\ \bibnamefont
  {Millis}},\ }\bibfield  {title} {\bibinfo {title} {{Transient trapping into
  metastable states in systems with competing orders}},\ }\href
  {https://doi.org/10.1103/PhysRevX.10.021028} {\bibfield  {journal} {\bibinfo
  {journal} {Phys. Rev. X}\ }\textbf {\bibinfo {volume} {10}},\ \bibinfo
  {pages} {021028} (\bibinfo {year} {2020})}\BibitemShut {NoStop}%
\bibitem [{\citenamefont {Kogar}\ \emph {et~al.}(2020)\citenamefont {Kogar},
  \citenamefont {Zong}, \citenamefont {Dolgirev}, \citenamefont {Shen},
  \citenamefont {Straquadine}, \citenamefont {Bie}, \citenamefont {Wang},
  \citenamefont {Rohwer}, \citenamefont {Tung}, \citenamefont {Yang},
  \citenamefont {Li}, \citenamefont {Yang}, \citenamefont {Weathersby},
  \citenamefont {Park}, \citenamefont {Kozina}, \citenamefont {Sie},
  \citenamefont {Wen}, \citenamefont {Jarillo-Herrero}, \citenamefont {Fisher},
  \citenamefont {Wang},\ and\ \citenamefont {Gedik}}]{Kogar2020}%
  \BibitemOpen
  \bibfield  {author} {\bibinfo {author} {\bibfnamefont {A.}~\bibnamefont
  {Kogar}}, \bibinfo {author} {\bibfnamefont {A.}~\bibnamefont {Zong}},
  \bibinfo {author} {\bibfnamefont {P.~E.}\ \bibnamefont {Dolgirev}}, \bibinfo
  {author} {\bibfnamefont {X.}~\bibnamefont {Shen}}, \bibinfo {author}
  {\bibfnamefont {J.}~\bibnamefont {Straquadine}}, \bibinfo {author}
  {\bibfnamefont {Y.-Q.}\ \bibnamefont {Bie}}, \bibinfo {author} {\bibfnamefont
  {X.}~\bibnamefont {Wang}}, \bibinfo {author} {\bibfnamefont {T.}~\bibnamefont
  {Rohwer}}, \bibinfo {author} {\bibfnamefont {I.-C.}\ \bibnamefont {Tung}},
  \bibinfo {author} {\bibfnamefont {Y.}~\bibnamefont {Yang}}, \bibinfo {author}
  {\bibfnamefont {R.}~\bibnamefont {Li}}, \bibinfo {author} {\bibfnamefont
  {J.}~\bibnamefont {Yang}}, \bibinfo {author} {\bibfnamefont {S.}~\bibnamefont
  {Weathersby}}, \bibinfo {author} {\bibfnamefont {S.}~\bibnamefont {Park}},
  \bibinfo {author} {\bibfnamefont {M.~E.}\ \bibnamefont {Kozina}}, \bibinfo
  {author} {\bibfnamefont {E.~J.}\ \bibnamefont {Sie}}, \bibinfo {author}
  {\bibfnamefont {H.}~\bibnamefont {Wen}}, \bibinfo {author} {\bibfnamefont
  {P.}~\bibnamefont {Jarillo-Herrero}}, \bibinfo {author} {\bibfnamefont
  {I.~R.}\ \bibnamefont {Fisher}}, \bibinfo {author} {\bibfnamefont
  {X.}~\bibnamefont {Wang}},\ and\ \bibinfo {author} {\bibfnamefont
  {N.}~\bibnamefont {Gedik}},\ }\bibfield  {title} {\bibinfo {title}
  {{Light-induced charge density wave in LaTe$_3$}},\ }\href
  {https://doi.org/10.1038/s41567-019-0705-3} {\bibfield  {journal} {\bibinfo
  {journal} {Nat. Phys.}\ }\textbf {\bibinfo {volume} {16}},\ \bibinfo {pages}
  {159} (\bibinfo {year} {2020})}\BibitemShut {NoStop}%
\bibitem [{\citenamefont {Zhou}\ \emph {et~al.}(2021)\citenamefont {Zhou},
  \citenamefont {Williams}, \citenamefont {Sun}, \citenamefont {Malliakas},
  \citenamefont {Kanatzidis}, \citenamefont {Kemper},\ and\ \citenamefont
  {Ruan}}]{Zhou2021}%
  \BibitemOpen
  \bibfield  {author} {\bibinfo {author} {\bibfnamefont {F.}~\bibnamefont
  {Zhou}}, \bibinfo {author} {\bibfnamefont {J.}~\bibnamefont {Williams}},
  \bibinfo {author} {\bibfnamefont {S.}~\bibnamefont {Sun}}, \bibinfo {author}
  {\bibfnamefont {C.~D.}\ \bibnamefont {Malliakas}}, \bibinfo {author}
  {\bibfnamefont {M.~G.}\ \bibnamefont {Kanatzidis}}, \bibinfo {author}
  {\bibfnamefont {A.~F.}\ \bibnamefont {Kemper}},\ and\ \bibinfo {author}
  {\bibfnamefont {C.-Y.}\ \bibnamefont {Ruan}},\ }\bibfield  {title} {\bibinfo
  {title} {{Nonequilibrium dynamics of spontaneous symmetry breaking into a
  hidden state of charge-density wave}},\ }\href
  {https://doi.org/10.1038/s41467-020-20834-5} {\bibfield  {journal} {\bibinfo
  {journal} {Nat. Commun.}\ }\textbf {\bibinfo {volume} {12}},\ \bibinfo
  {pages} {566} (\bibinfo {year} {2021})}\BibitemShut {NoStop}%
\bibitem [{\citenamefont {Ru}(2008)}]{RuThesis}%
  \BibitemOpen
  \bibfield  {author} {\bibinfo {author} {\bibfnamefont {N.}~\bibnamefont
  {Ru}},\ }\emph {\bibinfo {title} {{Charge Density Wave Formation in
  Rare-earth Tellurides}}},\ \href
  {https://web.stanford.edu/group/fisher/people/Nancy_Ru_thesis.pdf} {\bibinfo
  {type} {{Ph.D. thesis}}},\ \bibinfo  {school} {Stanford University}, \bibinfo
  {address} {Stanford} (\bibinfo {year} {2008})\BibitemShut {NoStop}%
\bibitem [{\citenamefont {Yusupov}\ \emph {et~al.}(2008)\citenamefont
  {Yusupov}, \citenamefont {Mertelj}, \citenamefont {Chu}, \citenamefont
  {Fisher},\ and\ \citenamefont {Mihailovic}}]{Yusupov2008}%
  \BibitemOpen
  \bibfield  {author} {\bibinfo {author} {\bibfnamefont {R.~V.}\ \bibnamefont
  {Yusupov}}, \bibinfo {author} {\bibfnamefont {T.}~\bibnamefont {Mertelj}},
  \bibinfo {author} {\bibfnamefont {J.~H.}\ \bibnamefont {Chu}}, \bibinfo
  {author} {\bibfnamefont {I.~R.}\ \bibnamefont {Fisher}},\ and\ \bibinfo
  {author} {\bibfnamefont {D.}~\bibnamefont {Mihailovic}},\ }\bibfield  {title}
  {\bibinfo {title} {{Single-particle and collective mode couplings associated
  with 1- and 2-directional electronic ordering in metallic \textit{R}Te$_3$
  (\textit{R}=Ho, Dy, Tb)}},\ }\href
  {https://doi.org/10.1103/PhysRevLett.101.246402} {\bibfield  {journal}
  {\bibinfo  {journal} {Phys. Rev. Lett.}\ }\textbf {\bibinfo {volume} {101}},\
  \bibinfo {pages} {24602} (\bibinfo {year} {2008})}\BibitemShut {NoStop}%
\bibitem [{\citenamefont {Yusupov}\ \emph {et~al.}(2010)\citenamefont
  {Yusupov}, \citenamefont {Mertelj}, \citenamefont {Kabanov}, \citenamefont
  {Brazovskii}, \citenamefont {Kusar}, \citenamefont {Chu}, \citenamefont
  {Fisher},\ and\ \citenamefont {Mihailovic}}]{Yusupov2010}%
  \BibitemOpen
  \bibfield  {author} {\bibinfo {author} {\bibfnamefont {R.}~\bibnamefont
  {Yusupov}}, \bibinfo {author} {\bibfnamefont {T.}~\bibnamefont {Mertelj}},
  \bibinfo {author} {\bibfnamefont {V.~V.}\ \bibnamefont {Kabanov}}, \bibinfo
  {author} {\bibfnamefont {S.}~\bibnamefont {Brazovskii}}, \bibinfo {author}
  {\bibfnamefont {P.}~\bibnamefont {Kusar}}, \bibinfo {author} {\bibfnamefont
  {J.-H.}\ \bibnamefont {Chu}}, \bibinfo {author} {\bibfnamefont {I.~R.}\
  \bibnamefont {Fisher}},\ and\ \bibinfo {author} {\bibfnamefont
  {D.}~\bibnamefont {Mihailovic}},\ }\bibfield  {title} {\bibinfo {title}
  {{Coherent dynamics of macroscopic electronic order through a symmetry
  breaking transition}},\ }\href {https://doi.org/10.1038/nphys1738} {\bibfield
   {journal} {\bibinfo  {journal} {Nat. Phys.}\ }\textbf {\bibinfo {volume}
  {6}},\ \bibinfo {pages} {681} (\bibinfo {year} {2010})}\BibitemShut {NoStop}%
\bibitem [{\citenamefont {Zong}\ \emph
  {et~al.}(2019{\natexlab{a}})\citenamefont {Zong}, \citenamefont {Kogar},
  \citenamefont {Bie}, \citenamefont {Rohwer}, \citenamefont {Lee},
  \citenamefont {Baldini}, \citenamefont {Erge{\c{c}}en}, \citenamefont
  {Yilmaz}, \citenamefont {Freelon}, \citenamefont {Sie}, \citenamefont {Zhou},
  \citenamefont {Straquadine}, \citenamefont {Walmsley}, \citenamefont
  {Dolgirev}, \citenamefont {Rozhkov}, \citenamefont {Fisher}, \citenamefont
  {Jarillo-Herrero}, \citenamefont {Fine},\ and\ \citenamefont
  {Gedik}}]{Zong2019a}%
  \BibitemOpen
  \bibfield  {author} {\bibinfo {author} {\bibfnamefont {A.}~\bibnamefont
  {Zong}}, \bibinfo {author} {\bibfnamefont {A.}~\bibnamefont {Kogar}},
  \bibinfo {author} {\bibfnamefont {Y.-Q.}\ \bibnamefont {Bie}}, \bibinfo
  {author} {\bibfnamefont {T.}~\bibnamefont {Rohwer}}, \bibinfo {author}
  {\bibfnamefont {C.}~\bibnamefont {Lee}}, \bibinfo {author} {\bibfnamefont
  {E.}~\bibnamefont {Baldini}}, \bibinfo {author} {\bibfnamefont
  {E.}~\bibnamefont {Erge{\c{c}}en}}, \bibinfo {author} {\bibfnamefont {M.~B.}\
  \bibnamefont {Yilmaz}}, \bibinfo {author} {\bibfnamefont {B.}~\bibnamefont
  {Freelon}}, \bibinfo {author} {\bibfnamefont {E.~J.}\ \bibnamefont {Sie}},
  \bibinfo {author} {\bibfnamefont {H.}~\bibnamefont {Zhou}}, \bibinfo {author}
  {\bibfnamefont {J.}~\bibnamefont {Straquadine}}, \bibinfo {author}
  {\bibfnamefont {P.}~\bibnamefont {Walmsley}}, \bibinfo {author}
  {\bibfnamefont {P.~E.}\ \bibnamefont {Dolgirev}}, \bibinfo {author}
  {\bibfnamefont {A.~V.}\ \bibnamefont {Rozhkov}}, \bibinfo {author}
  {\bibfnamefont {I.~R.}\ \bibnamefont {Fisher}}, \bibinfo {author}
  {\bibfnamefont {P.}~\bibnamefont {Jarillo-Herrero}}, \bibinfo {author}
  {\bibfnamefont {B.~V.}\ \bibnamefont {Fine}},\ and\ \bibinfo {author}
  {\bibfnamefont {N.}~\bibnamefont {Gedik}},\ }\bibfield  {title} {\bibinfo
  {title} {{Evidence for topological defects in a photoinduced phase
  transition}},\ }\href {https://doi.org/10.1038/s41567-018-0311-9} {\bibfield
  {journal} {\bibinfo  {journal} {Nat. Phys.}\ }\textbf {\bibinfo {volume}
  {15}},\ \bibinfo {pages} {27} (\bibinfo {year}
  {2019}{\natexlab{a}})}\BibitemShut {NoStop}%
\bibitem [{\citenamefont {Zong}\ \emph
  {et~al.}(2019{\natexlab{b}})\citenamefont {Zong}, \citenamefont {Dolgirev},
  \citenamefont {Kogar}, \citenamefont {Erge{\c{c}}en}, \citenamefont {Yilmaz},
  \citenamefont {Bie}, \citenamefont {Rohwer}, \citenamefont {Tung},
  \citenamefont {Straquadine}, \citenamefont {Wang}, \citenamefont {Yang},
  \citenamefont {Shen}, \citenamefont {Li}, \citenamefont {Yang}, \citenamefont
  {Park}, \citenamefont {Hoffmann}, \citenamefont {Ofori-Okai}, \citenamefont
  {Kozina}, \citenamefont {Wen}, \citenamefont {Wang}, \citenamefont {Fisher},
  \citenamefont {Jarillo-Herrero},\ and\ \citenamefont {Gedik}}]{Zong2019b}%
  \BibitemOpen
  \bibfield  {author} {\bibinfo {author} {\bibfnamefont {A.}~\bibnamefont
  {Zong}}, \bibinfo {author} {\bibfnamefont {P.~E.}\ \bibnamefont {Dolgirev}},
  \bibinfo {author} {\bibfnamefont {A.}~\bibnamefont {Kogar}}, \bibinfo
  {author} {\bibfnamefont {E.}~\bibnamefont {Erge{\c{c}}en}}, \bibinfo {author}
  {\bibfnamefont {M.~B.}\ \bibnamefont {Yilmaz}}, \bibinfo {author}
  {\bibfnamefont {Y.-Q.}\ \bibnamefont {Bie}}, \bibinfo {author} {\bibfnamefont
  {T.}~\bibnamefont {Rohwer}}, \bibinfo {author} {\bibfnamefont {I.-C.}\
  \bibnamefont {Tung}}, \bibinfo {author} {\bibfnamefont {J.}~\bibnamefont
  {Straquadine}}, \bibinfo {author} {\bibfnamefont {X.}~\bibnamefont {Wang}},
  \bibinfo {author} {\bibfnamefont {Y.}~\bibnamefont {Yang}}, \bibinfo {author}
  {\bibfnamefont {X.}~\bibnamefont {Shen}}, \bibinfo {author} {\bibfnamefont
  {R.}~\bibnamefont {Li}}, \bibinfo {author} {\bibfnamefont {J.}~\bibnamefont
  {Yang}}, \bibinfo {author} {\bibfnamefont {S.}~\bibnamefont {Park}}, \bibinfo
  {author} {\bibfnamefont {M.~C.}\ \bibnamefont {Hoffmann}}, \bibinfo {author}
  {\bibfnamefont {B.~K.}\ \bibnamefont {Ofori-Okai}}, \bibinfo {author}
  {\bibfnamefont {M.~E.}\ \bibnamefont {Kozina}}, \bibinfo {author}
  {\bibfnamefont {H.}~\bibnamefont {Wen}}, \bibinfo {author} {\bibfnamefont
  {X.}~\bibnamefont {Wang}}, \bibinfo {author} {\bibfnamefont {I.~R.}\
  \bibnamefont {Fisher}}, \bibinfo {author} {\bibfnamefont {P.}~\bibnamefont
  {Jarillo-Herrero}},\ and\ \bibinfo {author} {\bibfnamefont {N.}~\bibnamefont
  {Gedik}},\ }\bibfield  {title} {\bibinfo {title} {{Dynamical slowing-down in
  an ultrafast photoinduced phase transition}},\ }\href
  {https://doi.org/10.1103/PhysRevLett.123.097601} {\bibfield  {journal}
  {\bibinfo  {journal} {Phys. Rev. Lett.}\ }\textbf {\bibinfo {volume} {123}},\
  \bibinfo {pages} {097601} (\bibinfo {year} {2019}{\natexlab{b}})}\BibitemShut
  {NoStop}%
\bibitem [{\citenamefont {Schmitt}\ \emph {et~al.}(2008)\citenamefont
  {Schmitt}, \citenamefont {Kirchmann}, \citenamefont {Bovensiepen},
  \citenamefont {Moore}, \citenamefont {Rettig}, \citenamefont {Krenz},
  \citenamefont {Chu}, \citenamefont {Ru}, \citenamefont {Perfetti},
  \citenamefont {Lu}, \citenamefont {Wolf}, \citenamefont {Fisher},\ and\
  \citenamefont {Shen}}]{Schmitt2008}%
  \BibitemOpen
  \bibfield  {author} {\bibinfo {author} {\bibfnamefont {F.}~\bibnamefont
  {Schmitt}}, \bibinfo {author} {\bibfnamefont {P.~S.}\ \bibnamefont
  {Kirchmann}}, \bibinfo {author} {\bibfnamefont {U.}~\bibnamefont
  {Bovensiepen}}, \bibinfo {author} {\bibfnamefont {R.~G.}\ \bibnamefont
  {Moore}}, \bibinfo {author} {\bibfnamefont {L.}~\bibnamefont {Rettig}},
  \bibinfo {author} {\bibfnamefont {M.}~\bibnamefont {Krenz}}, \bibinfo
  {author} {\bibfnamefont {J.-H.}\ \bibnamefont {Chu}}, \bibinfo {author}
  {\bibfnamefont {N.}~\bibnamefont {Ru}}, \bibinfo {author} {\bibfnamefont
  {L.}~\bibnamefont {Perfetti}}, \bibinfo {author} {\bibfnamefont {D.~H.}\
  \bibnamefont {Lu}}, \bibinfo {author} {\bibfnamefont {M.}~\bibnamefont
  {Wolf}}, \bibinfo {author} {\bibfnamefont {I.~R.}\ \bibnamefont {Fisher}},\
  and\ \bibinfo {author} {\bibfnamefont {Z.-X.}\ \bibnamefont {Shen}},\
  }\bibfield  {title} {\bibinfo {title} {{Transient electronic structure and
  melting of a charge density wave in TbTe$_3$}},\ }\href
  {https://doi.org/10.1126/science.1160778} {\bibfield  {journal} {\bibinfo
  {journal} {Science}\ }\textbf {\bibinfo {volume} {321}},\ \bibinfo {pages}
  {1649} (\bibinfo {year} {2008})}\BibitemShut {NoStop}%
\bibitem [{\citenamefont {Schmitt}\ \emph {et~al.}(2011)\citenamefont
  {Schmitt}, \citenamefont {Kirchmann}, \citenamefont {Bovensiepen},
  \citenamefont {Moore}, \citenamefont {Chu}, \citenamefont {Lu}, \citenamefont
  {Rettig}, \citenamefont {Wolf}, \citenamefont {Fisher},\ and\ \citenamefont
  {Shen}}]{Schmitt2011}%
  \BibitemOpen
  \bibfield  {author} {\bibinfo {author} {\bibfnamefont {F.}~\bibnamefont
  {Schmitt}}, \bibinfo {author} {\bibfnamefont {P.~S.}\ \bibnamefont
  {Kirchmann}}, \bibinfo {author} {\bibfnamefont {U.}~\bibnamefont
  {Bovensiepen}}, \bibinfo {author} {\bibfnamefont {R.~G.}\ \bibnamefont
  {Moore}}, \bibinfo {author} {\bibfnamefont {J.-H.}\ \bibnamefont {Chu}},
  \bibinfo {author} {\bibfnamefont {D.~H.}\ \bibnamefont {Lu}}, \bibinfo
  {author} {\bibfnamefont {L.}~\bibnamefont {Rettig}}, \bibinfo {author}
  {\bibfnamefont {M.}~\bibnamefont {Wolf}}, \bibinfo {author} {\bibfnamefont
  {I.~R.}\ \bibnamefont {Fisher}},\ and\ \bibinfo {author} {\bibfnamefont
  {Z.-X.}\ \bibnamefont {Shen}},\ }\bibfield  {title} {\bibinfo {title}
  {{Ultrafast electron dynamics in the charge density wave material
  TbTe$_3$}},\ }\href {https://doi.org/10.1088/1367-2630/13/6/063022}
  {\bibfield  {journal} {\bibinfo  {journal} {New J. Phys.}\ }\textbf {\bibinfo
  {volume} {13}},\ \bibinfo {pages} {063022} (\bibinfo {year}
  {2011})}\BibitemShut {NoStop}%
\bibitem [{\citenamefont {Leuenberger}\ \emph {et~al.}(2015)\citenamefont
  {Leuenberger}, \citenamefont {Sobota}, \citenamefont {Yang}, \citenamefont
  {Kemper}, \citenamefont {Giraldo-Gallo}, \citenamefont {Moore}, \citenamefont
  {Fisher}, \citenamefont {Kirchmann}, \citenamefont {Devereaux},\ and\
  \citenamefont {Shen}}]{Leuenberger2015}%
  \BibitemOpen
  \bibfield  {author} {\bibinfo {author} {\bibfnamefont {D.}~\bibnamefont
  {Leuenberger}}, \bibinfo {author} {\bibfnamefont {J.~A.}\ \bibnamefont
  {Sobota}}, \bibinfo {author} {\bibfnamefont {S.-L.}\ \bibnamefont {Yang}},
  \bibinfo {author} {\bibfnamefont {A.~F.}\ \bibnamefont {Kemper}}, \bibinfo
  {author} {\bibfnamefont {P.}~\bibnamefont {Giraldo-Gallo}}, \bibinfo {author}
  {\bibfnamefont {R.~G.}\ \bibnamefont {Moore}}, \bibinfo {author}
  {\bibfnamefont {I.~R.}\ \bibnamefont {Fisher}}, \bibinfo {author}
  {\bibfnamefont {P.~S.}\ \bibnamefont {Kirchmann}}, \bibinfo {author}
  {\bibfnamefont {T.~P.}\ \bibnamefont {Devereaux}},\ and\ \bibinfo {author}
  {\bibfnamefont {Z.-X.}\ \bibnamefont {Shen}},\ }\bibfield  {title} {\bibinfo
  {title} {{Classification of collective modes in a charge density wave by
  momentum-dependent modulation of the electronic band structure}},\ }\href
  {https://doi.org/10.1103/PhysRevB.91.201106} {\bibfield  {journal} {\bibinfo
  {journal} {Phys. Rev. B}\ }\textbf {\bibinfo {volume} {91}},\ \bibinfo
  {pages} {201106} (\bibinfo {year} {2015})}\BibitemShut {NoStop}%
\bibitem [{\citenamefont {Rettig}\ \emph {et~al.}(2016)\citenamefont {Rettig},
  \citenamefont {Cort{\'{e}}s}, \citenamefont {Chu}, \citenamefont {Fisher},
  \citenamefont {Schmitt}, \citenamefont {Moore}, \citenamefont {Shen},
  \citenamefont {Kirchmann}, \citenamefont {Wolf},\ and\ \citenamefont
  {Bovensiepen}}]{Rettig2016}%
  \BibitemOpen
  \bibfield  {author} {\bibinfo {author} {\bibfnamefont {L.}~\bibnamefont
  {Rettig}}, \bibinfo {author} {\bibfnamefont {R.}~\bibnamefont
  {Cort{\'{e}}s}}, \bibinfo {author} {\bibfnamefont {J.-H.}\ \bibnamefont
  {Chu}}, \bibinfo {author} {\bibfnamefont {I.~R.}\ \bibnamefont {Fisher}},
  \bibinfo {author} {\bibfnamefont {F.}~\bibnamefont {Schmitt}}, \bibinfo
  {author} {\bibfnamefont {R.~G.}\ \bibnamefont {Moore}}, \bibinfo {author}
  {\bibfnamefont {Z.-X.}\ \bibnamefont {Shen}}, \bibinfo {author}
  {\bibfnamefont {P.~S.}\ \bibnamefont {Kirchmann}}, \bibinfo {author}
  {\bibfnamefont {M.}~\bibnamefont {Wolf}},\ and\ \bibinfo {author}
  {\bibfnamefont {U.}~\bibnamefont {Bovensiepen}},\ }\bibfield  {title}
  {\bibinfo {title} {{Persistent order due to transiently enhanced nesting in
  an electronically excited charge density wave}},\ }\href
  {https://doi.org/10.1038/ncomms10459} {\bibfield  {journal} {\bibinfo
  {journal} {Nat. Commun.}\ }\textbf {\bibinfo {volume} {7}},\ \bibinfo {pages}
  {10459} (\bibinfo {year} {2016})}\BibitemShut {NoStop}%
\bibitem [{\citenamefont {Maklar}\ \emph {et~al.}(2021)\citenamefont {Maklar},
  \citenamefont {Windsor}, \citenamefont {Nicholson}, \citenamefont {Puppin},
  \citenamefont {Walmsley}, \citenamefont {Esposito}, \citenamefont {Porer},
  \citenamefont {Rittmann}, \citenamefont {Leuenberger}, \citenamefont {Kubli},
  \citenamefont {Savoini}, \citenamefont {Abreu}, \citenamefont {Johnson},
  \citenamefont {Beaud}, \citenamefont {Ingold}, \citenamefont {Staub},
  \citenamefont {Fisher}, \citenamefont {Ernstorfer}, \citenamefont {Wolf},\
  and\ \citenamefont {Rettig}}]{Maklar2021}%
  \BibitemOpen
  \bibfield  {author} {\bibinfo {author} {\bibfnamefont {J.}~\bibnamefont
  {Maklar}}, \bibinfo {author} {\bibfnamefont {Y.~W.}\ \bibnamefont {Windsor}},
  \bibinfo {author} {\bibfnamefont {C.~W.}\ \bibnamefont {Nicholson}}, \bibinfo
  {author} {\bibfnamefont {M.}~\bibnamefont {Puppin}}, \bibinfo {author}
  {\bibfnamefont {P.}~\bibnamefont {Walmsley}}, \bibinfo {author}
  {\bibfnamefont {V.}~\bibnamefont {Esposito}}, \bibinfo {author}
  {\bibfnamefont {M.}~\bibnamefont {Porer}}, \bibinfo {author} {\bibfnamefont
  {J.}~\bibnamefont {Rittmann}}, \bibinfo {author} {\bibfnamefont
  {D.}~\bibnamefont {Leuenberger}}, \bibinfo {author} {\bibfnamefont
  {M.}~\bibnamefont {Kubli}}, \bibinfo {author} {\bibfnamefont
  {M.}~\bibnamefont {Savoini}}, \bibinfo {author} {\bibfnamefont
  {E.}~\bibnamefont {Abreu}}, \bibinfo {author} {\bibfnamefont {S.~L.}\
  \bibnamefont {Johnson}}, \bibinfo {author} {\bibfnamefont {P.}~\bibnamefont
  {Beaud}}, \bibinfo {author} {\bibfnamefont {G.}~\bibnamefont {Ingold}},
  \bibinfo {author} {\bibfnamefont {U.}~\bibnamefont {Staub}}, \bibinfo
  {author} {\bibfnamefont {I.~R.}\ \bibnamefont {Fisher}}, \bibinfo {author}
  {\bibfnamefont {R.}~\bibnamefont {Ernstorfer}}, \bibinfo {author}
  {\bibfnamefont {M.}~\bibnamefont {Wolf}},\ and\ \bibinfo {author}
  {\bibfnamefont {L.}~\bibnamefont {Rettig}},\ }\bibfield  {title} {\bibinfo
  {title} {{Nonequilibrium charge-density-wave order beyond the thermal
  limit}},\ }\href {https://doi.org/10.1038/s41467-021-22778-w} {\bibfield
  {journal} {\bibinfo  {journal} {Nat. Commun.}\ }\textbf {\bibinfo {volume}
  {12}},\ \bibinfo {pages} {2499} (\bibinfo {year} {2021})}\BibitemShut
  {NoStop}%
\bibitem [{\citenamefont {Moore}\ \emph {et~al.}(2016)\citenamefont {Moore},
  \citenamefont {Lee}, \citenamefont {Kirchman}, \citenamefont {Chuang},
  \citenamefont {Kemper}, \citenamefont {Trigo}, \citenamefont {Patthey},
  \citenamefont {Lu}, \citenamefont {Krupin}, \citenamefont {Yi}, \citenamefont
  {Reis}, \citenamefont {Doering}, \citenamefont {Denes}, \citenamefont
  {Schlotter}, \citenamefont {Turner}, \citenamefont {Hays}, \citenamefont
  {Hering}, \citenamefont {Benson}, \citenamefont {Chu}, \citenamefont
  {Devereaux}, \citenamefont {Fisher}, \citenamefont {Hussain},\ and\
  \citenamefont {Shen}}]{Moore2016}%
  \BibitemOpen
  \bibfield  {author} {\bibinfo {author} {\bibfnamefont {R.~G.}\ \bibnamefont
  {Moore}}, \bibinfo {author} {\bibfnamefont {W.~S.}\ \bibnamefont {Lee}},
  \bibinfo {author} {\bibfnamefont {P.~S.}\ \bibnamefont {Kirchman}}, \bibinfo
  {author} {\bibfnamefont {Y.~D.}\ \bibnamefont {Chuang}}, \bibinfo {author}
  {\bibfnamefont {A.~F.}\ \bibnamefont {Kemper}}, \bibinfo {author}
  {\bibfnamefont {M.}~\bibnamefont {Trigo}}, \bibinfo {author} {\bibfnamefont
  {L.}~\bibnamefont {Patthey}}, \bibinfo {author} {\bibfnamefont {D.~H.}\
  \bibnamefont {Lu}}, \bibinfo {author} {\bibfnamefont {O.}~\bibnamefont
  {Krupin}}, \bibinfo {author} {\bibfnamefont {M.}~\bibnamefont {Yi}}, \bibinfo
  {author} {\bibfnamefont {D.~A.}\ \bibnamefont {Reis}}, \bibinfo {author}
  {\bibfnamefont {D.}~\bibnamefont {Doering}}, \bibinfo {author} {\bibfnamefont
  {P.}~\bibnamefont {Denes}}, \bibinfo {author} {\bibfnamefont {W.~F.}\
  \bibnamefont {Schlotter}}, \bibinfo {author} {\bibfnamefont {J.~J.}\
  \bibnamefont {Turner}}, \bibinfo {author} {\bibfnamefont {G.}~\bibnamefont
  {Hays}}, \bibinfo {author} {\bibfnamefont {P.}~\bibnamefont {Hering}},
  \bibinfo {author} {\bibfnamefont {T.}~\bibnamefont {Benson}}, \bibinfo
  {author} {\bibfnamefont {J.-H.}\ \bibnamefont {Chu}}, \bibinfo {author}
  {\bibfnamefont {T.~P.}\ \bibnamefont {Devereaux}}, \bibinfo {author}
  {\bibfnamefont {I.~R.}\ \bibnamefont {Fisher}}, \bibinfo {author}
  {\bibfnamefont {Z.}~\bibnamefont {Hussain}},\ and\ \bibinfo {author}
  {\bibfnamefont {Z.-X.}\ \bibnamefont {Shen}},\ }\bibfield  {title} {\bibinfo
  {title} {{Ultrafast resonant soft X-ray diffraction dynamics of the charge
  density wave in TbTe$_3$}},\ }\href
  {https://doi.org/10.1103/PhysRevB.93.024304} {\bibfield  {journal} {\bibinfo
  {journal} {Phys. Rev. B}\ }\textbf {\bibinfo {volume} {93}},\ \bibinfo
  {pages} {024304} (\bibinfo {year} {2016})}\BibitemShut {NoStop}%
\bibitem [{\citenamefont {Trigo}\ \emph {et~al.}(2019)\citenamefont {Trigo},
  \citenamefont {Giraldo-Gallo}, \citenamefont {Kozina}, \citenamefont
  {Henighan}, \citenamefont {Jiang}, \citenamefont {Liu}, \citenamefont
  {Clark}, \citenamefont {Chollet}, \citenamefont {Glownia}, \citenamefont
  {Zhu}, \citenamefont {Katayama}, \citenamefont {Leuenberger}, \citenamefont
  {Kirchmann}, \citenamefont {Fisher}, \citenamefont {Shen},\ and\
  \citenamefont {Reis}}]{Trigo2019}%
  \BibitemOpen
  \bibfield  {author} {\bibinfo {author} {\bibfnamefont {M.}~\bibnamefont
  {Trigo}}, \bibinfo {author} {\bibfnamefont {P.}~\bibnamefont
  {Giraldo-Gallo}}, \bibinfo {author} {\bibfnamefont {M.~E.}\ \bibnamefont
  {Kozina}}, \bibinfo {author} {\bibfnamefont {T.}~\bibnamefont {Henighan}},
  \bibinfo {author} {\bibfnamefont {M.~P.}\ \bibnamefont {Jiang}}, \bibinfo
  {author} {\bibfnamefont {H.}~\bibnamefont {Liu}}, \bibinfo {author}
  {\bibfnamefont {J.~N.}\ \bibnamefont {Clark}}, \bibinfo {author}
  {\bibfnamefont {M.}~\bibnamefont {Chollet}}, \bibinfo {author} {\bibfnamefont
  {J.~M.}\ \bibnamefont {Glownia}}, \bibinfo {author} {\bibfnamefont
  {D.}~\bibnamefont {Zhu}}, \bibinfo {author} {\bibfnamefont {T.}~\bibnamefont
  {Katayama}}, \bibinfo {author} {\bibfnamefont {D.}~\bibnamefont
  {Leuenberger}}, \bibinfo {author} {\bibfnamefont {P.~S.}\ \bibnamefont
  {Kirchmann}}, \bibinfo {author} {\bibfnamefont {I.~R.}\ \bibnamefont
  {Fisher}}, \bibinfo {author} {\bibfnamefont {Z.~X.}\ \bibnamefont {Shen}},\
  and\ \bibinfo {author} {\bibfnamefont {D.~A.}\ \bibnamefont {Reis}},\
  }\bibfield  {title} {\bibinfo {title} {{Coherent order parameter dynamics in
  SmTe$_3$}},\ }\href {https://doi.org/10.1103/PhysRevB.99.104111} {\bibfield
  {journal} {\bibinfo  {journal} {Phys. Rev. B}\ }\textbf {\bibinfo {volume}
  {99}},\ \bibinfo {pages} {104111} (\bibinfo {year} {2019})}\BibitemShut
  {NoStop}%
\bibitem [{\citenamefont {Trigo}\ \emph {et~al.}(2021)\citenamefont {Trigo},
  \citenamefont {Giraldo-Gallo}, \citenamefont {Clark}, \citenamefont {Kozina},
  \citenamefont {Henighan}, \citenamefont {Jiang}, \citenamefont {Chollet},
  \citenamefont {Fisher}, \citenamefont {Glownia}, \citenamefont {Katayama},
  \citenamefont {Kirchmann}, \citenamefont {Leuenberger}, \citenamefont {Liu},
  \citenamefont {Reis}, \citenamefont {Shen},\ and\ \citenamefont
  {Zhu}}]{Trigo2021}%
  \BibitemOpen
  \bibfield  {author} {\bibinfo {author} {\bibfnamefont {M.}~\bibnamefont
  {Trigo}}, \bibinfo {author} {\bibfnamefont {P.}~\bibnamefont
  {Giraldo-Gallo}}, \bibinfo {author} {\bibfnamefont {J.~N.}\ \bibnamefont
  {Clark}}, \bibinfo {author} {\bibfnamefont {M.~E.}\ \bibnamefont {Kozina}},
  \bibinfo {author} {\bibfnamefont {T.}~\bibnamefont {Henighan}}, \bibinfo
  {author} {\bibfnamefont {M.~P.}\ \bibnamefont {Jiang}}, \bibinfo {author}
  {\bibfnamefont {M.}~\bibnamefont {Chollet}}, \bibinfo {author} {\bibfnamefont
  {I.~R.}\ \bibnamefont {Fisher}}, \bibinfo {author} {\bibfnamefont {J.~M.}\
  \bibnamefont {Glownia}}, \bibinfo {author} {\bibfnamefont {T.}~\bibnamefont
  {Katayama}}, \bibinfo {author} {\bibfnamefont {P.~S.}\ \bibnamefont
  {Kirchmann}}, \bibinfo {author} {\bibfnamefont {D.}~\bibnamefont
  {Leuenberger}}, \bibinfo {author} {\bibfnamefont {H.}~\bibnamefont {Liu}},
  \bibinfo {author} {\bibfnamefont {D.~A.}\ \bibnamefont {Reis}}, \bibinfo
  {author} {\bibfnamefont {Z.~X.}\ \bibnamefont {Shen}},\ and\ \bibinfo
  {author} {\bibfnamefont {D.}~\bibnamefont {Zhu}},\ }\bibfield  {title}
  {\bibinfo {title} {{Ultrafast formation of domain walls of a charge density
  wave in SmTe$_3$}},\ }\href {https://doi.org/10.1103/PhysRevB.103.054109}
  {\bibfield  {journal} {\bibinfo  {journal} {Phys. Rev. B}\ }\textbf {\bibinfo
  {volume} {103}},\ \bibinfo {pages} {054109} (\bibinfo {year}
  {2021})}\BibitemShut {NoStop}%
\bibitem [{\citenamefont {Han}\ \emph {et~al.}(2012)\citenamefont {Han},
  \citenamefont {Tao}, \citenamefont {Mahanti}, \citenamefont {Chang},
  \citenamefont {Ruan}, \citenamefont {Malliakas},\ and\ \citenamefont
  {Kanatzidis}}]{Han2012}%
  \BibitemOpen
  \bibfield  {author} {\bibinfo {author} {\bibfnamefont {T.-R.~T.}\
  \bibnamefont {Han}}, \bibinfo {author} {\bibfnamefont {Z.}~\bibnamefont
  {Tao}}, \bibinfo {author} {\bibfnamefont {S.~D.}\ \bibnamefont {Mahanti}},
  \bibinfo {author} {\bibfnamefont {K.}~\bibnamefont {Chang}}, \bibinfo
  {author} {\bibfnamefont {C.-Y.}\ \bibnamefont {Ruan}}, \bibinfo {author}
  {\bibfnamefont {C.~D.}\ \bibnamefont {Malliakas}},\ and\ \bibinfo {author}
  {\bibfnamefont {M.~G.}\ \bibnamefont {Kanatzidis}},\ }\bibfield  {title}
  {\bibinfo {title} {{Structural dynamics of two-dimensional charge-density
  waves in CeTe$3$ investigated by ultrafast electron crystallography}},\
  }\href {https://doi.org/10.1103/PhysRevB.86.075145} {\bibfield  {journal}
  {\bibinfo  {journal} {Phys. Rev. B}\ }\textbf {\bibinfo {volume} {86}},\
  \bibinfo {pages} {075145} (\bibinfo {year} {2012})}\BibitemShut {NoStop}%
\bibitem [{\citenamefont {Budden}\ \emph {et~al.}(2021)\citenamefont {Budden},
  \citenamefont {Gebert}, \citenamefont {Buzzi}, \citenamefont {Jotzu},
  \citenamefont {Wang}, \citenamefont {Matsuyama}, \citenamefont {Meier},
  \citenamefont {Laplace}, \citenamefont {Pontiroli}, \citenamefont
  {Ricc{\`{o}}}, \citenamefont {Schlawin}, \citenamefont {Jaksch},\ and\
  \citenamefont {Cavalleri}}]{Budden2021}%
  \BibitemOpen
  \bibfield  {author} {\bibinfo {author} {\bibfnamefont {M.}~\bibnamefont
  {Budden}}, \bibinfo {author} {\bibfnamefont {T.}~\bibnamefont {Gebert}},
  \bibinfo {author} {\bibfnamefont {M.}~\bibnamefont {Buzzi}}, \bibinfo
  {author} {\bibfnamefont {G.}~\bibnamefont {Jotzu}}, \bibinfo {author}
  {\bibfnamefont {E.}~\bibnamefont {Wang}}, \bibinfo {author} {\bibfnamefont
  {T.}~\bibnamefont {Matsuyama}}, \bibinfo {author} {\bibfnamefont
  {G.}~\bibnamefont {Meier}}, \bibinfo {author} {\bibfnamefont
  {Y.}~\bibnamefont {Laplace}}, \bibinfo {author} {\bibfnamefont
  {D.}~\bibnamefont {Pontiroli}}, \bibinfo {author} {\bibfnamefont
  {M.}~\bibnamefont {Ricc{\`{o}}}}, \bibinfo {author} {\bibfnamefont
  {F.}~\bibnamefont {Schlawin}}, \bibinfo {author} {\bibfnamefont
  {D.}~\bibnamefont {Jaksch}},\ and\ \bibinfo {author} {\bibfnamefont
  {A.}~\bibnamefont {Cavalleri}},\ }\bibfield  {title} {\bibinfo {title}
  {{Evidence for metastable photo-induced superconductivity in
  K$_3$C$_{60}$}},\ }\href {https://doi.org/10.1038/s41567-020-01148-1}
  {\bibfield  {journal} {\bibinfo  {journal} {Nat. Phys.}\ }\textbf {\bibinfo
  {volume} {17}},\ \bibinfo {pages} {611} (\bibinfo {year} {2021})}\BibitemShut
  {NoStop}%
\bibitem [{\citenamefont {Morrison}\ \emph {et~al.}(2014)\citenamefont
  {Morrison}, \citenamefont {Chatelain}, \citenamefont {Tiwari}, \citenamefont
  {Hendaoui}, \citenamefont {Bruh{\'{a}}cs}, \citenamefont {Chaker},\ and\
  \citenamefont {Siwick}}]{Morrison2014}%
  \BibitemOpen
  \bibfield  {author} {\bibinfo {author} {\bibfnamefont {V.~R.}\ \bibnamefont
  {Morrison}}, \bibinfo {author} {\bibfnamefont {R.~P.}\ \bibnamefont
  {Chatelain}}, \bibinfo {author} {\bibfnamefont {K.~L.}\ \bibnamefont
  {Tiwari}}, \bibinfo {author} {\bibfnamefont {A.}~\bibnamefont {Hendaoui}},
  \bibinfo {author} {\bibfnamefont {A.}~\bibnamefont {Bruh{\'{a}}cs}}, \bibinfo
  {author} {\bibfnamefont {M.}~\bibnamefont {Chaker}},\ and\ \bibinfo {author}
  {\bibfnamefont {B.~J.}\ \bibnamefont {Siwick}},\ }\bibfield  {title}
  {\bibinfo {title} {{A photoinduced metal-like phase of monoclinic VO$_2$
  revealed by ultrafast electron diffraction}},\ }\href
  {https://doi.org/10.1126/science.1253779} {\bibfield  {journal} {\bibinfo
  {journal} {Science}\ }\textbf {\bibinfo {volume} {346}},\ \bibinfo {pages}
  {445} (\bibinfo {year} {2014})}\BibitemShut {NoStop}%
\bibitem [{\citenamefont {Otto}\ \emph {et~al.}(2019)\citenamefont {Otto},
  \citenamefont {{Ren{\'{e}} de Cotret}}, \citenamefont {Valverde-Chavez},
  \citenamefont {Tiwari}, \citenamefont {{\'{E}}mond}, \citenamefont {Chaker},
  \citenamefont {Cooke},\ and\ \citenamefont {Siwick}}]{Otto2019}%
  \BibitemOpen
  \bibfield  {author} {\bibinfo {author} {\bibfnamefont {M.~R.}\ \bibnamefont
  {Otto}}, \bibinfo {author} {\bibfnamefont {L.~P.}\ \bibnamefont {{Ren{\'{e}}
  de Cotret}}}, \bibinfo {author} {\bibfnamefont {D.~A.}\ \bibnamefont
  {Valverde-Chavez}}, \bibinfo {author} {\bibfnamefont {K.~L.}\ \bibnamefont
  {Tiwari}}, \bibinfo {author} {\bibfnamefont {N.}~\bibnamefont {{\'{E}}mond}},
  \bibinfo {author} {\bibfnamefont {M.}~\bibnamefont {Chaker}}, \bibinfo
  {author} {\bibfnamefont {D.~G.}\ \bibnamefont {Cooke}},\ and\ \bibinfo
  {author} {\bibfnamefont {B.~J.}\ \bibnamefont {Siwick}},\ }\bibfield  {title}
  {\bibinfo {title} {{How optical excitation controls the structure and
  properties of vanadium dioxide}},\ }\href
  {https://doi.org/10.1073/pnas.1808414115} {\bibfield  {journal} {\bibinfo
  {journal} {Proc. Natl. Acad. Sci. U.S.A.}\ }\textbf {\bibinfo {volume}
  {116}},\ \bibinfo {pages} {450} (\bibinfo {year} {2019})}\BibitemShut
  {NoStop}%
\bibitem [{\citenamefont {Zhang}\ and\ \citenamefont
  {Averitt}(2014)}]{Zhang2014}%
  \BibitemOpen
  \bibfield  {author} {\bibinfo {author} {\bibfnamefont {J.}~\bibnamefont
  {Zhang}}\ and\ \bibinfo {author} {\bibfnamefont {R.}~\bibnamefont
  {Averitt}},\ }\bibfield  {title} {\bibinfo {title} {{Dynamics and control in
  complex transition metal oxides}},\ }\href
  {https://doi.org/10.1146/annurev-matsci-070813-113258} {\bibfield  {journal}
  {\bibinfo  {journal} {Annu. Rev. Mater. Res.}\ }\textbf {\bibinfo {volume}
  {44}},\ \bibinfo {pages} {19} (\bibinfo {year} {2014})}\BibitemShut {NoStop}%
\bibitem [{\citenamefont {Shao}\ \emph {et~al.}(2018)\citenamefont {Shao},
  \citenamefont {Cao}, \citenamefont {Luo},\ and\ \citenamefont
  {Jin}}]{Shao2018}%
  \BibitemOpen
  \bibfield  {author} {\bibinfo {author} {\bibfnamefont {Z.}~\bibnamefont
  {Shao}}, \bibinfo {author} {\bibfnamefont {X.}~\bibnamefont {Cao}}, \bibinfo
  {author} {\bibfnamefont {H.}~\bibnamefont {Luo}},\ and\ \bibinfo {author}
  {\bibfnamefont {P.}~\bibnamefont {Jin}},\ }\bibfield  {title} {\bibinfo
  {title} {{Recent progress in the phase-transition mechanism and modulation of
  vanadium dioxide materials}},\ }\href
  {https://doi.org/10.1038/s41427-018-0061-2} {\bibfield  {journal} {\bibinfo
  {journal} {NPG Asia Mater.}\ }\textbf {\bibinfo {volume} {10}},\ \bibinfo
  {pages} {581} (\bibinfo {year} {2018})}\BibitemShut {NoStop}%
\bibitem [{\citenamefont {Becker}\ \emph {et~al.}(1996)\citenamefont {Becker},
  \citenamefont {Buckman}, \citenamefont {Walser}, \citenamefont
  {L{\'{e}}pine}, \citenamefont {Georges},\ and\ \citenamefont
  {Brun}}]{Becker1996}%
  \BibitemOpen
  \bibfield  {author} {\bibinfo {author} {\bibfnamefont {M.~F.}\ \bibnamefont
  {Becker}}, \bibinfo {author} {\bibfnamefont {A.~B.}\ \bibnamefont {Buckman}},
  \bibinfo {author} {\bibfnamefont {R.~M.}\ \bibnamefont {Walser}}, \bibinfo
  {author} {\bibfnamefont {T.}~\bibnamefont {L{\'{e}}pine}}, \bibinfo {author}
  {\bibfnamefont {P.}~\bibnamefont {Georges}},\ and\ \bibinfo {author}
  {\bibfnamefont {A.}~\bibnamefont {Brun}},\ }\bibfield  {title} {\bibinfo
  {title} {{Femtosecond laser excitation dynamics of the semiconductor‐metal
  phase transition in VO$_2$}},\ }\href {https://doi.org/10.1063/1.361167}
  {\bibfield  {journal} {\bibinfo  {journal} {J. Appl. Phys.}\ }\textbf
  {\bibinfo {volume} {79}},\ \bibinfo {pages} {2404} (\bibinfo {year}
  {1996})}\BibitemShut {NoStop}%
\bibitem [{\citenamefont {Cavalleri}\ \emph {et~al.}(2001)\citenamefont
  {Cavalleri}, \citenamefont {T{\'{o}}th}, \citenamefont {Siders},
  \citenamefont {Squier}, \citenamefont {R{\'{a}}ksi}, \citenamefont {Forget},\
  and\ \citenamefont {Kieffer}}]{Cavalleri2001}%
  \BibitemOpen
  \bibfield  {author} {\bibinfo {author} {\bibfnamefont {A.}~\bibnamefont
  {Cavalleri}}, \bibinfo {author} {\bibfnamefont {C.}~\bibnamefont
  {T{\'{o}}th}}, \bibinfo {author} {\bibfnamefont {C.~W.}\ \bibnamefont
  {Siders}}, \bibinfo {author} {\bibfnamefont {J.~A.}\ \bibnamefont {Squier}},
  \bibinfo {author} {\bibfnamefont {F.}~\bibnamefont {R{\'{a}}ksi}}, \bibinfo
  {author} {\bibfnamefont {P.}~\bibnamefont {Forget}},\ and\ \bibinfo {author}
  {\bibfnamefont {J.~C.}\ \bibnamefont {Kieffer}},\ }\bibfield  {title}
  {\bibinfo {title} {{Femtosecond structural dynamics in VO$_2$ during an
  ultrafast solid-solid phase transition}},\ }\href
  {https://doi.org/10.1103/PhysRevLett.87.237401} {\bibfield  {journal}
  {\bibinfo  {journal} {Phys. Rev. Lett.}\ }\textbf {\bibinfo {volume} {87}},\
  \bibinfo {pages} {237401} (\bibinfo {year} {2001})}\BibitemShut {NoStop}%
\bibitem [{\citenamefont {Kim}\ \emph {et~al.}(2006)\citenamefont {Kim},
  \citenamefont {Lee}, \citenamefont {Kim}, \citenamefont {Chae}, \citenamefont
  {Yun}, \citenamefont {Kang}, \citenamefont {Han}, \citenamefont {Yee},\ and\
  \citenamefont {Lim}}]{Kim2006}%
  \BibitemOpen
  \bibfield  {author} {\bibinfo {author} {\bibfnamefont {H.-T.}\ \bibnamefont
  {Kim}}, \bibinfo {author} {\bibfnamefont {Y.~W.}\ \bibnamefont {Lee}},
  \bibinfo {author} {\bibfnamefont {B.-J.}\ \bibnamefont {Kim}}, \bibinfo
  {author} {\bibfnamefont {B.-G.}\ \bibnamefont {Chae}}, \bibinfo {author}
  {\bibfnamefont {S.~J.}\ \bibnamefont {Yun}}, \bibinfo {author} {\bibfnamefont
  {K.-Y.}\ \bibnamefont {Kang}}, \bibinfo {author} {\bibfnamefont {K.-J.}\
  \bibnamefont {Han}}, \bibinfo {author} {\bibfnamefont {K.-J.}\ \bibnamefont
  {Yee}},\ and\ \bibinfo {author} {\bibfnamefont {Y.-S.}\ \bibnamefont {Lim}},\
  }\bibfield  {title} {\bibinfo {title} {{Monoclinic and correlated metal phase
  in VO$_2$ as evidence of the Mott transition: Coherent phonon analysis}},\
  }\href {https://doi.org/10.1103/PhysRevLett.97.266401} {\bibfield  {journal}
  {\bibinfo  {journal} {Phys. Rev. Lett.}\ }\textbf {\bibinfo {volume} {97}},\
  \bibinfo {pages} {266401} (\bibinfo {year} {2006})}\BibitemShut {NoStop}%
\bibitem [{\citenamefont {Baum}\ \emph {et~al.}(2007)\citenamefont {Baum},
  \citenamefont {Yang},\ and\ \citenamefont {Zewail}}]{Baum2007}%
  \BibitemOpen
  \bibfield  {author} {\bibinfo {author} {\bibfnamefont {P.}~\bibnamefont
  {Baum}}, \bibinfo {author} {\bibfnamefont {D.-S.}\ \bibnamefont {Yang}},\
  and\ \bibinfo {author} {\bibfnamefont {A.~H.}\ \bibnamefont {Zewail}},\
  }\bibfield  {title} {\bibinfo {title} {{4D visualization of transitional
  structures in phase transformations by electron diffraction}},\ }\href
  {https://doi.org/10.1126/science.1147724} {\bibfield  {journal} {\bibinfo
  {journal} {Science}\ }\textbf {\bibinfo {volume} {318}},\ \bibinfo {pages}
  {788} (\bibinfo {year} {2007})}\BibitemShut {NoStop}%
\bibitem [{\citenamefont {Sood}\ \emph {et~al.}(2021)\citenamefont {Sood},
  \citenamefont {Shen}, \citenamefont {Shi}, \citenamefont {Kumar},
  \citenamefont {Park}, \citenamefont {Zajac}, \citenamefont {Sun},
  \citenamefont {Chen}, \citenamefont {Ramanathan}, \citenamefont {Wang},
  \citenamefont {Chueh},\ and\ \citenamefont {Lindenberg}}]{Sood2021}%
  \BibitemOpen
  \bibfield  {author} {\bibinfo {author} {\bibfnamefont {A.}~\bibnamefont
  {Sood}}, \bibinfo {author} {\bibfnamefont {X.}~\bibnamefont {Shen}}, \bibinfo
  {author} {\bibfnamefont {Y.}~\bibnamefont {Shi}}, \bibinfo {author}
  {\bibfnamefont {S.}~\bibnamefont {Kumar}}, \bibinfo {author} {\bibfnamefont
  {S.~J.}\ \bibnamefont {Park}}, \bibinfo {author} {\bibfnamefont
  {M.}~\bibnamefont {Zajac}}, \bibinfo {author} {\bibfnamefont
  {Y.}~\bibnamefont {Sun}}, \bibinfo {author} {\bibfnamefont {L.-Q.}\
  \bibnamefont {Chen}}, \bibinfo {author} {\bibfnamefont {S.}~\bibnamefont
  {Ramanathan}}, \bibinfo {author} {\bibfnamefont {X.}~\bibnamefont {Wang}},
  \bibinfo {author} {\bibfnamefont {W.~C.}\ \bibnamefont {Chueh}},\ and\
  \bibinfo {author} {\bibfnamefont {A.~M.}\ \bibnamefont {Lindenberg}},\
  }\bibfield  {title} {\bibinfo {title} {{Universal phase dynamics in VO$_2$
  switches revealed by ultrafast operando diffraction}},\ }\href
  {https://doi.org/10.1126/science.abc0652} {\bibfield  {journal} {\bibinfo
  {journal} {Science}\ }\textbf {\bibinfo {volume} {373}},\ \bibinfo {pages}
  {352} (\bibinfo {year} {2021})}\BibitemShut {NoStop}%
\bibitem [{\citenamefont {Zheng}\ \emph {et~al.}(2005)\citenamefont {Zheng},
  \citenamefont {Zhu}, \citenamefont {Wu},\ and\ \citenamefont
  {Davenport}}]{Zheng2005}%
  \BibitemOpen
  \bibfield  {author} {\bibinfo {author} {\bibfnamefont {J.~C.}\ \bibnamefont
  {Zheng}}, \bibinfo {author} {\bibfnamefont {Y.}~\bibnamefont {Zhu}}, \bibinfo
  {author} {\bibfnamefont {L.}~\bibnamefont {Wu}},\ and\ \bibinfo {author}
  {\bibfnamefont {J.~W.}\ \bibnamefont {Davenport}},\ }\bibfield  {title}
  {\bibinfo {title} {{On the sensitivity of electron and X-ray scattering
  factors to valence charge distributions}},\ }\href
  {https://doi.org/10.1107/S0021889805016109} {\bibfield  {journal} {\bibinfo
  {journal} {J. Appl. Cryst.}\ }\textbf {\bibinfo {volume} {38}},\ \bibinfo
  {pages} {648} (\bibinfo {year} {2005})}\BibitemShut {NoStop}%
\bibitem [{\citenamefont {Gerber}\ \emph {et~al.}(2017)\citenamefont {Gerber},
  \citenamefont {Yang}, \citenamefont {Zhu}, \citenamefont {Soifer},
  \citenamefont {Sobota}, \citenamefont {Rebec}, \citenamefont {Lee},
  \citenamefont {Jia}, \citenamefont {Moritz}, \citenamefont {Jia},
  \citenamefont {Gauthier}, \citenamefont {Li}, \citenamefont {Leuenberger},
  \citenamefont {Zhang}, \citenamefont {Chaix}, \citenamefont {Li},
  \citenamefont {Jang}, \citenamefont {Lee}, \citenamefont {Yi}, \citenamefont
  {Dakovski}, \citenamefont {Song}, \citenamefont {Glownia}, \citenamefont
  {Nelson}, \citenamefont {Kim}, \citenamefont {Chuang}, \citenamefont
  {Hussain}, \citenamefont {Moore}, \citenamefont {Devereaux}, \citenamefont
  {Lee}, \citenamefont {Kirchmann},\ and\ \citenamefont {Shen}}]{Gerber2017}%
  \BibitemOpen
  \bibfield  {author} {\bibinfo {author} {\bibfnamefont {S.}~\bibnamefont
  {Gerber}}, \bibinfo {author} {\bibfnamefont {S.-L.}\ \bibnamefont {Yang}},
  \bibinfo {author} {\bibfnamefont {D.}~\bibnamefont {Zhu}}, \bibinfo {author}
  {\bibfnamefont {H.}~\bibnamefont {Soifer}}, \bibinfo {author} {\bibfnamefont
  {J.~a.}\ \bibnamefont {Sobota}}, \bibinfo {author} {\bibfnamefont
  {S.}~\bibnamefont {Rebec}}, \bibinfo {author} {\bibfnamefont {J.~J.}\
  \bibnamefont {Lee}}, \bibinfo {author} {\bibfnamefont {T.}~\bibnamefont
  {Jia}}, \bibinfo {author} {\bibfnamefont {B.}~\bibnamefont {Moritz}},
  \bibinfo {author} {\bibfnamefont {C.}~\bibnamefont {Jia}}, \bibinfo {author}
  {\bibfnamefont {A.}~\bibnamefont {Gauthier}}, \bibinfo {author}
  {\bibfnamefont {Y.}~\bibnamefont {Li}}, \bibinfo {author} {\bibfnamefont
  {D.}~\bibnamefont {Leuenberger}}, \bibinfo {author} {\bibfnamefont
  {Y.}~\bibnamefont {Zhang}}, \bibinfo {author} {\bibfnamefont
  {L.}~\bibnamefont {Chaix}}, \bibinfo {author} {\bibfnamefont
  {W.}~\bibnamefont {Li}}, \bibinfo {author} {\bibfnamefont {H.}~\bibnamefont
  {Jang}}, \bibinfo {author} {\bibfnamefont {J.-S.}\ \bibnamefont {Lee}},
  \bibinfo {author} {\bibfnamefont {M.}~\bibnamefont {Yi}}, \bibinfo {author}
  {\bibfnamefont {G.~L.}\ \bibnamefont {Dakovski}}, \bibinfo {author}
  {\bibfnamefont {S.}~\bibnamefont {Song}}, \bibinfo {author} {\bibfnamefont
  {J.~M.}\ \bibnamefont {Glownia}}, \bibinfo {author} {\bibfnamefont
  {S.}~\bibnamefont {Nelson}}, \bibinfo {author} {\bibfnamefont {K.~W.}\
  \bibnamefont {Kim}}, \bibinfo {author} {\bibfnamefont {Y.-D.}\ \bibnamefont
  {Chuang}}, \bibinfo {author} {\bibfnamefont {Z.}~\bibnamefont {Hussain}},
  \bibinfo {author} {\bibfnamefont {R.~G.}\ \bibnamefont {Moore}}, \bibinfo
  {author} {\bibfnamefont {T.~P.}\ \bibnamefont {Devereaux}}, \bibinfo {author}
  {\bibfnamefont {W.-S.}\ \bibnamefont {Lee}}, \bibinfo {author} {\bibfnamefont
  {P.~S.}\ \bibnamefont {Kirchmann}},\ and\ \bibinfo {author} {\bibfnamefont
  {Z.-X.}\ \bibnamefont {Shen}},\ }\bibfield  {title} {\bibinfo {title}
  {{Femtosecond electron-phonon lock-in by photoemission and x-ray
  free-electron laser}},\ }\href {https://doi.org/10.1126/science.aak9946}
  {\bibfield  {journal} {\bibinfo  {journal} {Science}\ }\textbf {\bibinfo
  {volume} {357}},\ \bibinfo {pages} {71} (\bibinfo {year} {2017})}\BibitemShut
  {NoStop}%
\bibitem [{\citenamefont {Sie}\ \emph {et~al.}(2019)\citenamefont {Sie},
  \citenamefont {Nyby}, \citenamefont {Pemmaraju}, \citenamefont {Park},
  \citenamefont {Shen}, \citenamefont {Yang}, \citenamefont {Hoffmann},
  \citenamefont {Ofori-Okai}, \citenamefont {Li}, \citenamefont {Reid},
  \citenamefont {Weathersby}, \citenamefont {Mannebach}, \citenamefont
  {Finney}, \citenamefont {Rhodes}, \citenamefont {Chenet}, \citenamefont
  {Antony}, \citenamefont {Balicas}, \citenamefont {Hone}, \citenamefont
  {Devereaux}, \citenamefont {Heinz}, \citenamefont {Wang},\ and\ \citenamefont
  {Lindenberg}}]{Sie2019}%
  \BibitemOpen
  \bibfield  {author} {\bibinfo {author} {\bibfnamefont {E.~J.}\ \bibnamefont
  {Sie}}, \bibinfo {author} {\bibfnamefont {C.~M.}\ \bibnamefont {Nyby}},
  \bibinfo {author} {\bibfnamefont {C.~D.}\ \bibnamefont {Pemmaraju}}, \bibinfo
  {author} {\bibfnamefont {S.~J.}\ \bibnamefont {Park}}, \bibinfo {author}
  {\bibfnamefont {X.}~\bibnamefont {Shen}}, \bibinfo {author} {\bibfnamefont
  {J.}~\bibnamefont {Yang}}, \bibinfo {author} {\bibfnamefont {M.~C.}\
  \bibnamefont {Hoffmann}}, \bibinfo {author} {\bibfnamefont {B.~K.}\
  \bibnamefont {Ofori-Okai}}, \bibinfo {author} {\bibfnamefont
  {R.}~\bibnamefont {Li}}, \bibinfo {author} {\bibfnamefont {A.~H.}\
  \bibnamefont {Reid}}, \bibinfo {author} {\bibfnamefont {S.}~\bibnamefont
  {Weathersby}}, \bibinfo {author} {\bibfnamefont {E.}~\bibnamefont
  {Mannebach}}, \bibinfo {author} {\bibfnamefont {N.}~\bibnamefont {Finney}},
  \bibinfo {author} {\bibfnamefont {D.}~\bibnamefont {Rhodes}}, \bibinfo
  {author} {\bibfnamefont {D.}~\bibnamefont {Chenet}}, \bibinfo {author}
  {\bibfnamefont {A.}~\bibnamefont {Antony}}, \bibinfo {author} {\bibfnamefont
  {L.}~\bibnamefont {Balicas}}, \bibinfo {author} {\bibfnamefont
  {J.}~\bibnamefont {Hone}}, \bibinfo {author} {\bibfnamefont {T.~P.}\
  \bibnamefont {Devereaux}}, \bibinfo {author} {\bibfnamefont {T.~F.}\
  \bibnamefont {Heinz}}, \bibinfo {author} {\bibfnamefont {X.}~\bibnamefont
  {Wang}},\ and\ \bibinfo {author} {\bibfnamefont {A.~M.}\ \bibnamefont
  {Lindenberg}},\ }\bibfield  {title} {\bibinfo {title} {{An ultrafast symmetry
  switch in a Weyl semimetal}},\ }\href
  {https://doi.org/10.1038/s41586-018-0809-4} {\bibfield  {journal} {\bibinfo
  {journal} {Nature}\ }\textbf {\bibinfo {volume} {565}},\ \bibinfo {pages}
  {61} (\bibinfo {year} {2019})}\BibitemShut {NoStop}%
\bibitem [{\citenamefont {Zong}\ \emph {et~al.}(2018)\citenamefont {Zong},
  \citenamefont {Shen}, \citenamefont {Kogar}, \citenamefont {Ye},
  \citenamefont {Marks}, \citenamefont {Chowdhury}, \citenamefont {Rohwer},
  \citenamefont {Freelon}, \citenamefont {Weathersby}, \citenamefont {Li},
  \citenamefont {Yang}, \citenamefont {Checkelsky}, \citenamefont {Wang},\ and\
  \citenamefont {Gedik}}]{Zong2018}%
  \BibitemOpen
  \bibfield  {author} {\bibinfo {author} {\bibfnamefont {A.}~\bibnamefont
  {Zong}}, \bibinfo {author} {\bibfnamefont {X.}~\bibnamefont {Shen}}, \bibinfo
  {author} {\bibfnamefont {A.}~\bibnamefont {Kogar}}, \bibinfo {author}
  {\bibfnamefont {L.}~\bibnamefont {Ye}}, \bibinfo {author} {\bibfnamefont
  {C.}~\bibnamefont {Marks}}, \bibinfo {author} {\bibfnamefont
  {D.}~\bibnamefont {Chowdhury}}, \bibinfo {author} {\bibfnamefont
  {T.}~\bibnamefont {Rohwer}}, \bibinfo {author} {\bibfnamefont
  {B.}~\bibnamefont {Freelon}}, \bibinfo {author} {\bibfnamefont
  {S.}~\bibnamefont {Weathersby}}, \bibinfo {author} {\bibfnamefont
  {R.}~\bibnamefont {Li}}, \bibinfo {author} {\bibfnamefont {J.}~\bibnamefont
  {Yang}}, \bibinfo {author} {\bibfnamefont {J.}~\bibnamefont {Checkelsky}},
  \bibinfo {author} {\bibfnamefont {X.}~\bibnamefont {Wang}},\ and\ \bibinfo
  {author} {\bibfnamefont {N.}~\bibnamefont {Gedik}},\ }\bibfield  {title}
  {\bibinfo {title} {{Ultrafast manipulation of mirror domain walls in a charge
  density wave}},\ }\href {https://doi.org/10.1126/sciadv.aau5501} {\bibfield
  {journal} {\bibinfo  {journal} {Sci. Adv.}\ }\textbf {\bibinfo {volume}
  {4}},\ \bibinfo {pages} {eaau5501} (\bibinfo {year} {2018})}\BibitemShut
  {NoStop}%
\bibitem [{\citenamefont {Horstmann}\ \emph {et~al.}(2020)\citenamefont
  {Horstmann}, \citenamefont {B{\"{o}}ckmann}, \citenamefont {Wit},
  \citenamefont {Kurtz}, \citenamefont {Storeck},\ and\ \citenamefont
  {Ropers}}]{Horstmann2020}%
  \BibitemOpen
  \bibfield  {author} {\bibinfo {author} {\bibfnamefont {J.~G.}\ \bibnamefont
  {Horstmann}}, \bibinfo {author} {\bibfnamefont {H.}~\bibnamefont
  {B{\"{o}}ckmann}}, \bibinfo {author} {\bibfnamefont {B.}~\bibnamefont {Wit}},
  \bibinfo {author} {\bibfnamefont {F.}~\bibnamefont {Kurtz}}, \bibinfo
  {author} {\bibfnamefont {G.}~\bibnamefont {Storeck}},\ and\ \bibinfo {author}
  {\bibfnamefont {C.}~\bibnamefont {Ropers}},\ }\bibfield  {title} {\bibinfo
  {title} {{Coherent control of a surface structural phase transition}},\
  }\href {https://doi.org/10.1038/s41586-020-2440-4} {\bibfield  {journal}
  {\bibinfo  {journal} {Nature}\ }\textbf {\bibinfo {volume} {583}},\ \bibinfo
  {pages} {232} (\bibinfo {year} {2020})}\BibitemShut {NoStop}%
\bibitem [{\citenamefont {Zhang}\ and\ \citenamefont
  {Flannigan}(2019)}]{Zhang2019}%
  \BibitemOpen
  \bibfield  {author} {\bibinfo {author} {\bibfnamefont {Y.}~\bibnamefont
  {Zhang}}\ and\ \bibinfo {author} {\bibfnamefont {D.~J.}\ \bibnamefont
  {Flannigan}},\ }\bibfield  {title} {\bibinfo {title} {{Observation of
  anisotropic strain-wave dynamics and few-layer dephasing in MoS$_2$ with
  ultrafast electron microscopy}},\ }\href
  {https://doi.org/10.1021/acs.nanolett.9b03596} {\bibfield  {journal}
  {\bibinfo  {journal} {Nano Lett.}\ }\textbf {\bibinfo {volume} {19}},\
  \bibinfo {pages} {8216} (\bibinfo {year} {2019})}\BibitemShut {NoStop}%
\bibitem [{\citenamefont {Subedi}\ \emph {et~al.}(2014)\citenamefont {Subedi},
  \citenamefont {Cavalleri},\ and\ \citenamefont {Georges}}]{Subedi2014}%
  \BibitemOpen
  \bibfield  {author} {\bibinfo {author} {\bibfnamefont {A.}~\bibnamefont
  {Subedi}}, \bibinfo {author} {\bibfnamefont {A.}~\bibnamefont {Cavalleri}},\
  and\ \bibinfo {author} {\bibfnamefont {A.}~\bibnamefont {Georges}},\
  }\bibfield  {title} {\bibinfo {title} {{Theory of nonlinear phononics for
  coherent light control of solids}},\ }\href
  {https://doi.org/10.1103/PhysRevB.89.220301} {\bibfield  {journal} {\bibinfo
  {journal} {Phys. Rev. B}\ }\textbf {\bibinfo {volume} {89}},\ \bibinfo
  {pages} {220301} (\bibinfo {year} {2014})}\BibitemShut {NoStop}%
\bibitem [{\citenamefont {Stevens}\ \emph {et~al.}(2002)\citenamefont
  {Stevens}, \citenamefont {Kuhl},\ and\ \citenamefont {Merlin}}]{Stevens2002}%
  \BibitemOpen
  \bibfield  {author} {\bibinfo {author} {\bibfnamefont {T.~E.}\ \bibnamefont
  {Stevens}}, \bibinfo {author} {\bibfnamefont {J.}~\bibnamefont {Kuhl}},\ and\
  \bibinfo {author} {\bibfnamefont {R.}~\bibnamefont {Merlin}},\ }\bibfield
  {title} {\bibinfo {title} {{Coherent phonon generation and the two stimulated
  Raman tensors}},\ }\href {https://doi.org/10.1103/PhysRevB.65.144304}
  {\bibfield  {journal} {\bibinfo  {journal} {Phys. Rev. B}\ }\textbf {\bibinfo
  {volume} {65}},\ \bibinfo {pages} {144304} (\bibinfo {year}
  {2002})}\BibitemShut {NoStop}%
\bibitem [{\citenamefont {Zeiger}\ \emph {et~al.}(1992)\citenamefont {Zeiger},
  \citenamefont {Vidal}, \citenamefont {Cheng}, \citenamefont {Ippen},
  \citenamefont {Dresselhaus},\ and\ \citenamefont {Dresselhaus}}]{Zeiger1992}%
  \BibitemOpen
  \bibfield  {author} {\bibinfo {author} {\bibfnamefont {H.~J.}\ \bibnamefont
  {Zeiger}}, \bibinfo {author} {\bibfnamefont {J.}~\bibnamefont {Vidal}},
  \bibinfo {author} {\bibfnamefont {T.~K.}\ \bibnamefont {Cheng}}, \bibinfo
  {author} {\bibfnamefont {E.~P.}\ \bibnamefont {Ippen}}, \bibinfo {author}
  {\bibfnamefont {G.}~\bibnamefont {Dresselhaus}},\ and\ \bibinfo {author}
  {\bibfnamefont {M.~S.}\ \bibnamefont {Dresselhaus}},\ }\bibfield  {title}
  {\bibinfo {title} {{Theory for displacive excitation of coherent phonons}},\
  }\href {https://doi.org/10.1103/PhysRevB.45.768} {\bibfield  {journal}
  {\bibinfo  {journal} {Phys. Rev. B}\ }\textbf {\bibinfo {volume} {45}},\
  \bibinfo {pages} {768} (\bibinfo {year} {1992})}\BibitemShut {NoStop}%
\bibitem [{\citenamefont {F\"{o}rst}\ and\ \citenamefont
  {Dekorsy}(2007)}]{Forst2007}%
  \BibitemOpen
  \bibfield  {author} {\bibinfo {author} {\bibfnamefont {M.}~\bibnamefont
  {F\"{o}rst}}\ and\ \bibinfo {author} {\bibfnamefont {T.}~\bibnamefont
  {Dekorsy}},\ }\bibfield  {title} {\bibinfo {title} {{Coherent phonons in bulk
  and low-dimensional semiconductors}},\ }in\ \href
  {https://doi.org/10.1201/9781420017519} {\emph {\bibinfo {booktitle}
  {{Coherent Vibrational Dynamics}}}},\ \bibinfo {editor} {edited by\ \bibinfo
  {editor} {\bibfnamefont {G.}~\bibnamefont {Lanzani}}, \bibinfo {editor}
  {\bibfnamefont {G.}~\bibnamefont {Cerullo}},\ and\ \bibinfo {editor}
  {\bibfnamefont {S.}~\bibnamefont {De~Silvestri}}}\ (\bibinfo  {publisher}
  {CRC Press},\ \bibinfo {address} {Boca Raton},\ \bibinfo {year} {2007})\ pp.\
  \bibinfo {pages} {129--172}\BibitemShut {NoStop}%
\bibitem [{\citenamefont {Ruello}\ and\ \citenamefont
  {Gusev}(2015)}]{Ruello2015}%
  \BibitemOpen
  \bibfield  {author} {\bibinfo {author} {\bibfnamefont {P.}~\bibnamefont
  {Ruello}}\ and\ \bibinfo {author} {\bibfnamefont {V.~E.}\ \bibnamefont
  {Gusev}},\ }\bibfield  {title} {\bibinfo {title} {{Physical mechanisms of
  coherent acoustic phonons generation by ultrafast laser action}},\ }\href
  {https://doi.org/10.1016/j.ultras.2014.06.004} {\bibfield  {journal}
  {\bibinfo  {journal} {Ultrasonics}\ }\textbf {\bibinfo {volume} {56}},\
  \bibinfo {pages} {21} (\bibinfo {year} {2015})}\BibitemShut {NoStop}%
\bibitem [{\citenamefont {{Korff Schmising}}\ \emph {et~al.}(2008)\citenamefont
  {{Korff Schmising}}, \citenamefont {Harpoeth}, \citenamefont {Zhavoronkov},
  \citenamefont {Ansari}, \citenamefont {Aku-Leh}, \citenamefont {Woerner},
  \citenamefont {Elsaesser}, \citenamefont {Bargheer}, \citenamefont
  {Schmidbauer}, \citenamefont {Vrejoiu}, \citenamefont {Hesse},\ and\
  \citenamefont {Alexe}}]{KorffSchmising2008}%
  \BibitemOpen
  \bibfield  {author} {\bibinfo {author} {\bibfnamefont {C.~V.}\ \bibnamefont
  {{Korff Schmising}}}, \bibinfo {author} {\bibfnamefont {A.}~\bibnamefont
  {Harpoeth}}, \bibinfo {author} {\bibfnamefont {N.}~\bibnamefont
  {Zhavoronkov}}, \bibinfo {author} {\bibfnamefont {Z.}~\bibnamefont {Ansari}},
  \bibinfo {author} {\bibfnamefont {C.}~\bibnamefont {Aku-Leh}}, \bibinfo
  {author} {\bibfnamefont {M.}~\bibnamefont {Woerner}}, \bibinfo {author}
  {\bibfnamefont {T.}~\bibnamefont {Elsaesser}}, \bibinfo {author}
  {\bibfnamefont {M.}~\bibnamefont {Bargheer}}, \bibinfo {author}
  {\bibfnamefont {M.}~\bibnamefont {Schmidbauer}}, \bibinfo {author}
  {\bibfnamefont {I.}~\bibnamefont {Vrejoiu}}, \bibinfo {author} {\bibfnamefont
  {D.}~\bibnamefont {Hesse}},\ and\ \bibinfo {author} {\bibfnamefont
  {M.}~\bibnamefont {Alexe}},\ }\bibfield  {title} {\bibinfo {title}
  {{Ultrafast magnetostriction and phonon-mediated stress in a photoexcited
  ferromagnet}},\ }\href {https://doi.org/10.1103/PhysRevB.78.060404}
  {\bibfield  {journal} {\bibinfo  {journal} {Phys. Rev. B}\ }\textbf {\bibinfo
  {volume} {78}},\ \bibinfo {pages} {060404} (\bibinfo {year}
  {2008})}\BibitemShut {NoStop}%
\bibitem [{\citenamefont {Nie}\ \emph {et~al.}(2006)\citenamefont {Nie},
  \citenamefont {Wang}, \citenamefont {Park}, \citenamefont {Clinite},\ and\
  \citenamefont {Cao}}]{Nie2006}%
  \BibitemOpen
  \bibfield  {author} {\bibinfo {author} {\bibfnamefont {S.}~\bibnamefont
  {Nie}}, \bibinfo {author} {\bibfnamefont {X.}~\bibnamefont {Wang}}, \bibinfo
  {author} {\bibfnamefont {H.}~\bibnamefont {Park}}, \bibinfo {author}
  {\bibfnamefont {R.}~\bibnamefont {Clinite}},\ and\ \bibinfo {author}
  {\bibfnamefont {J.}~\bibnamefont {Cao}},\ }\bibfield  {title} {\bibinfo
  {title} {{Measurement of the electronic Gr\"{u}neisen constant using
  femtosecond electron diffraction}},\ }\href
  {https://doi.org/10.1103/PhysRevLett.96.025901} {\bibfield  {journal}
  {\bibinfo  {journal} {Phys. Rev. Lett.}\ }\textbf {\bibinfo {volume} {96}},\
  \bibinfo {pages} {15} (\bibinfo {year} {2006})}\BibitemShut {NoStop}%
\bibitem [{\citenamefont {Harb}\ \emph {et~al.}(2009)\citenamefont {Harb},
  \citenamefont {Peng}, \citenamefont {Sciaini}, \citenamefont {Hebeisen},
  \citenamefont {Ernstorfer}, \citenamefont {Eriksson}, \citenamefont
  {Lagally}, \citenamefont {Kruglik},\ and\ \citenamefont {Miller}}]{Harb2009}%
  \BibitemOpen
  \bibfield  {author} {\bibinfo {author} {\bibfnamefont {M.}~\bibnamefont
  {Harb}}, \bibinfo {author} {\bibfnamefont {W.}~\bibnamefont {Peng}}, \bibinfo
  {author} {\bibfnamefont {G.}~\bibnamefont {Sciaini}}, \bibinfo {author}
  {\bibfnamefont {C.~T.}\ \bibnamefont {Hebeisen}}, \bibinfo {author}
  {\bibfnamefont {R.}~\bibnamefont {Ernstorfer}}, \bibinfo {author}
  {\bibfnamefont {M.~A.}\ \bibnamefont {Eriksson}}, \bibinfo {author}
  {\bibfnamefont {M.~G.}\ \bibnamefont {Lagally}}, \bibinfo {author}
  {\bibfnamefont {S.~G.}\ \bibnamefont {Kruglik}},\ and\ \bibinfo {author}
  {\bibfnamefont {R.~J.~D.}\ \bibnamefont {Miller}},\ }\bibfield  {title}
  {\bibinfo {title} {{Excitation of longitudinal and transverse coherent
  acoustic phonons in nanometer free-standing films of (001) Si}},\ }\href
  {https://doi.org/10.1103/PhysRevB.79.094301} {\bibfield  {journal} {\bibinfo
  {journal} {Phys. Rev. B}\ }\textbf {\bibinfo {volume} {79}},\ \bibinfo
  {pages} {094301} (\bibinfo {year} {2009})}\BibitemShut {NoStop}%
\bibitem [{\citenamefont {Bugayev}\ \emph {et~al.}(2011)\citenamefont
  {Bugayev}, \citenamefont {Esmail}, \citenamefont {Abdel-Fattah},\ and\
  \citenamefont {Elsayed-Ali}}]{Bugayev2011}%
  \BibitemOpen
  \bibfield  {author} {\bibinfo {author} {\bibfnamefont {A.}~\bibnamefont
  {Bugayev}}, \bibinfo {author} {\bibfnamefont {A.}~\bibnamefont {Esmail}},
  \bibinfo {author} {\bibfnamefont {M.}~\bibnamefont {Abdel-Fattah}},\ and\
  \bibinfo {author} {\bibfnamefont {H.~E.}\ \bibnamefont {Elsayed-Ali}},\
  }\bibfield  {title} {\bibinfo {title} {{Coherent phonons in bismuth film
  observed by ultrafast electron diffraction}},\ }\href
  {https://doi.org/10.1063/1.3574888} {\bibfield  {journal} {\bibinfo
  {journal} {AIP Adv.}\ }\textbf {\bibinfo {volume} {1}},\ \bibinfo {pages}
  {012117} (\bibinfo {year} {2011})}\BibitemShut {NoStop}%
\bibitem [{\citenamefont {Moriena}\ \emph {et~al.}(2012)\citenamefont
  {Moriena}, \citenamefont {Hada}, \citenamefont {Sciaini}, \citenamefont
  {Matsuo},\ and\ \citenamefont {{Dwayne Miller}}}]{Moriena2012}%
  \BibitemOpen
  \bibfield  {author} {\bibinfo {author} {\bibfnamefont {G.}~\bibnamefont
  {Moriena}}, \bibinfo {author} {\bibfnamefont {M.}~\bibnamefont {Hada}},
  \bibinfo {author} {\bibfnamefont {G.}~\bibnamefont {Sciaini}}, \bibinfo
  {author} {\bibfnamefont {J.}~\bibnamefont {Matsuo}},\ and\ \bibinfo {author}
  {\bibfnamefont {R.~J.}\ \bibnamefont {{Dwayne Miller}}},\ }\bibfield  {title}
  {\bibinfo {title} {{Femtosecond electron diffraction: Preparation and
  characterization of (110)-oriented bismuth films}},\ }\href
  {https://doi.org/10.1063/1.3684975} {\bibfield  {journal} {\bibinfo
  {journal} {J. Appl. Phys.}\ }\textbf {\bibinfo {volume} {111}},\ \bibinfo
  {pages} {043504} (\bibinfo {year} {2012})}\BibitemShut {NoStop}%
\bibitem [{\citenamefont {Park}\ \emph {et~al.}(2009)\citenamefont {Park},
  \citenamefont {Baskin}, \citenamefont {Barwick}, \citenamefont {Kwon},\ and\
  \citenamefont {Zewail}}]{Park2009}%
  \BibitemOpen
  \bibfield  {author} {\bibinfo {author} {\bibfnamefont {H.~S.}\ \bibnamefont
  {Park}}, \bibinfo {author} {\bibfnamefont {J.~S.}\ \bibnamefont {Baskin}},
  \bibinfo {author} {\bibfnamefont {B.}~\bibnamefont {Barwick}}, \bibinfo
  {author} {\bibfnamefont {O.-H.}\ \bibnamefont {Kwon}},\ and\ \bibinfo
  {author} {\bibfnamefont {A.~H.}\ \bibnamefont {Zewail}},\ }\bibfield  {title}
  {\bibinfo {title} {{4D ultrafast electron microscopy: Imaging of atomic
  motions, acoustic resonances, and moir{\'{e}} fringe dynamics}},\ }\href
  {https://doi.org/10.1016/j.ultramic.2009.08.005} {\bibfield  {journal}
  {\bibinfo  {journal} {Ultramicroscopy}\ }\textbf {\bibinfo {volume} {110}},\
  \bibinfo {pages} {7} (\bibinfo {year} {2009})}\BibitemShut {NoStop}%
\bibitem [{\citenamefont {Chatelain}\ \emph {et~al.}(2014)\citenamefont
  {Chatelain}, \citenamefont {Morrison}, \citenamefont {Klarenaar},\ and\
  \citenamefont {Siwick}}]{Chatelain2014}%
  \BibitemOpen
  \bibfield  {author} {\bibinfo {author} {\bibfnamefont {R.~P.}\ \bibnamefont
  {Chatelain}}, \bibinfo {author} {\bibfnamefont {V.~R.}\ \bibnamefont
  {Morrison}}, \bibinfo {author} {\bibfnamefont {B.~L.~M.}\ \bibnamefont
  {Klarenaar}},\ and\ \bibinfo {author} {\bibfnamefont {B.~J.}\ \bibnamefont
  {Siwick}},\ }\bibfield  {title} {\bibinfo {title} {{Coherent and incoherent
  electron-phonon coupling in graphite observed with radio-frequency compressed
  ultrafast electron diffraction}},\ }\href
  {https://doi.org/10.1103/PhysRevLett.113.235502} {\bibfield  {journal}
  {\bibinfo  {journal} {Phys. Rev. Lett.}\ }\textbf {\bibinfo {volume} {113}},\
  \bibinfo {pages} {235502} (\bibinfo {year} {2014})}\BibitemShut {NoStop}%
\bibitem [{\citenamefont {Zhang}\ \emph {et~al.}(2019)\citenamefont {Zhang},
  \citenamefont {Wang}, \citenamefont {Li}, \citenamefont {Shi}, \citenamefont
  {Wu}, \citenamefont {Lin}, \citenamefont {Zhang}, \citenamefont {Liu},
  \citenamefont {Liu}, \citenamefont {Wang}, \citenamefont {Dong},\ and\
  \citenamefont {Wang}}]{Zhang2019b}%
  \BibitemOpen
  \bibfield  {author} {\bibinfo {author} {\bibfnamefont {M.~Y.}\ \bibnamefont
  {Zhang}}, \bibinfo {author} {\bibfnamefont {Z.~X.}\ \bibnamefont {Wang}},
  \bibinfo {author} {\bibfnamefont {Y.~N.}\ \bibnamefont {Li}}, \bibinfo
  {author} {\bibfnamefont {L.~Y.}\ \bibnamefont {Shi}}, \bibinfo {author}
  {\bibfnamefont {D.}~\bibnamefont {Wu}}, \bibinfo {author} {\bibfnamefont
  {T.}~\bibnamefont {Lin}}, \bibinfo {author} {\bibfnamefont {S.~J.}\
  \bibnamefont {Zhang}}, \bibinfo {author} {\bibfnamefont {Y.~Q.}\ \bibnamefont
  {Liu}}, \bibinfo {author} {\bibfnamefont {Q.~M.}\ \bibnamefont {Liu}},
  \bibinfo {author} {\bibfnamefont {J.}~\bibnamefont {Wang}}, \bibinfo {author}
  {\bibfnamefont {T.}~\bibnamefont {Dong}},\ and\ \bibinfo {author}
  {\bibfnamefont {N.~L.}\ \bibnamefont {Wang}},\ }\bibfield  {title} {\bibinfo
  {title} {{Light-induced subpicosecond lattice symmetry switch in MoTe$_2$}},\
  }\href {https://doi.org/10.1103/PhysRevX.9.021036} {\bibfield  {journal}
  {\bibinfo  {journal} {Phys. Rev. X}\ }\textbf {\bibinfo {volume} {9}},\
  \bibinfo {pages} {021036} (\bibinfo {year} {2019})}\BibitemShut {NoStop}%
\bibitem [{\citenamefont {Huber}\ \emph {et~al.}(2014)\citenamefont {Huber},
  \citenamefont {Mariager}, \citenamefont {Ferrer}, \citenamefont
  {Sch{\"{a}}fer}, \citenamefont {Johnson}, \citenamefont {Gr{\"{u}}bel},
  \citenamefont {L{\"{u}}bcke}, \citenamefont {Huber}, \citenamefont {Kubacka},
  \citenamefont {Dornes}, \citenamefont {Laulhe}, \citenamefont {Ravy},
  \citenamefont {Ingold}, \citenamefont {Beaud}, \citenamefont {Demsar},\ and\
  \citenamefont {Johnson}}]{Huber2014}%
  \BibitemOpen
  \bibfield  {author} {\bibinfo {author} {\bibfnamefont {T.}~\bibnamefont
  {Huber}}, \bibinfo {author} {\bibfnamefont {S.~O.}\ \bibnamefont {Mariager}},
  \bibinfo {author} {\bibfnamefont {A.}~\bibnamefont {Ferrer}}, \bibinfo
  {author} {\bibfnamefont {H.}~\bibnamefont {Sch{\"{a}}fer}}, \bibinfo {author}
  {\bibfnamefont {J.~A.}\ \bibnamefont {Johnson}}, \bibinfo {author}
  {\bibfnamefont {S.}~\bibnamefont {Gr{\"{u}}bel}}, \bibinfo {author}
  {\bibfnamefont {A.}~\bibnamefont {L{\"{u}}bcke}}, \bibinfo {author}
  {\bibfnamefont {L.}~\bibnamefont {Huber}}, \bibinfo {author} {\bibfnamefont
  {T.}~\bibnamefont {Kubacka}}, \bibinfo {author} {\bibfnamefont
  {C.}~\bibnamefont {Dornes}}, \bibinfo {author} {\bibfnamefont
  {C.}~\bibnamefont {Laulhe}}, \bibinfo {author} {\bibfnamefont
  {S.}~\bibnamefont {Ravy}}, \bibinfo {author} {\bibfnamefont {G.}~\bibnamefont
  {Ingold}}, \bibinfo {author} {\bibfnamefont {P.}~\bibnamefont {Beaud}},
  \bibinfo {author} {\bibfnamefont {J.}~\bibnamefont {Demsar}},\ and\ \bibinfo
  {author} {\bibfnamefont {S.~L.}\ \bibnamefont {Johnson}},\ }\bibfield
  {title} {\bibinfo {title} {{Coherent structural dynamics of a prototypical
  charge-density-wave-to-metal transition}},\ }\href
  {https://doi.org/10.1103/PhysRevLett.113.026401} {\bibfield  {journal}
  {\bibinfo  {journal} {Phys. Rev. Lett.}\ }\textbf {\bibinfo {volume} {113}},\
  \bibinfo {pages} {026401} (\bibinfo {year} {2014})}\BibitemShut {NoStop}%
\bibitem [{\citenamefont {Zhang}\ \emph {et~al.}(2020)\citenamefont {Zhang},
  \citenamefont {Shi}, \citenamefont {Guan}, \citenamefont {You}, \citenamefont
  {Zhong}, \citenamefont {Kafle}, \citenamefont {Huang}, \citenamefont {Ding},
  \citenamefont {Bauer}, \citenamefont {Rossnagel}, \citenamefont {Meng},
  \citenamefont {Kapteyn},\ and\ \citenamefont {Murnane}}]{Zhang2020}%
  \BibitemOpen
  \bibfield  {author} {\bibinfo {author} {\bibfnamefont {Y.}~\bibnamefont
  {Zhang}}, \bibinfo {author} {\bibfnamefont {X.}~\bibnamefont {Shi}}, \bibinfo
  {author} {\bibfnamefont {M.}~\bibnamefont {Guan}}, \bibinfo {author}
  {\bibfnamefont {W.}~\bibnamefont {You}}, \bibinfo {author} {\bibfnamefont
  {Y.}~\bibnamefont {Zhong}}, \bibinfo {author} {\bibfnamefont {T.~R.}\
  \bibnamefont {Kafle}}, \bibinfo {author} {\bibfnamefont {Y.}~\bibnamefont
  {Huang}}, \bibinfo {author} {\bibfnamefont {H.}~\bibnamefont {Ding}},
  \bibinfo {author} {\bibfnamefont {M.}~\bibnamefont {Bauer}}, \bibinfo
  {author} {\bibfnamefont {K.}~\bibnamefont {Rossnagel}}, \bibinfo {author}
  {\bibfnamefont {S.}~\bibnamefont {Meng}}, \bibinfo {author} {\bibfnamefont
  {H.~C.}\ \bibnamefont {Kapteyn}},\ and\ \bibinfo {author} {\bibfnamefont
  {M.~M.}\ \bibnamefont {Murnane}},\ }\href@noop {} {\bibinfo {title}
  {{Creation of a novel inverted charge density wave state}}} (\bibinfo {year}
  {2020}),\ \Eprint {https://arxiv.org/abs/2011.07623} {arXiv:2011.07623}
  \BibitemShut {NoStop}%
\bibitem [{\citenamefont {Wang}\ and\ \citenamefont {Wang}(2019)}]{Wang2019}%
  \BibitemOpen
  \bibfield  {author} {\bibinfo {author} {\bibfnamefont {Y.}~\bibnamefont
  {Wang}}\ and\ \bibinfo {author} {\bibfnamefont {D.}~\bibnamefont {Wang}},\
  }\bibfield  {title} {\bibinfo {title} {{Writing and erasing topological
  defects in charge density wave materials with femtosecond laser pulses}},\
  }\href {https://doi.org/10.1364/OL.44.002939} {\bibfield  {journal} {\bibinfo
   {journal} {Opt. Lett.}\ }\textbf {\bibinfo {volume} {44}},\ \bibinfo {pages}
  {2939} (\bibinfo {year} {2019})}\BibitemShut {NoStop}%
\bibitem [{\citenamefont {Duan}\ \emph {et~al.}(2021)\citenamefont {Duan},
  \citenamefont {Cheng}, \citenamefont {Xia}, \citenamefont {Yang},
  \citenamefont {Xu}, \citenamefont {Qi}, \citenamefont {Huang}, \citenamefont
  {Tang}, \citenamefont {Guo}, \citenamefont {Luo}, \citenamefont {Qian},
  \citenamefont {Xiang}, \citenamefont {Zhang},\ and\ \citenamefont
  {Zhang}}]{Duan2021}%
  \BibitemOpen
  \bibfield  {author} {\bibinfo {author} {\bibfnamefont {S.}~\bibnamefont
  {Duan}}, \bibinfo {author} {\bibfnamefont {Y.}~\bibnamefont {Cheng}},
  \bibinfo {author} {\bibfnamefont {W.}~\bibnamefont {Xia}}, \bibinfo {author}
  {\bibfnamefont {Y.}~\bibnamefont {Yang}}, \bibinfo {author} {\bibfnamefont
  {C.}~\bibnamefont {Xu}}, \bibinfo {author} {\bibfnamefont {F.}~\bibnamefont
  {Qi}}, \bibinfo {author} {\bibfnamefont {C.}~\bibnamefont {Huang}}, \bibinfo
  {author} {\bibfnamefont {T.}~\bibnamefont {Tang}}, \bibinfo {author}
  {\bibfnamefont {Y.}~\bibnamefont {Guo}}, \bibinfo {author} {\bibfnamefont
  {W.}~\bibnamefont {Luo}}, \bibinfo {author} {\bibfnamefont {D.}~\bibnamefont
  {Qian}}, \bibinfo {author} {\bibfnamefont {D.}~\bibnamefont {Xiang}},
  \bibinfo {author} {\bibfnamefont {J.}~\bibnamefont {Zhang}},\ and\ \bibinfo
  {author} {\bibfnamefont {W.}~\bibnamefont {Zhang}},\ }\bibfield  {title}
  {\bibinfo {title} {{Optical manipulation of electronic dimensionality in a
  quantum material}},\ }\href {https://doi.org/10.1038/s41586-021-03643-8}
  {\bibfield  {journal} {\bibinfo  {journal} {Nature}\ }\textbf {\bibinfo
  {volume} {595}},\ \bibinfo {pages} {239} (\bibinfo {year}
  {2021})}\BibitemShut {NoStop}%
\bibitem [{\citenamefont {Snijders}\ and\ \citenamefont
  {Weitering}(2010)}]{Snijders2010}%
  \BibitemOpen
  \bibfield  {author} {\bibinfo {author} {\bibfnamefont {P.~C.}\ \bibnamefont
  {Snijders}}\ and\ \bibinfo {author} {\bibfnamefont {H.~H.}\ \bibnamefont
  {Weitering}},\ }\bibfield  {title} {\bibinfo {title} {{Colloquium: Electronic
  instabilities in self-assembled atom wires}},\ }\href
  {https://doi.org/10.1103/RevModPhys.82.307} {\bibfield  {journal} {\bibinfo
  {journal} {Rev. Mod. Phys.}\ }\textbf {\bibinfo {volume} {82}},\ \bibinfo
  {pages} {307} (\bibinfo {year} {2010})}\BibitemShut {NoStop}%
\bibitem [{\citenamefont {Wall}\ \emph {et~al.}(2012)\citenamefont {Wall},
  \citenamefont {Krenzer}, \citenamefont {Wippermann}, \citenamefont {Sanna},
  \citenamefont {Klasing}, \citenamefont {Hanisch-Blicharski}, \citenamefont
  {Kammler}, \citenamefont {Schmidt},\ and\ \citenamefont {{Horn-von
  Hoegen}}}]{Wall2012}%
  \BibitemOpen
  \bibfield  {author} {\bibinfo {author} {\bibfnamefont {S.}~\bibnamefont
  {Wall}}, \bibinfo {author} {\bibfnamefont {B.}~\bibnamefont {Krenzer}},
  \bibinfo {author} {\bibfnamefont {S.}~\bibnamefont {Wippermann}}, \bibinfo
  {author} {\bibfnamefont {S.}~\bibnamefont {Sanna}}, \bibinfo {author}
  {\bibfnamefont {F.}~\bibnamefont {Klasing}}, \bibinfo {author} {\bibfnamefont
  {A.}~\bibnamefont {Hanisch-Blicharski}}, \bibinfo {author} {\bibfnamefont
  {M.}~\bibnamefont {Kammler}}, \bibinfo {author} {\bibfnamefont {W.~G.}\
  \bibnamefont {Schmidt}},\ and\ \bibinfo {author} {\bibfnamefont
  {M.}~\bibnamefont {{Horn-von Hoegen}}},\ }\bibfield  {title} {\bibinfo
  {title} {{Atomistic Picture of Charge Density Wave Formation at Surfaces}},\
  }\href {https://doi.org/10.1103/PhysRevLett.109.186101} {\bibfield  {journal}
  {\bibinfo  {journal} {Phys. Rev. Lett.}\ }\textbf {\bibinfo {volume} {109}},\
  \bibinfo {pages} {186101} (\bibinfo {year} {2012})}\BibitemShut {NoStop}%
\bibitem [{\citenamefont {Frigge}\ \emph {et~al.}(2017)\citenamefont {Frigge},
  \citenamefont {Hafke}, \citenamefont {Witte}, \citenamefont {Krenzer},
  \citenamefont {Streub{\"{u}}hr}, \citenamefont {{Samad Syed}}, \citenamefont
  {{Mik{\v{s}}i{\'{c}} Trontl}}, \citenamefont {Avigo}, \citenamefont {Zhou},
  \citenamefont {Ligges}, \citenamefont {von~der Linde}, \citenamefont
  {Bovensiepen}, \citenamefont {{Horn-von Hoegen}}, \citenamefont {Wippermann},
  \citenamefont {L{\"{u}}cke}, \citenamefont {Sanna}, \citenamefont
  {Gerstmann},\ and\ \citenamefont {Schmidt}}]{Frigge2017}%
  \BibitemOpen
  \bibfield  {author} {\bibinfo {author} {\bibfnamefont {T.}~\bibnamefont
  {Frigge}}, \bibinfo {author} {\bibfnamefont {B.}~\bibnamefont {Hafke}},
  \bibinfo {author} {\bibfnamefont {T.}~\bibnamefont {Witte}}, \bibinfo
  {author} {\bibfnamefont {B.}~\bibnamefont {Krenzer}}, \bibinfo {author}
  {\bibfnamefont {C.}~\bibnamefont {Streub{\"{u}}hr}}, \bibinfo {author}
  {\bibfnamefont {A.}~\bibnamefont {{Samad Syed}}}, \bibinfo {author}
  {\bibfnamefont {V.}~\bibnamefont {{Mik{\v{s}}i{\'{c}} Trontl}}}, \bibinfo
  {author} {\bibfnamefont {I.}~\bibnamefont {Avigo}}, \bibinfo {author}
  {\bibfnamefont {P.}~\bibnamefont {Zhou}}, \bibinfo {author} {\bibfnamefont
  {M.}~\bibnamefont {Ligges}}, \bibinfo {author} {\bibfnamefont
  {D.}~\bibnamefont {von~der Linde}}, \bibinfo {author} {\bibfnamefont
  {U.}~\bibnamefont {Bovensiepen}}, \bibinfo {author} {\bibfnamefont
  {M.}~\bibnamefont {{Horn-von Hoegen}}}, \bibinfo {author} {\bibfnamefont
  {S.}~\bibnamefont {Wippermann}}, \bibinfo {author} {\bibfnamefont
  {A.}~\bibnamefont {L{\"{u}}cke}}, \bibinfo {author} {\bibfnamefont
  {S.}~\bibnamefont {Sanna}}, \bibinfo {author} {\bibfnamefont
  {U.}~\bibnamefont {Gerstmann}},\ and\ \bibinfo {author} {\bibfnamefont
  {W.~G.}\ \bibnamefont {Schmidt}},\ }\bibfield  {title} {\bibinfo {title}
  {{Optically excited structural transition in atomic wires on surfaces at the
  quantum limit}},\ }\href {https://doi.org/10.1038/nature21432} {\bibfield
  {journal} {\bibinfo  {journal} {Nature}\ }\textbf {\bibinfo {volume} {544}},\
  \bibinfo {pages} {207} (\bibinfo {year} {2017})}\BibitemShut {NoStop}%
\bibitem [{\citenamefont {Zewail}(2010)}]{Zewail2010}%
  \BibitemOpen
  \bibfield  {author} {\bibinfo {author} {\bibfnamefont {A.~H.}\ \bibnamefont
  {Zewail}},\ }\bibfield  {title} {\bibinfo {title} {{Four-Dimensional Electron
  Microscopy}},\ }\href {https://doi.org/10.1126/science.1166135} {\bibfield
  {journal} {\bibinfo  {journal} {Science}\ }\textbf {\bibinfo {volume}
  {328}},\ \bibinfo {pages} {187} (\bibinfo {year} {2010})}\BibitemShut
  {NoStop}%
\bibitem [{\citenamefont {Cremons}\ \emph {et~al.}(2016)\citenamefont
  {Cremons}, \citenamefont {Plemmons},\ and\ \citenamefont
  {Flannigan}}]{Cremons2016}%
  \BibitemOpen
  \bibfield  {author} {\bibinfo {author} {\bibfnamefont {D.~R.}\ \bibnamefont
  {Cremons}}, \bibinfo {author} {\bibfnamefont {D.~A.}\ \bibnamefont
  {Plemmons}},\ and\ \bibinfo {author} {\bibfnamefont {D.~J.}\ \bibnamefont
  {Flannigan}},\ }\bibfield  {title} {\bibinfo {title} {{Femtosecond electron
  imaging of defect-modulated phonon dynamics}},\ }\href
  {https://doi.org/10.1038/ncomms11230} {\bibfield  {journal} {\bibinfo
  {journal} {Nat. Commun.}\ }\textbf {\bibinfo {volume} {7}},\ \bibinfo {pages}
  {11230} (\bibinfo {year} {2016})}\BibitemShut {NoStop}%
\bibitem [{\citenamefont {McKenna}\ \emph {et~al.}(2017)\citenamefont
  {McKenna}, \citenamefont {Eliason},\ and\ \citenamefont
  {Flannigan}}]{McKenna2017}%
  \BibitemOpen
  \bibfield  {author} {\bibinfo {author} {\bibfnamefont {A.~J.}\ \bibnamefont
  {McKenna}}, \bibinfo {author} {\bibfnamefont {J.~K.}\ \bibnamefont
  {Eliason}},\ and\ \bibinfo {author} {\bibfnamefont {D.~J.}\ \bibnamefont
  {Flannigan}},\ }\bibfield  {title} {\bibinfo {title} {{Spatiotemporal
  evolution of coherent elastic strain waves in a single MoS$_2$ flake}},\
  }\href {https://doi.org/10.1021/acs.nanolett.7b01565} {\bibfield  {journal}
  {\bibinfo  {journal} {Nano Lett.}\ }\textbf {\bibinfo {volume} {17}},\
  \bibinfo {pages} {3952} (\bibinfo {year} {2017})}\BibitemShut {NoStop}%
\bibitem [{\citenamefont {Feist}\ \emph {et~al.}(2018)\citenamefont {Feist},
  \citenamefont {{Rubiano da Silva}}, \citenamefont {Liang}, \citenamefont
  {Ropers},\ and\ \citenamefont {Sch{\"{a}}fer}}]{Feist2018}%
  \BibitemOpen
  \bibfield  {author} {\bibinfo {author} {\bibfnamefont {A.}~\bibnamefont
  {Feist}}, \bibinfo {author} {\bibfnamefont {N.}~\bibnamefont {{Rubiano da
  Silva}}}, \bibinfo {author} {\bibfnamefont {W.}~\bibnamefont {Liang}},
  \bibinfo {author} {\bibfnamefont {C.}~\bibnamefont {Ropers}},\ and\ \bibinfo
  {author} {\bibfnamefont {S.}~\bibnamefont {Sch{\"{a}}fer}},\ }\bibfield
  {title} {\bibinfo {title} {{Nanoscale diffractive probing of strain dynamics
  in ultrafast transmission electron microscopy}},\ }\href
  {https://doi.org/10.1063/1.5009822} {\bibfield  {journal} {\bibinfo
  {journal} {Struct. Dynam.}\ }\textbf {\bibinfo {volume} {5}},\ \bibinfo
  {pages} {014302} (\bibinfo {year} {2018})}\BibitemShut {NoStop}%
\bibitem [{\citenamefont {Kim}\ \emph {et~al.}(2019{\natexlab{a}})\citenamefont
  {Kim}, \citenamefont {Jung}, \citenamefont {Han},\ and\ \citenamefont
  {Kwon}}]{Kim2019}%
  \BibitemOpen
  \bibfield  {author} {\bibinfo {author} {\bibfnamefont {Y.-J.}\ \bibnamefont
  {Kim}}, \bibinfo {author} {\bibfnamefont {H.}~\bibnamefont {Jung}}, \bibinfo
  {author} {\bibfnamefont {S.~W.}\ \bibnamefont {Han}},\ and\ \bibinfo {author}
  {\bibfnamefont {O.-H.}\ \bibnamefont {Kwon}},\ }\bibfield  {title} {\bibinfo
  {title} {{Ultrafast electron microscopy visualizes acoustic vibrations of
  plasmonic nanorods at the interfaces}},\ }\href
  {https://doi.org/10.1016/j.matt.2019.03.004} {\bibfield  {journal} {\bibinfo
  {journal} {Matter}\ }\textbf {\bibinfo {volume} {1}},\ \bibinfo {pages} {481}
  (\bibinfo {year} {2019}{\natexlab{a}})}\BibitemShut {NoStop}%
\bibitem [{\citenamefont {Nakamura}\ \emph {et~al.}(2020)\citenamefont
  {Nakamura}, \citenamefont {Shimojima}, \citenamefont {Chiashi}, \citenamefont
  {Kamitani}, \citenamefont {Sakai}, \citenamefont {Ishiwata}, \citenamefont
  {Li},\ and\ \citenamefont {Ishizaka}}]{Nakamura2020}%
  \BibitemOpen
  \bibfield  {author} {\bibinfo {author} {\bibfnamefont {A.}~\bibnamefont
  {Nakamura}}, \bibinfo {author} {\bibfnamefont {T.}~\bibnamefont {Shimojima}},
  \bibinfo {author} {\bibfnamefont {Y.}~\bibnamefont {Chiashi}}, \bibinfo
  {author} {\bibfnamefont {M.}~\bibnamefont {Kamitani}}, \bibinfo {author}
  {\bibfnamefont {H.}~\bibnamefont {Sakai}}, \bibinfo {author} {\bibfnamefont
  {S.}~\bibnamefont {Ishiwata}}, \bibinfo {author} {\bibfnamefont
  {H.}~\bibnamefont {Li}},\ and\ \bibinfo {author} {\bibfnamefont
  {K.}~\bibnamefont {Ishizaka}},\ }\bibfield  {title} {\bibinfo {title}
  {{Nanoscale imaging of unusual photoacoustic waves in thin flake VTe$_2$}},\
  }\href {https://doi.org/10.1021/acs.nanolett.0c01006} {\bibfield  {journal}
  {\bibinfo  {journal} {Nano Lett.}\ }\textbf {\bibinfo {volume} {20}},\
  \bibinfo {pages} {4932} (\bibinfo {year} {2020})}\BibitemShut {NoStop}%
\bibitem [{\citenamefont {Kim}\ \emph {et~al.}(2020)\citenamefont {Kim},
  \citenamefont {Lee}, \citenamefont {Kim},\ and\ \citenamefont
  {Kwon}}]{Kim2020}%
  \BibitemOpen
  \bibfield  {author} {\bibinfo {author} {\bibfnamefont {Y.-J.}\ \bibnamefont
  {Kim}}, \bibinfo {author} {\bibfnamefont {Y.}~\bibnamefont {Lee}}, \bibinfo
  {author} {\bibfnamefont {K.}~\bibnamefont {Kim}},\ and\ \bibinfo {author}
  {\bibfnamefont {O.-H.}\ \bibnamefont {Kwon}},\ }\bibfield  {title} {\bibinfo
  {title} {{Light-induced anisotropic morphological dynamics of black
  phosphorus membranes visualized by dark-field ultrafast electron
  microscopy}},\ }\href {https://doi.org/10.1021/acsnano.0c03644} {\bibfield
  {journal} {\bibinfo  {journal} {ACS Nano}\ }\textbf {\bibinfo {volume}
  {14}},\ \bibinfo {pages} {11383} (\bibinfo {year} {2020})}\BibitemShut
  {NoStop}%
\bibitem [{\citenamefont {Srivastava}(2001)}]{Srivastava2001}%
  \BibitemOpen
  \bibfield  {author} {\bibinfo {author} {\bibfnamefont {A.~M.}\ \bibnamefont
  {Srivastava}},\ }\bibinfo {title} {Topological defects in condensed matter
  physics},\ in\ \href {https://doi.org/10.1007/978-93-86279-07-1_5} {\emph
  {\bibinfo {booktitle} {Field Theories in Condensed Matter Physics}}}\
  (\bibinfo  {publisher} {Hindustan Book Agency},\ \bibinfo {address}
  {Gurgaon},\ \bibinfo {year} {2001})\ pp.\ \bibinfo {pages}
  {189--237}\BibitemShut {NoStop}%
\bibitem [{\citenamefont {Stojchevska}\ \emph {et~al.}(2014)\citenamefont
  {Stojchevska}, \citenamefont {Vaskivskyi}, \citenamefont {Mertelj},
  \citenamefont {Kusar}, \citenamefont {Svetin}, \citenamefont {Brazovskii},\
  and\ \citenamefont {Mihailovic}}]{Stojchevska2014}%
  \BibitemOpen
  \bibfield  {author} {\bibinfo {author} {\bibfnamefont {L.}~\bibnamefont
  {Stojchevska}}, \bibinfo {author} {\bibfnamefont {I.}~\bibnamefont
  {Vaskivskyi}}, \bibinfo {author} {\bibfnamefont {T.}~\bibnamefont {Mertelj}},
  \bibinfo {author} {\bibfnamefont {P.}~\bibnamefont {Kusar}}, \bibinfo
  {author} {\bibfnamefont {D.}~\bibnamefont {Svetin}}, \bibinfo {author}
  {\bibfnamefont {S.}~\bibnamefont {Brazovskii}},\ and\ \bibinfo {author}
  {\bibfnamefont {D.}~\bibnamefont {Mihailovic}},\ }\bibfield  {title}
  {\bibinfo {title} {{Ultrafast switching to a stable hidden quantum state in
  an electronic crystal}},\ }\href {https://doi.org/10.1126/science.1241591}
  {\bibfield  {journal} {\bibinfo  {journal} {Science}\ }\textbf {\bibinfo
  {volume} {344}},\ \bibinfo {pages} {177} (\bibinfo {year}
  {2014})}\BibitemShut {NoStop}%
\bibitem [{\citenamefont {Gerasimenko}\ \emph {et~al.}(2019)\citenamefont
  {Gerasimenko}, \citenamefont {Vaskivskyi}, \citenamefont {Litskevich},
  \citenamefont {Ravnik}, \citenamefont {Vodeb}, \citenamefont {Diego},
  \citenamefont {Kabanov},\ and\ \citenamefont {Mihailovic}}]{Gerasimenko2019}%
  \BibitemOpen
  \bibfield  {author} {\bibinfo {author} {\bibfnamefont {Y.~A.}\ \bibnamefont
  {Gerasimenko}}, \bibinfo {author} {\bibfnamefont {I.}~\bibnamefont
  {Vaskivskyi}}, \bibinfo {author} {\bibfnamefont {M.}~\bibnamefont
  {Litskevich}}, \bibinfo {author} {\bibfnamefont {J.}~\bibnamefont {Ravnik}},
  \bibinfo {author} {\bibfnamefont {J.}~\bibnamefont {Vodeb}}, \bibinfo
  {author} {\bibfnamefont {M.}~\bibnamefont {Diego}}, \bibinfo {author}
  {\bibfnamefont {V.}~\bibnamefont {Kabanov}},\ and\ \bibinfo {author}
  {\bibfnamefont {D.}~\bibnamefont {Mihailovic}},\ }\bibfield  {title}
  {\bibinfo {title} {{Quantum jamming transition to a correlated electron glass
  in 1\textit{T}-TaS$_2$}},\ }\href {https://doi.org/10.1038/s41563-019-0423-3}
  {\bibfield  {journal} {\bibinfo  {journal} {Nat. Mater.}\ }\textbf {\bibinfo
  {volume} {18}},\ \bibinfo {pages} {1078} (\bibinfo {year}
  {2019})}\BibitemShut {NoStop}%
\bibitem [{\citenamefont {Stoica}\ \emph {et~al.}(2019)\citenamefont {Stoica},
  \citenamefont {Laanait}, \citenamefont {Dai}, \citenamefont {Hong},
  \citenamefont {Yuan}, \citenamefont {Zhang}, \citenamefont {Lei},
  \citenamefont {McCarter}, \citenamefont {Yadav}, \citenamefont {Damodaran},
  \citenamefont {Das}, \citenamefont {Stone}, \citenamefont {Karapetrova},
  \citenamefont {Walko}, \citenamefont {Zhang}, \citenamefont {Martin},
  \citenamefont {Ramesh}, \citenamefont {Chen}, \citenamefont {Wen},
  \citenamefont {Gopalan},\ and\ \citenamefont {Freeland}}]{Stoica2019}%
  \BibitemOpen
  \bibfield  {author} {\bibinfo {author} {\bibfnamefont {V.~A.}\ \bibnamefont
  {Stoica}}, \bibinfo {author} {\bibfnamefont {N.}~\bibnamefont {Laanait}},
  \bibinfo {author} {\bibfnamefont {C.}~\bibnamefont {Dai}}, \bibinfo {author}
  {\bibfnamefont {Z.}~\bibnamefont {Hong}}, \bibinfo {author} {\bibfnamefont
  {Y.}~\bibnamefont {Yuan}}, \bibinfo {author} {\bibfnamefont {Z.}~\bibnamefont
  {Zhang}}, \bibinfo {author} {\bibfnamefont {S.}~\bibnamefont {Lei}}, \bibinfo
  {author} {\bibfnamefont {M.~R.}\ \bibnamefont {McCarter}}, \bibinfo {author}
  {\bibfnamefont {A.}~\bibnamefont {Yadav}}, \bibinfo {author} {\bibfnamefont
  {A.~R.}\ \bibnamefont {Damodaran}}, \bibinfo {author} {\bibfnamefont
  {S.}~\bibnamefont {Das}}, \bibinfo {author} {\bibfnamefont {G.~A.}\
  \bibnamefont {Stone}}, \bibinfo {author} {\bibfnamefont {J.}~\bibnamefont
  {Karapetrova}}, \bibinfo {author} {\bibfnamefont {D.~A.}\ \bibnamefont
  {Walko}}, \bibinfo {author} {\bibfnamefont {X.}~\bibnamefont {Zhang}},
  \bibinfo {author} {\bibfnamefont {L.~W.}\ \bibnamefont {Martin}}, \bibinfo
  {author} {\bibfnamefont {R.}~\bibnamefont {Ramesh}}, \bibinfo {author}
  {\bibfnamefont {L.-Q.}\ \bibnamefont {Chen}}, \bibinfo {author}
  {\bibfnamefont {H.}~\bibnamefont {Wen}}, \bibinfo {author} {\bibfnamefont
  {V.}~\bibnamefont {Gopalan}},\ and\ \bibinfo {author} {\bibfnamefont {J.~W.}\
  \bibnamefont {Freeland}},\ }\bibfield  {title} {\bibinfo {title} {{Optical
  creation of a supercrystal with three-dimensional nanoscale periodicity}},\
  }\href {https://doi.org/10.1038/s41563-019-0311-x} {\bibfield  {journal}
  {\bibinfo  {journal} {Nat. Mater.}\ }\textbf {\bibinfo {volume} {18}},\
  \bibinfo {pages} {377} (\bibinfo {year} {2019})}\BibitemShut {NoStop}%
\bibitem [{\citenamefont {B{\"{u}}ttner}\ \emph {et~al.}(2021)\citenamefont
  {B{\"{u}}ttner}, \citenamefont {Pfau}, \citenamefont {B{\"{o}}ttcher},
  \citenamefont {Schneider}, \citenamefont {Mercurio}, \citenamefont
  {G{\"{u}}nther}, \citenamefont {Hessing}, \citenamefont {Klose},
  \citenamefont {Wittmann}, \citenamefont {Gerlinger}, \citenamefont {Kern},
  \citenamefont {Str{\"{u}}ber}, \citenamefont {{von Korff Schmising}},
  \citenamefont {Fuchs}, \citenamefont {Engel}, \citenamefont {Churikova},
  \citenamefont {Huang}, \citenamefont {Suzuki}, \citenamefont {Lemesh},
  \citenamefont {Huang}, \citenamefont {Caretta}, \citenamefont {Weder},
  \citenamefont {Gaida}, \citenamefont {M{\"{o}}ller}, \citenamefont {Harvey},
  \citenamefont {Zayko}, \citenamefont {Bagschik}, \citenamefont {Carley},
  \citenamefont {Mercadier}, \citenamefont {Schlappa}, \citenamefont
  {Yaroslavtsev}, \citenamefont {{Le Guyarder}}, \citenamefont {Gerasimova},
  \citenamefont {Scherz}, \citenamefont {Deiter}, \citenamefont {Gort},
  \citenamefont {Hickin}, \citenamefont {Zhu}, \citenamefont {Turcato},
  \citenamefont {Lomidze}, \citenamefont {Erdinger}, \citenamefont {Castoldi},
  \citenamefont {Maffessanti}, \citenamefont {Porro}, \citenamefont
  {Samartsev}, \citenamefont {Sinova}, \citenamefont {Ropers}, \citenamefont
  {Mentink}, \citenamefont {Dup{\'{e}}}, \citenamefont {Beach},\ and\
  \citenamefont {Eisebitt}}]{Buttner2021}%
  \BibitemOpen
  \bibfield  {author} {\bibinfo {author} {\bibfnamefont {F.}~\bibnamefont
  {B{\"{u}}ttner}}, \bibinfo {author} {\bibfnamefont {B.}~\bibnamefont {Pfau}},
  \bibinfo {author} {\bibfnamefont {M.}~\bibnamefont {B{\"{o}}ttcher}},
  \bibinfo {author} {\bibfnamefont {M.}~\bibnamefont {Schneider}}, \bibinfo
  {author} {\bibfnamefont {G.}~\bibnamefont {Mercurio}}, \bibinfo {author}
  {\bibfnamefont {C.~M.}\ \bibnamefont {G{\"{u}}nther}}, \bibinfo {author}
  {\bibfnamefont {P.}~\bibnamefont {Hessing}}, \bibinfo {author} {\bibfnamefont
  {C.}~\bibnamefont {Klose}}, \bibinfo {author} {\bibfnamefont
  {A.}~\bibnamefont {Wittmann}}, \bibinfo {author} {\bibfnamefont
  {K.}~\bibnamefont {Gerlinger}}, \bibinfo {author} {\bibfnamefont {L.-M.}\
  \bibnamefont {Kern}}, \bibinfo {author} {\bibfnamefont {C.}~\bibnamefont
  {Str{\"{u}}ber}}, \bibinfo {author} {\bibfnamefont {C.}~\bibnamefont {{von
  Korff Schmising}}}, \bibinfo {author} {\bibfnamefont {J.}~\bibnamefont
  {Fuchs}}, \bibinfo {author} {\bibfnamefont {D.}~\bibnamefont {Engel}},
  \bibinfo {author} {\bibfnamefont {A.}~\bibnamefont {Churikova}}, \bibinfo
  {author} {\bibfnamefont {S.}~\bibnamefont {Huang}}, \bibinfo {author}
  {\bibfnamefont {D.}~\bibnamefont {Suzuki}}, \bibinfo {author} {\bibfnamefont
  {I.}~\bibnamefont {Lemesh}}, \bibinfo {author} {\bibfnamefont
  {M.}~\bibnamefont {Huang}}, \bibinfo {author} {\bibfnamefont
  {L.}~\bibnamefont {Caretta}}, \bibinfo {author} {\bibfnamefont
  {D.}~\bibnamefont {Weder}}, \bibinfo {author} {\bibfnamefont {J.~H.}\
  \bibnamefont {Gaida}}, \bibinfo {author} {\bibfnamefont {M.}~\bibnamefont
  {M{\"{o}}ller}}, \bibinfo {author} {\bibfnamefont {T.~R.}\ \bibnamefont
  {Harvey}}, \bibinfo {author} {\bibfnamefont {S.}~\bibnamefont {Zayko}},
  \bibinfo {author} {\bibfnamefont {K.}~\bibnamefont {Bagschik}}, \bibinfo
  {author} {\bibfnamefont {R.}~\bibnamefont {Carley}}, \bibinfo {author}
  {\bibfnamefont {L.}~\bibnamefont {Mercadier}}, \bibinfo {author}
  {\bibfnamefont {J.}~\bibnamefont {Schlappa}}, \bibinfo {author}
  {\bibfnamefont {A.}~\bibnamefont {Yaroslavtsev}}, \bibinfo {author}
  {\bibfnamefont {L.}~\bibnamefont {{Le Guyarder}}}, \bibinfo {author}
  {\bibfnamefont {N.}~\bibnamefont {Gerasimova}}, \bibinfo {author}
  {\bibfnamefont {A.}~\bibnamefont {Scherz}}, \bibinfo {author} {\bibfnamefont
  {C.}~\bibnamefont {Deiter}}, \bibinfo {author} {\bibfnamefont
  {R.}~\bibnamefont {Gort}}, \bibinfo {author} {\bibfnamefont {D.}~\bibnamefont
  {Hickin}}, \bibinfo {author} {\bibfnamefont {J.}~\bibnamefont {Zhu}},
  \bibinfo {author} {\bibfnamefont {M.}~\bibnamefont {Turcato}}, \bibinfo
  {author} {\bibfnamefont {D.}~\bibnamefont {Lomidze}}, \bibinfo {author}
  {\bibfnamefont {F.}~\bibnamefont {Erdinger}}, \bibinfo {author}
  {\bibfnamefont {A.}~\bibnamefont {Castoldi}}, \bibinfo {author}
  {\bibfnamefont {S.}~\bibnamefont {Maffessanti}}, \bibinfo {author}
  {\bibfnamefont {M.}~\bibnamefont {Porro}}, \bibinfo {author} {\bibfnamefont
  {A.}~\bibnamefont {Samartsev}}, \bibinfo {author} {\bibfnamefont
  {J.}~\bibnamefont {Sinova}}, \bibinfo {author} {\bibfnamefont
  {C.}~\bibnamefont {Ropers}}, \bibinfo {author} {\bibfnamefont {J.~H.}\
  \bibnamefont {Mentink}}, \bibinfo {author} {\bibfnamefont {B.}~\bibnamefont
  {Dup{\'{e}}}}, \bibinfo {author} {\bibfnamefont {G.~S.~D.}\ \bibnamefont
  {Beach}},\ and\ \bibinfo {author} {\bibfnamefont {S.}~\bibnamefont
  {Eisebitt}},\ }\bibfield  {title} {\bibinfo {title} {{Observation of
  fluctuation-mediated picosecond nucleation of a topological phase}},\ }\href
  {https://doi.org/10.1038/s41563-020-00807-1} {\bibfield  {journal} {\bibinfo
  {journal} {Nat. Mater.}\ }\textbf {\bibinfo {volume} {20}},\ \bibinfo {pages}
  {30} (\bibinfo {year} {2021})}\BibitemShut {NoStop}%
\bibitem [{\citenamefont {Kibble}(2007)}]{Kibble2007}%
  \BibitemOpen
  \bibfield  {author} {\bibinfo {author} {\bibfnamefont {T.}~\bibnamefont
  {Kibble}},\ }\bibfield  {title} {\bibinfo {title} {{Phase-transition dynamics
  in the lab and the universe}},\ }\href {https://doi.org/10.1063/1.2784684}
  {\bibfield  {journal} {\bibinfo  {journal} {Phys. Today}\ }\textbf {\bibinfo
  {volume} {60}},\ \bibinfo {pages} {47} (\bibinfo {year} {2007})}\BibitemShut
  {NoStop}%
\bibitem [{\citenamefont {Sun}\ \emph {et~al.}(2018)\citenamefont {Sun},
  \citenamefont {Sun}, \citenamefont {Zhu}, \citenamefont {Tian}, \citenamefont
  {Yang},\ and\ \citenamefont {Li}}]{Sun2018}%
  \BibitemOpen
  \bibfield  {author} {\bibinfo {author} {\bibfnamefont {K.}~\bibnamefont
  {Sun}}, \bibinfo {author} {\bibfnamefont {S.}~\bibnamefont {Sun}}, \bibinfo
  {author} {\bibfnamefont {C.}~\bibnamefont {Zhu}}, \bibinfo {author}
  {\bibfnamefont {H.}~\bibnamefont {Tian}}, \bibinfo {author} {\bibfnamefont
  {H.}~\bibnamefont {Yang}},\ and\ \bibinfo {author} {\bibfnamefont
  {J.}~\bibnamefont {Li}},\ }\bibfield  {title} {\bibinfo {title} {{Hidden CDW
  states and insulator-to-metal transition after a pulsed femtosecond laser
  excitation in layered chalcogenide 1$T$-TaS$_{2-x}$Se$_x$}},\ }\href
  {https://doi.org/10.1126/sciadv.aas9660} {\bibfield  {journal} {\bibinfo
  {journal} {Sci. Adv.}\ }\textbf {\bibinfo {volume} {4}},\ \bibinfo {pages}
  {eaas9660} (\bibinfo {year} {2018})}\BibitemShut {NoStop}%
\bibitem [{\citenamefont {Vogelgesang}\ \emph {et~al.}(2018)\citenamefont
  {Vogelgesang}, \citenamefont {Storeck}, \citenamefont {Horstmann},
  \citenamefont {Diekmann}, \citenamefont {Sivis}, \citenamefont {Schramm},
  \citenamefont {Rossnagel}, \citenamefont {Sch{\"a}fer},\ and\ \citenamefont
  {Ropers}}]{Vogelgesang2018}%
  \BibitemOpen
  \bibfield  {author} {\bibinfo {author} {\bibfnamefont {S.}~\bibnamefont
  {Vogelgesang}}, \bibinfo {author} {\bibfnamefont {G.}~\bibnamefont
  {Storeck}}, \bibinfo {author} {\bibfnamefont {J.~G.}\ \bibnamefont
  {Horstmann}}, \bibinfo {author} {\bibfnamefont {T.}~\bibnamefont {Diekmann}},
  \bibinfo {author} {\bibfnamefont {M.}~\bibnamefont {Sivis}}, \bibinfo
  {author} {\bibfnamefont {S.}~\bibnamefont {Schramm}}, \bibinfo {author}
  {\bibfnamefont {K.}~\bibnamefont {Rossnagel}}, \bibinfo {author}
  {\bibfnamefont {S.}~\bibnamefont {Sch{\"a}fer}},\ and\ \bibinfo {author}
  {\bibfnamefont {C.}~\bibnamefont {Ropers}},\ }\bibfield  {title} {\bibinfo
  {title} {{Phase ordering of charge density waves traced by ultrafast
  low-energy electron diffraction}},\ }\href
  {https://doi.org/10.1038/nphys4309} {\bibfield  {journal} {\bibinfo
  {journal} {Nat. Phys.}\ }\textbf {\bibinfo {volume} {14}},\ \bibinfo {pages}
  {184} (\bibinfo {year} {2018})}\BibitemShut {NoStop}%
\bibitem [{\citenamefont {Danz}\ \emph {et~al.}(2021)\citenamefont {Danz},
  \citenamefont {Domr{\"{o}}se},\ and\ \citenamefont {Ropers}}]{Danz2021}%
  \BibitemOpen
  \bibfield  {author} {\bibinfo {author} {\bibfnamefont {T.}~\bibnamefont
  {Danz}}, \bibinfo {author} {\bibfnamefont {T.}~\bibnamefont
  {Domr{\"{o}}se}},\ and\ \bibinfo {author} {\bibfnamefont {C.}~\bibnamefont
  {Ropers}},\ }\bibfield  {title} {\bibinfo {title} {{Ultrafast nanoimaging of
  the order parameter in a structural phase transition}},\ }\href
  {https://doi.org/10.1126/science.abd2774} {\bibfield  {journal} {\bibinfo
  {journal} {Science}\ }\textbf {\bibinfo {volume} {371}},\ \bibinfo {pages}
  {371} (\bibinfo {year} {2021})}\BibitemShut {NoStop}%
\bibitem [{\citenamefont {Erasmus}\ \emph {et~al.}(2012)\citenamefont
  {Erasmus}, \citenamefont {Eichberger}, \citenamefont {Haupt}, \citenamefont
  {Boshoff}, \citenamefont {Kassier}, \citenamefont {Birmurske}, \citenamefont
  {Berger}, \citenamefont {Demsar},\ and\ \citenamefont
  {Schwoerer}}]{Erasmus2012}%
  \BibitemOpen
  \bibfield  {author} {\bibinfo {author} {\bibfnamefont {N.}~\bibnamefont
  {Erasmus}}, \bibinfo {author} {\bibfnamefont {M.}~\bibnamefont {Eichberger}},
  \bibinfo {author} {\bibfnamefont {K.}~\bibnamefont {Haupt}}, \bibinfo
  {author} {\bibfnamefont {I.}~\bibnamefont {Boshoff}}, \bibinfo {author}
  {\bibfnamefont {G.}~\bibnamefont {Kassier}}, \bibinfo {author} {\bibfnamefont
  {R.}~\bibnamefont {Birmurske}}, \bibinfo {author} {\bibfnamefont
  {H.}~\bibnamefont {Berger}}, \bibinfo {author} {\bibfnamefont
  {J.}~\bibnamefont {Demsar}},\ and\ \bibinfo {author} {\bibfnamefont
  {H.}~\bibnamefont {Schwoerer}},\ }\bibfield  {title} {\bibinfo {title}
  {{Ultrafast dynamics of charge density waves in 4$H_b$-TaSe$_2$ probed by
  femtosecond electron diffraction}},\ }\href
  {https://doi.org/10.1103/PhysRevLett.109.167402} {\bibfield  {journal}
  {\bibinfo  {journal} {Phys. Rev. Lett.}\ }\textbf {\bibinfo {volume} {109}},\
  \bibinfo {pages} {1} (\bibinfo {year} {2012})}\BibitemShut {NoStop}%
\bibitem [{\citenamefont {Sun}\ \emph {et~al.}(2015)\citenamefont {Sun},
  \citenamefont {Wei}, \citenamefont {Li}, \citenamefont {Cao}, \citenamefont
  {Liu}, \citenamefont {Lu}, \citenamefont {Sun}, \citenamefont {Tian},
  \citenamefont {Yang},\ and\ \citenamefont {Li}}]{Sun2015}%
  \BibitemOpen
  \bibfield  {author} {\bibinfo {author} {\bibfnamefont {S.}~\bibnamefont
  {Sun}}, \bibinfo {author} {\bibfnamefont {L.}~\bibnamefont {Wei}}, \bibinfo
  {author} {\bibfnamefont {Z.}~\bibnamefont {Li}}, \bibinfo {author}
  {\bibfnamefont {G.}~\bibnamefont {Cao}}, \bibinfo {author} {\bibfnamefont
  {Y.}~\bibnamefont {Liu}}, \bibinfo {author} {\bibfnamefont {W.~J.}\
  \bibnamefont {Lu}}, \bibinfo {author} {\bibfnamefont {Y.~P.}\ \bibnamefont
  {Sun}}, \bibinfo {author} {\bibfnamefont {H.}~\bibnamefont {Tian}}, \bibinfo
  {author} {\bibfnamefont {H.}~\bibnamefont {Yang}},\ and\ \bibinfo {author}
  {\bibfnamefont {J.}~\bibnamefont {Li}},\ }\bibfield  {title} {\bibinfo
  {title} {{Direct observation of an optically induced charge density wave
  transition in 1$T$-TaSe$_2$}},\ }\href
  {https://doi.org/10.1103/PhysRevB.92.224303} {\bibfield  {journal} {\bibinfo
  {journal} {Phys. Rev. B}\ }\textbf {\bibinfo {volume} {92}},\ \bibinfo
  {pages} {224303} (\bibinfo {year} {2015})}\BibitemShut {NoStop}%
\bibitem [{\citenamefont {Han}\ \emph {et~al.}(2015)\citenamefont {Han},
  \citenamefont {Zhou}, \citenamefont {Malliakas}, \citenamefont {Duxbury},
  \citenamefont {Mahanti}, \citenamefont {Kanatzidis},\ and\ \citenamefont
  {Ruan}}]{Han2015}%
  \BibitemOpen
  \bibfield  {author} {\bibinfo {author} {\bibfnamefont {T.-R.~T.}\
  \bibnamefont {Han}}, \bibinfo {author} {\bibfnamefont {F.}~\bibnamefont
  {Zhou}}, \bibinfo {author} {\bibfnamefont {C.~D.}\ \bibnamefont {Malliakas}},
  \bibinfo {author} {\bibfnamefont {P.~M.}\ \bibnamefont {Duxbury}}, \bibinfo
  {author} {\bibfnamefont {S.~D.}\ \bibnamefont {Mahanti}}, \bibinfo {author}
  {\bibfnamefont {M.~G.}\ \bibnamefont {Kanatzidis}},\ and\ \bibinfo {author}
  {\bibfnamefont {C.-Y.}\ \bibnamefont {Ruan}},\ }\bibfield  {title} {\bibinfo
  {title} {{Exploration of metastability and hidden phases in correlated
  electron crystals visualized by femtosecond optical doping and electron
  crystallography}},\ }\href {https://doi.org/10.1126/sciadv.1400173}
  {\bibfield  {journal} {\bibinfo  {journal} {Sci. Adv.}\ }\textbf {\bibinfo
  {volume} {1}},\ \bibinfo {pages} {e1400173} (\bibinfo {year}
  {2015})}\BibitemShut {NoStop}%
\bibitem [{\citenamefont {Haupt}\ \emph {et~al.}(2016)\citenamefont {Haupt},
  \citenamefont {Eichberger}, \citenamefont {Erasmus}, \citenamefont {Rohwer},
  \citenamefont {Demsar}, \citenamefont {Rossnagel},\ and\ \citenamefont
  {Schwoerer}}]{Haupt2016}%
  \BibitemOpen
  \bibfield  {author} {\bibinfo {author} {\bibfnamefont {K.}~\bibnamefont
  {Haupt}}, \bibinfo {author} {\bibfnamefont {M.}~\bibnamefont {Eichberger}},
  \bibinfo {author} {\bibfnamefont {N.}~\bibnamefont {Erasmus}}, \bibinfo
  {author} {\bibfnamefont {A.}~\bibnamefont {Rohwer}}, \bibinfo {author}
  {\bibfnamefont {J.}~\bibnamefont {Demsar}}, \bibinfo {author} {\bibfnamefont
  {K.}~\bibnamefont {Rossnagel}},\ and\ \bibinfo {author} {\bibfnamefont
  {H.}~\bibnamefont {Schwoerer}},\ }\bibfield  {title} {\bibinfo {title}
  {{Ultrafast metamorphosis of a complex charge-density wave}},\ }\href
  {https://doi.org/10.1103/PhysRevLett.116.016402} {\bibfield  {journal}
  {\bibinfo  {journal} {Phys. Rev. Lett.}\ }\textbf {\bibinfo {volume} {116}},\
  \bibinfo {pages} {016402} (\bibinfo {year} {2016})}\BibitemShut {NoStop}%
\bibitem [{\citenamefont {Wei}\ \emph {et~al.}(2017)\citenamefont {Wei},
  \citenamefont {Sun}, \citenamefont {Guo}, \citenamefont {Li}, \citenamefont
  {Sun}, \citenamefont {Liu}, \citenamefont {Lu}, \citenamefont {Sun},
  \citenamefont {Tian}, \citenamefont {Yang},\ and\ \citenamefont
  {Li}}]{Wei2017}%
  \BibitemOpen
  \bibfield  {author} {\bibinfo {author} {\bibfnamefont {L.}~\bibnamefont
  {Wei}}, \bibinfo {author} {\bibfnamefont {S.}~\bibnamefont {Sun}}, \bibinfo
  {author} {\bibfnamefont {C.}~\bibnamefont {Guo}}, \bibinfo {author}
  {\bibfnamefont {Z.}~\bibnamefont {Li}}, \bibinfo {author} {\bibfnamefont
  {K.}~\bibnamefont {Sun}}, \bibinfo {author} {\bibfnamefont {Y.}~\bibnamefont
  {Liu}}, \bibinfo {author} {\bibfnamefont {W.}~\bibnamefont {Lu}}, \bibinfo
  {author} {\bibfnamefont {Y.}~\bibnamefont {Sun}}, \bibinfo {author}
  {\bibfnamefont {H.}~\bibnamefont {Tian}}, \bibinfo {author} {\bibfnamefont
  {H.}~\bibnamefont {Yang}},\ and\ \bibinfo {author} {\bibfnamefont
  {J.}~\bibnamefont {Li}},\ }\bibfield  {title} {\bibinfo {title} {{Dynamic
  diffraction effects and coherent breathing oscillations in ultrafast electron
  diffraction in layered 1$T$-TaSeTe}},\ }\href
  {https://doi.org/10.1063/1.4979643} {\bibfield  {journal} {\bibinfo
  {journal} {Struct. Dynam.}\ }\textbf {\bibinfo {volume} {4}},\ \bibinfo
  {pages} {044012} (\bibinfo {year} {2017})}\BibitemShut {NoStop}%
\bibitem [{\citenamefont {{Le Guyader}}\ \emph {et~al.}(2017)\citenamefont {{Le
  Guyader}}, \citenamefont {Chase}, \citenamefont {Reid}, \citenamefont {Li},
  \citenamefont {Svetin}, \citenamefont {Shen}, \citenamefont {Vecchione},
  \citenamefont {Wang}, \citenamefont {Mihailovic},\ and\ \citenamefont
  {D{\"{u}}rr}}]{LeGuyader2017}%
  \BibitemOpen
  \bibfield  {author} {\bibinfo {author} {\bibfnamefont {L.}~\bibnamefont {{Le
  Guyader}}}, \bibinfo {author} {\bibfnamefont {T.}~\bibnamefont {Chase}},
  \bibinfo {author} {\bibfnamefont {A.~H.}\ \bibnamefont {Reid}}, \bibinfo
  {author} {\bibfnamefont {R.~K.}\ \bibnamefont {Li}}, \bibinfo {author}
  {\bibfnamefont {D.}~\bibnamefont {Svetin}}, \bibinfo {author} {\bibfnamefont
  {X.}~\bibnamefont {Shen}}, \bibinfo {author} {\bibfnamefont {T.}~\bibnamefont
  {Vecchione}}, \bibinfo {author} {\bibfnamefont {X.~J.}\ \bibnamefont {Wang}},
  \bibinfo {author} {\bibfnamefont {D.}~\bibnamefont {Mihailovic}},\ and\
  \bibinfo {author} {\bibfnamefont {H.~A.}\ \bibnamefont {D{\"{u}}rr}},\
  }\bibfield  {title} {\bibinfo {title} {{Stacking order dynamics in the
  quasi-two-dimensional dichalcogenide 1$T$-TaS$_2$ probed with MeV ultrafast
  electron diffraction}},\ }\href {https://doi.org/10.1063/1.4982918}
  {\bibfield  {journal} {\bibinfo  {journal} {Struct. Dynam.}\ }\textbf
  {\bibinfo {volume} {4}},\ \bibinfo {pages} {044020} (\bibinfo {year}
  {2017})}\BibitemShut {NoStop}%
\bibitem [{\citenamefont {Li}\ \emph {et~al.}(2019)\citenamefont {Li},
  \citenamefont {Li}, \citenamefont {Sun}, \citenamefont {Wu}, \citenamefont
  {Huang}, \citenamefont {Li}, \citenamefont {Yang}, \citenamefont {Shen},
  \citenamefont {Wang}, \citenamefont {Luo}, \citenamefont {Cava},
  \citenamefont {Robinson}, \citenamefont {Zhu}, \citenamefont {Yin},\ and\
  \citenamefont {Tao}}]{Li2019}%
  \BibitemOpen
  \bibfield  {author} {\bibinfo {author} {\bibfnamefont {J.}~\bibnamefont
  {Li}}, \bibinfo {author} {\bibfnamefont {J.}~\bibnamefont {Li}}, \bibinfo
  {author} {\bibfnamefont {K.}~\bibnamefont {Sun}}, \bibinfo {author}
  {\bibfnamefont {L.}~\bibnamefont {Wu}}, \bibinfo {author} {\bibfnamefont
  {H.}~\bibnamefont {Huang}}, \bibinfo {author} {\bibfnamefont
  {R.}~\bibnamefont {Li}}, \bibinfo {author} {\bibfnamefont {J.}~\bibnamefont
  {Yang}}, \bibinfo {author} {\bibfnamefont {X.}~\bibnamefont {Shen}}, \bibinfo
  {author} {\bibfnamefont {X.}~\bibnamefont {Wang}}, \bibinfo {author}
  {\bibfnamefont {H.}~\bibnamefont {Luo}}, \bibinfo {author} {\bibfnamefont
  {R.~J.}\ \bibnamefont {Cava}}, \bibinfo {author} {\bibfnamefont {I.~K.}\
  \bibnamefont {Robinson}}, \bibinfo {author} {\bibfnamefont {Y.}~\bibnamefont
  {Zhu}}, \bibinfo {author} {\bibfnamefont {W.}~\bibnamefont {Yin}},\ and\
  \bibinfo {author} {\bibfnamefont {J.}~\bibnamefont {Tao}},\ }\href@noop {}
  {\bibinfo {title} {{Ultrafast decoupling of atomic sublattices in a
  charge-density-wave material}}} (\bibinfo {year} {2019}),\ \Eprint
  {https://arxiv.org/abs/1903.09911} {arXiv:1903.09911} \BibitemShut {NoStop}%
\bibitem [{\citenamefont {Ji}\ \emph {et~al.}(2020)\citenamefont {Ji},
  \citenamefont {Gr{\aa}n{\"{a}}s}, \citenamefont {Rossnagel},\ and\
  \citenamefont {Weissenrieder}}]{Ji2020}%
  \BibitemOpen
  \bibfield  {author} {\bibinfo {author} {\bibfnamefont {S.}~\bibnamefont
  {Ji}}, \bibinfo {author} {\bibfnamefont {O.}~\bibnamefont
  {Gr{\aa}n{\"{a}}s}}, \bibinfo {author} {\bibfnamefont {K.}~\bibnamefont
  {Rossnagel}},\ and\ \bibinfo {author} {\bibfnamefont {J.}~\bibnamefont
  {Weissenrieder}},\ }\bibfield  {title} {\bibinfo {title} {{Transient
  three-dimensional structural dynamics in 1$T$-TaSe$_2$}},\ }\href
  {https://doi.org/10.1103/PhysRevB.101.094303} {\bibfield  {journal} {\bibinfo
   {journal} {Phys. Rev. B}\ }\textbf {\bibinfo {volume} {101}},\ \bibinfo
  {pages} {094303} (\bibinfo {year} {2020})}\BibitemShut {NoStop}%
\bibitem [{\citenamefont {Vaskivskyi}\ \emph {et~al.}(2015)\citenamefont
  {Vaskivskyi}, \citenamefont {Gospodaric}, \citenamefont {Brazovskii},
  \citenamefont {Svetin}, \citenamefont {Sutar}, \citenamefont {Goreshnik},
  \citenamefont {Mihailovic}, \citenamefont {Mertelj},\ and\ \citenamefont
  {Mihailovic}}]{Vaskivskyi2015}%
  \BibitemOpen
  \bibfield  {author} {\bibinfo {author} {\bibfnamefont {I.}~\bibnamefont
  {Vaskivskyi}}, \bibinfo {author} {\bibfnamefont {J.}~\bibnamefont
  {Gospodaric}}, \bibinfo {author} {\bibfnamefont {S.}~\bibnamefont
  {Brazovskii}}, \bibinfo {author} {\bibfnamefont {D.}~\bibnamefont {Svetin}},
  \bibinfo {author} {\bibfnamefont {P.}~\bibnamefont {Sutar}}, \bibinfo
  {author} {\bibfnamefont {E.}~\bibnamefont {Goreshnik}}, \bibinfo {author}
  {\bibfnamefont {I.~A.}\ \bibnamefont {Mihailovic}}, \bibinfo {author}
  {\bibfnamefont {T.}~\bibnamefont {Mertelj}},\ and\ \bibinfo {author}
  {\bibfnamefont {D.}~\bibnamefont {Mihailovic}},\ }\bibfield  {title}
  {\bibinfo {title} {{Controlling the metal-to-insulator relaxation of the
  metastable hidden quantum state in 1$T$-TaS$_2$}},\ }\href
  {https://doi.org/10.1126/sciadv.1500168} {\bibfield  {journal} {\bibinfo
  {journal} {Sci. Adv.}\ }\textbf {\bibinfo {volume} {1}},\ \bibinfo {pages}
  {e1500168} (\bibinfo {year} {2015})}\BibitemShut {NoStop}%
\bibitem [{\citenamefont {Cho}\ \emph {et~al.}(2016)\citenamefont {Cho},
  \citenamefont {Cheon}, \citenamefont {Kim}, \citenamefont {Lee},
  \citenamefont {Cho}, \citenamefont {Cheong},\ and\ \citenamefont
  {Yeom}}]{Cho2016}%
  \BibitemOpen
  \bibfield  {author} {\bibinfo {author} {\bibfnamefont {D.}~\bibnamefont
  {Cho}}, \bibinfo {author} {\bibfnamefont {S.}~\bibnamefont {Cheon}}, \bibinfo
  {author} {\bibfnamefont {K.-S.}\ \bibnamefont {Kim}}, \bibinfo {author}
  {\bibfnamefont {S.-H.}\ \bibnamefont {Lee}}, \bibinfo {author} {\bibfnamefont
  {Y.-H.}\ \bibnamefont {Cho}}, \bibinfo {author} {\bibfnamefont {S.-W.}\
  \bibnamefont {Cheong}},\ and\ \bibinfo {author} {\bibfnamefont {H.~W.}\
  \bibnamefont {Yeom}},\ }\bibfield  {title} {\bibinfo {title} {{Nanoscale
  manipulation of the Mott insulating state coupled to charge order in
  1\textit{T}-TaS$_2$}},\ }\href {https://doi.org/10.1038/ncomms10453}
  {\bibfield  {journal} {\bibinfo  {journal} {Nat. Commun.}\ }\textbf {\bibinfo
  {volume} {7}},\ \bibinfo {pages} {10453} (\bibinfo {year}
  {2016})}\BibitemShut {NoStop}%
\bibitem [{\citenamefont {Ma}\ \emph {et~al.}(2016)\citenamefont {Ma},
  \citenamefont {Ye}, \citenamefont {Yu}, \citenamefont {Lu}, \citenamefont
  {Niu}, \citenamefont {Kim}, \citenamefont {Feng}, \citenamefont
  {Tom{\'{a}}nek}, \citenamefont {Son}, \citenamefont {Chen},\ and\
  \citenamefont {Zhang}}]{Ma2016}%
  \BibitemOpen
  \bibfield  {author} {\bibinfo {author} {\bibfnamefont {L.}~\bibnamefont
  {Ma}}, \bibinfo {author} {\bibfnamefont {C.}~\bibnamefont {Ye}}, \bibinfo
  {author} {\bibfnamefont {Y.}~\bibnamefont {Yu}}, \bibinfo {author}
  {\bibfnamefont {X.~F.}\ \bibnamefont {Lu}}, \bibinfo {author} {\bibfnamefont
  {X.}~\bibnamefont {Niu}}, \bibinfo {author} {\bibfnamefont {S.}~\bibnamefont
  {Kim}}, \bibinfo {author} {\bibfnamefont {D.}~\bibnamefont {Feng}}, \bibinfo
  {author} {\bibfnamefont {D.}~\bibnamefont {Tom{\'{a}}nek}}, \bibinfo {author}
  {\bibfnamefont {Y.-W.}\ \bibnamefont {Son}}, \bibinfo {author} {\bibfnamefont
  {X.~H.}\ \bibnamefont {Chen}},\ and\ \bibinfo {author} {\bibfnamefont
  {Y.}~\bibnamefont {Zhang}},\ }\bibfield  {title} {\bibinfo {title} {{A
  metallic mosaic phase and the origin of Mott-insulating state in
  1\textit{T}-TaS$_2$}},\ }\href {https://doi.org/10.1038/ncomms10956}
  {\bibfield  {journal} {\bibinfo  {journal} {Nat. Commun.}\ }\textbf {\bibinfo
  {volume} {7}},\ \bibinfo {pages} {10956} (\bibinfo {year}
  {2016})}\BibitemShut {NoStop}%
\bibitem [{\citenamefont {Spijkerman}\ \emph {et~al.}(1997)\citenamefont
  {Spijkerman}, \citenamefont {de~Boer}, \citenamefont {Meetsma}, \citenamefont
  {Wiegers},\ and\ \citenamefont {van Smaalen}}]{Spijkerman1997}%
  \BibitemOpen
  \bibfield  {author} {\bibinfo {author} {\bibfnamefont {A.}~\bibnamefont
  {Spijkerman}}, \bibinfo {author} {\bibfnamefont {J.~L.}\ \bibnamefont
  {de~Boer}}, \bibinfo {author} {\bibfnamefont {A.}~\bibnamefont {Meetsma}},
  \bibinfo {author} {\bibfnamefont {G.~A.}\ \bibnamefont {Wiegers}},\ and\
  \bibinfo {author} {\bibfnamefont {S.}~\bibnamefont {van Smaalen}},\
  }\bibfield  {title} {\bibinfo {title} {{X-ray crystal-structure refinement of
  the nearly commensurate phase of 1$T$-TaS$_2$ in $(3+2)$-dimensional
  superspace}},\ }\href {https://doi.org/10.1103/PhysRevB.56.13757} {\bibfield
  {journal} {\bibinfo  {journal} {Phys. Rev. B}\ }\textbf {\bibinfo {volume}
  {56}},\ \bibinfo {pages} {13757} (\bibinfo {year} {1997})}\BibitemShut
  {NoStop}%
\bibitem [{\citenamefont {Laulh{\'{e}}}\ \emph {et~al.}(2017)\citenamefont
  {Laulh{\'{e}}}, \citenamefont {Huber}, \citenamefont {Lantz}, \citenamefont
  {Ferrer}, \citenamefont {Mariager}, \citenamefont {Gr{\"{u}}bel},
  \citenamefont {Rittmann}, \citenamefont {Johnson}, \citenamefont {Esposito},
  \citenamefont {L{\"{u}}bcke}, \citenamefont {Huber}, \citenamefont {Kubli},
  \citenamefont {Savoini}, \citenamefont {Jacques}, \citenamefont {Cario},
  \citenamefont {Corraze}, \citenamefont {Janod}, \citenamefont {Ingold},
  \citenamefont {Beaud}, \citenamefont {Johnson},\ and\ \citenamefont
  {Ravy}}]{Laulhe2017}%
  \BibitemOpen
  \bibfield  {author} {\bibinfo {author} {\bibfnamefont {C.}~\bibnamefont
  {Laulh{\'{e}}}}, \bibinfo {author} {\bibfnamefont {T.}~\bibnamefont {Huber}},
  \bibinfo {author} {\bibfnamefont {G.}~\bibnamefont {Lantz}}, \bibinfo
  {author} {\bibfnamefont {A.}~\bibnamefont {Ferrer}}, \bibinfo {author}
  {\bibfnamefont {S.~O.}\ \bibnamefont {Mariager}}, \bibinfo {author}
  {\bibfnamefont {S.}~\bibnamefont {Gr{\"{u}}bel}}, \bibinfo {author}
  {\bibfnamefont {J.}~\bibnamefont {Rittmann}}, \bibinfo {author}
  {\bibfnamefont {J.~A.}\ \bibnamefont {Johnson}}, \bibinfo {author}
  {\bibfnamefont {V.}~\bibnamefont {Esposito}}, \bibinfo {author}
  {\bibfnamefont {A.}~\bibnamefont {L{\"{u}}bcke}}, \bibinfo {author}
  {\bibfnamefont {L.}~\bibnamefont {Huber}}, \bibinfo {author} {\bibfnamefont
  {M.}~\bibnamefont {Kubli}}, \bibinfo {author} {\bibfnamefont
  {M.}~\bibnamefont {Savoini}}, \bibinfo {author} {\bibfnamefont {V.~L.~R.}\
  \bibnamefont {Jacques}}, \bibinfo {author} {\bibfnamefont {L.}~\bibnamefont
  {Cario}}, \bibinfo {author} {\bibfnamefont {B.}~\bibnamefont {Corraze}},
  \bibinfo {author} {\bibfnamefont {E.}~\bibnamefont {Janod}}, \bibinfo
  {author} {\bibfnamefont {G.}~\bibnamefont {Ingold}}, \bibinfo {author}
  {\bibfnamefont {P.}~\bibnamefont {Beaud}}, \bibinfo {author} {\bibfnamefont
  {S.~L.}\ \bibnamefont {Johnson}},\ and\ \bibinfo {author} {\bibfnamefont
  {S.}~\bibnamefont {Ravy}},\ }\bibfield  {title} {\bibinfo {title} {{Ultrafast
  formation of a charge density wave state in 1$T$-TaS$_2$: Observation at
  nanometer scales using time-resolved x-ray diffraction}},\ }\href
  {https://doi.org/10.1103/PhysRevLett.118.247401} {\bibfield  {journal}
  {\bibinfo  {journal} {Phys. Rev. Lett.}\ }\textbf {\bibinfo {volume} {118}},\
  \bibinfo {pages} {247401} (\bibinfo {year} {2017})}\BibitemShut {NoStop}%
\bibitem [{\citenamefont {Bray}(1994)}]{Bray1994}%
  \BibitemOpen
  \bibfield  {author} {\bibinfo {author} {\bibfnamefont {A.~J.}\ \bibnamefont
  {Bray}},\ }\bibfield  {title} {\bibinfo {title} {{Theory of phase-ordering
  kinetics}},\ }\href {https://doi.org/10.1080/00018739400101505} {\bibfield
  {journal} {\bibinfo  {journal} {Adv. Phys.}\ }\textbf {\bibinfo {volume}
  {43}},\ \bibinfo {pages} {357} (\bibinfo {year} {1994})}\BibitemShut
  {NoStop}%
\bibitem [{\citenamefont {Straquadine}\ \emph {et~al.}(2020)\citenamefont
  {Straquadine}, \citenamefont {Ikeda},\ and\ \citenamefont
  {Fisher}}]{Straquadine2020}%
  \BibitemOpen
  \bibfield  {author} {\bibinfo {author} {\bibfnamefont {J.~A.~W.}\
  \bibnamefont {Straquadine}}, \bibinfo {author} {\bibfnamefont {M.~S.}\
  \bibnamefont {Ikeda}},\ and\ \bibinfo {author} {\bibfnamefont {I.~R.}\
  \bibnamefont {Fisher}},\ }\href@noop {} {\bibinfo {title} {{Evidence for
  realignment of the charge density wave state in ErTe$_3$ and TmTe$_3$ under
  uniaxial stress via elastocaloric and elastoresistivity measurements}}}
  (\bibinfo {year} {2020}),\ \Eprint {https://arxiv.org/abs/2005.10461}
  {arXiv:2005.10461} \BibitemShut {NoStop}%
\bibitem [{\citenamefont {Siwick}\ \emph {et~al.}(2003)\citenamefont {Siwick},
  \citenamefont {Dwyer}, \citenamefont {Jordan},\ and\ \citenamefont
  {Miller}}]{Siwick2003}%
  \BibitemOpen
  \bibfield  {author} {\bibinfo {author} {\bibfnamefont {B.~J.}\ \bibnamefont
  {Siwick}}, \bibinfo {author} {\bibfnamefont {J.~R.}\ \bibnamefont {Dwyer}},
  \bibinfo {author} {\bibfnamefont {R.~E.}\ \bibnamefont {Jordan}},\ and\
  \bibinfo {author} {\bibfnamefont {R.~J.~D.}\ \bibnamefont {Miller}},\
  }\bibfield  {title} {\bibinfo {title} {{An Atomic-Level View of Melting Using
  Femtosecond Electron Diffraction}},\ }\href
  {https://doi.org/10.1126/science.1090052} {\bibfield  {journal} {\bibinfo
  {journal} {Science}\ }\textbf {\bibinfo {volume} {302}},\ \bibinfo {pages}
  {1382} (\bibinfo {year} {2003})}\BibitemShut {NoStop}%
\bibitem [{\citenamefont {Sciaini}\ \emph {et~al.}(2009)\citenamefont
  {Sciaini}, \citenamefont {Harb}, \citenamefont {Kruglik}, \citenamefont
  {Payer}, \citenamefont {Hebeisen}, \citenamefont {Heringdorf}, \citenamefont
  {Yamaguchi}, \citenamefont {Hoegen}, \citenamefont {Ernstorfer},\ and\
  \citenamefont {Miller}}]{Sciaini2009}%
  \BibitemOpen
  \bibfield  {author} {\bibinfo {author} {\bibfnamefont {G.}~\bibnamefont
  {Sciaini}}, \bibinfo {author} {\bibfnamefont {M.}~\bibnamefont {Harb}},
  \bibinfo {author} {\bibfnamefont {S.~G.}\ \bibnamefont {Kruglik}}, \bibinfo
  {author} {\bibfnamefont {T.}~\bibnamefont {Payer}}, \bibinfo {author}
  {\bibfnamefont {C.~T.}\ \bibnamefont {Hebeisen}}, \bibinfo {author}
  {\bibfnamefont {F.-J. M.~Z.}\ \bibnamefont {Heringdorf}}, \bibinfo {author}
  {\bibfnamefont {M.}~\bibnamefont {Yamaguchi}}, \bibinfo {author}
  {\bibfnamefont {M.~H.-v.}\ \bibnamefont {Hoegen}}, \bibinfo {author}
  {\bibfnamefont {R.}~\bibnamefont {Ernstorfer}},\ and\ \bibinfo {author}
  {\bibfnamefont {R.~J.~D.}\ \bibnamefont {Miller}},\ }\bibfield  {title}
  {\bibinfo {title} {{Electronic acceleration of atomic motions and disordering
  in bismuth}},\ }\href {https://doi.org/10.1038/nature07788} {\bibfield
  {journal} {\bibinfo  {journal} {Nature}\ }\textbf {\bibinfo {volume} {458}},\
  \bibinfo {pages} {56} (\bibinfo {year} {2009})}\BibitemShut {NoStop}%
\bibitem [{\citenamefont {Tokita}\ \emph {et~al.}(2010)\citenamefont {Tokita},
  \citenamefont {Hashida}, \citenamefont {Inoue}, \citenamefont {Nishoji},
  \citenamefont {Otani},\ and\ \citenamefont {Sakabe}}]{Tokita2010}%
  \BibitemOpen
  \bibfield  {author} {\bibinfo {author} {\bibfnamefont {S.}~\bibnamefont
  {Tokita}}, \bibinfo {author} {\bibfnamefont {M.}~\bibnamefont {Hashida}},
  \bibinfo {author} {\bibfnamefont {S.}~\bibnamefont {Inoue}}, \bibinfo
  {author} {\bibfnamefont {T.}~\bibnamefont {Nishoji}}, \bibinfo {author}
  {\bibfnamefont {K.}~\bibnamefont {Otani}},\ and\ \bibinfo {author}
  {\bibfnamefont {S.}~\bibnamefont {Sakabe}},\ }\bibfield  {title} {\bibinfo
  {title} {{Single-Shot Femtosecond Electron Diffraction with Laser-Accelerated
  Electrons: Experimental Demonstration of Electron Pulse Compression}},\
  }\href {https://doi.org/10.1103/PhysRevLett.105.215004} {\bibfield  {journal}
  {\bibinfo  {journal} {Phys. Rev. Lett.}\ }\textbf {\bibinfo {volume} {105}},\
  \bibinfo {pages} {215004} (\bibinfo {year} {2010})}\BibitemShut {NoStop}%
\bibitem [{\citenamefont {Musumeci}\ \emph {et~al.}(2010)\citenamefont
  {Musumeci}, \citenamefont {Moody}, \citenamefont {Scoby}, \citenamefont
  {Gutierrez}, \citenamefont {Bender},\ and\ \citenamefont
  {Wilcox}}]{Musumeci2010}%
  \BibitemOpen
  \bibfield  {author} {\bibinfo {author} {\bibfnamefont {P.}~\bibnamefont
  {Musumeci}}, \bibinfo {author} {\bibfnamefont {J.~T.}\ \bibnamefont {Moody}},
  \bibinfo {author} {\bibfnamefont {C.~M.}\ \bibnamefont {Scoby}}, \bibinfo
  {author} {\bibfnamefont {M.~S.}\ \bibnamefont {Gutierrez}}, \bibinfo {author}
  {\bibfnamefont {H.~A.}\ \bibnamefont {Bender}},\ and\ \bibinfo {author}
  {\bibfnamefont {N.~S.}\ \bibnamefont {Wilcox}},\ }\bibfield  {title}
  {\bibinfo {title} {{High quality single shot diffraction patterns using
  ultrashort megaelectron volt electron beams from a radio frequency
  photoinjector}},\ }\href {https://doi.org/10.1063/1.3292683} {\bibfield
  {journal} {\bibinfo  {journal} {Rev. Sci. Instrum.}\ }\textbf {\bibinfo
  {volume} {81}},\ \bibinfo {pages} {013306} (\bibinfo {year}
  {2010})}\BibitemShut {NoStop}%
\bibitem [{\citenamefont {Li}\ \emph {et~al.}(2010)\citenamefont {Li},
  \citenamefont {Huang}, \citenamefont {Du}, \citenamefont {Yan}, \citenamefont
  {Du}, \citenamefont {Shi}, \citenamefont {Hua}, \citenamefont {Chen},
  \citenamefont {Du}, \citenamefont {Xu},\ and\ \citenamefont {Tang}}]{Li2010}%
  \BibitemOpen
  \bibfield  {author} {\bibinfo {author} {\bibfnamefont {R.}~\bibnamefont
  {Li}}, \bibinfo {author} {\bibfnamefont {W.}~\bibnamefont {Huang}}, \bibinfo
  {author} {\bibfnamefont {Y.}~\bibnamefont {Du}}, \bibinfo {author}
  {\bibfnamefont {L.}~\bibnamefont {Yan}}, \bibinfo {author} {\bibfnamefont
  {Q.}~\bibnamefont {Du}}, \bibinfo {author} {\bibfnamefont {J.}~\bibnamefont
  {Shi}}, \bibinfo {author} {\bibfnamefont {J.}~\bibnamefont {Hua}}, \bibinfo
  {author} {\bibfnamefont {H.}~\bibnamefont {Chen}}, \bibinfo {author}
  {\bibfnamefont {T.}~\bibnamefont {Du}}, \bibinfo {author} {\bibfnamefont
  {H.}~\bibnamefont {Xu}},\ and\ \bibinfo {author} {\bibfnamefont
  {C.}~\bibnamefont {Tang}},\ }\bibfield  {title} {\bibinfo {title} {{Note:
  Single-shot continuously time-resolved MeV ultrafast electron diffraction}},\
  }\href {https://doi.org/10.1063/1.3361196} {\bibfield  {journal} {\bibinfo
  {journal} {Rev. Sci. Instrum.}\ }\textbf {\bibinfo {volume} {81}},\ \bibinfo
  {pages} {036110} (\bibinfo {year} {2010})}\BibitemShut {NoStop}%
\bibitem [{\citenamefont {Speirs}\ \emph {et~al.}(2015)\citenamefont {Speirs},
  \citenamefont {Putkunz}, \citenamefont {McCulloch}, \citenamefont {Nugent},
  \citenamefont {Sparkes},\ and\ \citenamefont {Scholten}}]{Speirs2015}%
  \BibitemOpen
  \bibfield  {author} {\bibinfo {author} {\bibfnamefont {R.~W.}\ \bibnamefont
  {Speirs}}, \bibinfo {author} {\bibfnamefont {C.~T.}\ \bibnamefont {Putkunz}},
  \bibinfo {author} {\bibfnamefont {A.~J.}\ \bibnamefont {McCulloch}}, \bibinfo
  {author} {\bibfnamefont {K.~A.}\ \bibnamefont {Nugent}}, \bibinfo {author}
  {\bibfnamefont {B.~M.}\ \bibnamefont {Sparkes}},\ and\ \bibinfo {author}
  {\bibfnamefont {R.~E.}\ \bibnamefont {Scholten}},\ }\bibfield  {title}
  {\bibinfo {title} {{Single-shot electron diffraction using a cold atom
  electron source}},\ }\href {https://doi.org/10.1088/0953-4075/48/21/214002}
  {\bibfield  {journal} {\bibinfo  {journal} {J. Phys. B}\ }\textbf {\bibinfo
  {volume} {48}},\ \bibinfo {pages} {214002} (\bibinfo {year} {2015})},\
  \Eprint {https://arxiv.org/abs/1506.07442} {1506.07442} \BibitemShut
  {NoStop}%
\bibitem [{\citenamefont {Mo}\ \emph {et~al.}(2018)\citenamefont {Mo},
  \citenamefont {Chen}, \citenamefont {Li}, \citenamefont {Dunning},
  \citenamefont {Witte}, \citenamefont {Baldwin}, \citenamefont {Fletcher},
  \citenamefont {Kim}, \citenamefont {Ng}, \citenamefont {Redmer},
  \citenamefont {Reid}, \citenamefont {Shekhar}, \citenamefont {Shen},
  \citenamefont {Shen}, \citenamefont {Sokolowski-Tinten}, \citenamefont
  {Tsui}, \citenamefont {Wang}, \citenamefont {Zheng}, \citenamefont {Wang},\
  and\ \citenamefont {Glenzer}}]{Mo2018}%
  \BibitemOpen
  \bibfield  {author} {\bibinfo {author} {\bibfnamefont {M.~Z.}\ \bibnamefont
  {Mo}}, \bibinfo {author} {\bibfnamefont {Z.}~\bibnamefont {Chen}}, \bibinfo
  {author} {\bibfnamefont {R.~K.}\ \bibnamefont {Li}}, \bibinfo {author}
  {\bibfnamefont {M.}~\bibnamefont {Dunning}}, \bibinfo {author} {\bibfnamefont
  {B.~B.~L.}\ \bibnamefont {Witte}}, \bibinfo {author} {\bibfnamefont {J.~K.}\
  \bibnamefont {Baldwin}}, \bibinfo {author} {\bibfnamefont {L.~B.}\
  \bibnamefont {Fletcher}}, \bibinfo {author} {\bibfnamefont {J.~B.}\
  \bibnamefont {Kim}}, \bibinfo {author} {\bibfnamefont {A.}~\bibnamefont
  {Ng}}, \bibinfo {author} {\bibfnamefont {R.}~\bibnamefont {Redmer}}, \bibinfo
  {author} {\bibfnamefont {A.~H.}\ \bibnamefont {Reid}}, \bibinfo {author}
  {\bibfnamefont {P.}~\bibnamefont {Shekhar}}, \bibinfo {author} {\bibfnamefont
  {X.~Z.}\ \bibnamefont {Shen}}, \bibinfo {author} {\bibfnamefont
  {M.}~\bibnamefont {Shen}}, \bibinfo {author} {\bibfnamefont {K.}~\bibnamefont
  {Sokolowski-Tinten}}, \bibinfo {author} {\bibfnamefont {Y.~Y.}\ \bibnamefont
  {Tsui}}, \bibinfo {author} {\bibfnamefont {Y.~Q.}\ \bibnamefont {Wang}},
  \bibinfo {author} {\bibfnamefont {Q.}~\bibnamefont {Zheng}}, \bibinfo
  {author} {\bibfnamefont {X.~J.}\ \bibnamefont {Wang}},\ and\ \bibinfo
  {author} {\bibfnamefont {S.~H.}\ \bibnamefont {Glenzer}},\ }\bibfield
  {title} {\bibinfo {title} {{Heterogeneous to homogeneous melting transition
  visualized with ultrafast electron diffraction}},\ }\href
  {https://doi.org/10.1126/science.aar2058} {\bibfield  {journal} {\bibinfo
  {journal} {Science}\ }\textbf {\bibinfo {volume} {360}},\ \bibinfo {pages}
  {1451} (\bibinfo {year} {2018})}\BibitemShut {NoStop}%
\bibitem [{\citenamefont {Wang}\ \emph {et~al.}(1996)\citenamefont {Wang},
  \citenamefont {Qiu},\ and\ \citenamefont {Ben-Zvi}}]{Wang1996}%
  \BibitemOpen
  \bibfield  {author} {\bibinfo {author} {\bibfnamefont {X.~J.}\ \bibnamefont
  {Wang}}, \bibinfo {author} {\bibfnamefont {X.}~\bibnamefont {Qiu}},\ and\
  \bibinfo {author} {\bibfnamefont {I.}~\bibnamefont {Ben-Zvi}},\ }\bibfield
  {title} {\bibinfo {title} {{Experimental observation of high-brightness
  microbunching in a photocathode rf electron gun}},\ }\href
  {https://doi.org/10.1103/PhysRevE.54.R3121} {\bibfield  {journal} {\bibinfo
  {journal} {Phys. Rev. E}\ }\textbf {\bibinfo {volume} {54}},\ \bibinfo
  {pages} {R3121} (\bibinfo {year} {1996})}\BibitemShut {NoStop}%
\bibitem [{\citenamefont {Wang}\ \emph {et~al.}(2003)\citenamefont {Wang},
  \citenamefont {Wu},\ and\ \citenamefont {Ihee}}]{Wang2003}%
  \BibitemOpen
  \bibfield  {author} {\bibinfo {author} {\bibfnamefont {X.~J.}\ \bibnamefont
  {Wang}}, \bibinfo {author} {\bibfnamefont {Z.}~\bibnamefont {Wu}},\ and\
  \bibinfo {author} {\bibfnamefont {H.}~\bibnamefont {Ihee}},\ }\bibfield
  {title} {\bibinfo {title} {{Femto-seconds electron beam diffraction using
  photocathode RF gun}},\ }in\ \href {https://doi.org/10.1109/PAC.2003.1288940}
  {\emph {\bibinfo {booktitle} {Proc. 2003 Part. Accel. Conf.}}},\
  Vol.~\bibinfo {volume} {1}\ (\bibinfo  {publisher} {IEEE},\ \bibinfo {year}
  {2003})\ pp.\ \bibinfo {pages} {420--422}\BibitemShut {NoStop}%
\bibitem [{\citenamefont {Weathersby}\ \emph {et~al.}(2015)\citenamefont
  {Weathersby}, \citenamefont {Brown}, \citenamefont {Centurion}, \citenamefont
  {Chase}, \citenamefont {Coffee}, \citenamefont {Corbett}, \citenamefont
  {Eichner}, \citenamefont {Frisch}, \citenamefont {Fry}, \citenamefont
  {G{\"{u}}hr}, \citenamefont {Hartmann}, \citenamefont {Hast}, \citenamefont
  {Hettel}, \citenamefont {Jobe}, \citenamefont {Jongewaard}, \citenamefont
  {Lewandowski}, \citenamefont {Li}, \citenamefont {Lindenberg}, \citenamefont
  {Makasyuk}, \citenamefont {May}, \citenamefont {McCormick}, \citenamefont
  {Nguyen}, \citenamefont {Reid}, \citenamefont {Shen}, \citenamefont
  {Sokolowski-Tinten}, \citenamefont {Vecchione}, \citenamefont {Vetter},
  \citenamefont {Wu}, \citenamefont {Yang}, \citenamefont {D{\"{u}}rr},\ and\
  \citenamefont {Wang}}]{Weathersby2015}%
  \BibitemOpen
  \bibfield  {author} {\bibinfo {author} {\bibfnamefont {S.~P.}\ \bibnamefont
  {Weathersby}}, \bibinfo {author} {\bibfnamefont {G.}~\bibnamefont {Brown}},
  \bibinfo {author} {\bibfnamefont {M.}~\bibnamefont {Centurion}}, \bibinfo
  {author} {\bibfnamefont {T.~F.}\ \bibnamefont {Chase}}, \bibinfo {author}
  {\bibfnamefont {R.}~\bibnamefont {Coffee}}, \bibinfo {author} {\bibfnamefont
  {J.}~\bibnamefont {Corbett}}, \bibinfo {author} {\bibfnamefont {J.~P.}\
  \bibnamefont {Eichner}}, \bibinfo {author} {\bibfnamefont {J.~C.}\
  \bibnamefont {Frisch}}, \bibinfo {author} {\bibfnamefont {A.~R.}\
  \bibnamefont {Fry}}, \bibinfo {author} {\bibfnamefont {M.}~\bibnamefont
  {G{\"{u}}hr}}, \bibinfo {author} {\bibfnamefont {N.}~\bibnamefont
  {Hartmann}}, \bibinfo {author} {\bibfnamefont {C.}~\bibnamefont {Hast}},
  \bibinfo {author} {\bibfnamefont {R.}~\bibnamefont {Hettel}}, \bibinfo
  {author} {\bibfnamefont {R.~K.}\ \bibnamefont {Jobe}}, \bibinfo {author}
  {\bibfnamefont {E.~N.}\ \bibnamefont {Jongewaard}}, \bibinfo {author}
  {\bibfnamefont {J.~R.}\ \bibnamefont {Lewandowski}}, \bibinfo {author}
  {\bibfnamefont {R.~K.}\ \bibnamefont {Li}}, \bibinfo {author} {\bibfnamefont
  {A.~M.}\ \bibnamefont {Lindenberg}}, \bibinfo {author} {\bibfnamefont
  {I.}~\bibnamefont {Makasyuk}}, \bibinfo {author} {\bibfnamefont {J.~E.}\
  \bibnamefont {May}}, \bibinfo {author} {\bibfnamefont {D.}~\bibnamefont
  {McCormick}}, \bibinfo {author} {\bibfnamefont {M.~N.}\ \bibnamefont
  {Nguyen}}, \bibinfo {author} {\bibfnamefont {A.~H.}\ \bibnamefont {Reid}},
  \bibinfo {author} {\bibfnamefont {X.}~\bibnamefont {Shen}}, \bibinfo {author}
  {\bibfnamefont {K.}~\bibnamefont {Sokolowski-Tinten}}, \bibinfo {author}
  {\bibfnamefont {T.}~\bibnamefont {Vecchione}}, \bibinfo {author}
  {\bibfnamefont {S.~L.}\ \bibnamefont {Vetter}}, \bibinfo {author}
  {\bibfnamefont {J.}~\bibnamefont {Wu}}, \bibinfo {author} {\bibfnamefont
  {J.}~\bibnamefont {Yang}}, \bibinfo {author} {\bibfnamefont {H.~A.}\
  \bibnamefont {D{\"{u}}rr}},\ and\ \bibinfo {author} {\bibfnamefont {X.~J.}\
  \bibnamefont {Wang}},\ }\bibfield  {title} {\bibinfo {title}
  {{Mega-electron-volt ultrafast electron diffraction at SLAC National
  Accelerator Laboratory}},\ }\href {https://doi.org/10.1063/1.4926994}
  {\bibfield  {journal} {\bibinfo  {journal} {Rev. Sci. Instrum.}\ }\textbf
  {\bibinfo {volume} {86}},\ \bibinfo {pages} {073702} (\bibinfo {year}
  {2015})}\BibitemShut {NoStop}%
\bibitem [{\citenamefont {Kim}\ \emph {et~al.}(2019{\natexlab{b}})\citenamefont
  {Kim}, \citenamefont {Vinokurov}, \citenamefont {Baek}, \citenamefont {Oang},
  \citenamefont {Kim}, \citenamefont {Kim}, \citenamefont {Jang}, \citenamefont
  {Lee}, \citenamefont {Park}, \citenamefont {Park}, \citenamefont {Shin},
  \citenamefont {Kim}, \citenamefont {Rotermund}, \citenamefont {Cho},
  \citenamefont {Feurer},\ and\ \citenamefont {Jeong}}]{Kim2019b}%
  \BibitemOpen
  \bibfield  {author} {\bibinfo {author} {\bibfnamefont {H.~W.}\ \bibnamefont
  {Kim}}, \bibinfo {author} {\bibfnamefont {N.~A.}\ \bibnamefont {Vinokurov}},
  \bibinfo {author} {\bibfnamefont {I.~H.}\ \bibnamefont {Baek}}, \bibinfo
  {author} {\bibfnamefont {K.~Y.}\ \bibnamefont {Oang}}, \bibinfo {author}
  {\bibfnamefont {M.~H.}\ \bibnamefont {Kim}}, \bibinfo {author} {\bibfnamefont
  {Y.~C.}\ \bibnamefont {Kim}}, \bibinfo {author} {\bibfnamefont {K.~H.}\
  \bibnamefont {Jang}}, \bibinfo {author} {\bibfnamefont {K.}~\bibnamefont
  {Lee}}, \bibinfo {author} {\bibfnamefont {S.~H.}\ \bibnamefont {Park}},
  \bibinfo {author} {\bibfnamefont {S.}~\bibnamefont {Park}}, \bibinfo {author}
  {\bibfnamefont {J.}~\bibnamefont {Shin}}, \bibinfo {author} {\bibfnamefont
  {J.}~\bibnamefont {Kim}}, \bibinfo {author} {\bibfnamefont {F.}~\bibnamefont
  {Rotermund}}, \bibinfo {author} {\bibfnamefont {S.}~\bibnamefont {Cho}},
  \bibinfo {author} {\bibfnamefont {T.}~\bibnamefont {Feurer}},\ and\ \bibinfo
  {author} {\bibfnamefont {Y.~U.}\ \bibnamefont {Jeong}},\ }\bibfield  {title}
  {\bibinfo {title} {{Towards jitter-free ultrafast electron diffraction
  technology}},\ }\href {https://doi.org/10.1038/s41566-019-0566-4} {\bibfield
  {journal} {\bibinfo  {journal} {Nat. Photon.}\ }\textbf {\bibinfo {volume}
  {14}},\ \bibinfo {pages} {245} (\bibinfo {year}
  {2019}{\natexlab{b}})}\BibitemShut {NoStop}%
\bibitem [{\citenamefont {Qi}\ \emph {et~al.}(2020)\citenamefont {Qi},
  \citenamefont {Ma}, \citenamefont {Zhao}, \citenamefont {Cheng},
  \citenamefont {Jiang}, \citenamefont {Lu}, \citenamefont {Jiang},
  \citenamefont {Qian}, \citenamefont {Wang}, \citenamefont {Zhang},
  \citenamefont {Zhu}, \citenamefont {Zou}, \citenamefont {Wan}, \citenamefont
  {Xiang},\ and\ \citenamefont {Zhang}}]{Qi2020}%
  \BibitemOpen
  \bibfield  {author} {\bibinfo {author} {\bibfnamefont {F.}~\bibnamefont
  {Qi}}, \bibinfo {author} {\bibfnamefont {Z.}~\bibnamefont {Ma}}, \bibinfo
  {author} {\bibfnamefont {L.}~\bibnamefont {Zhao}}, \bibinfo {author}
  {\bibfnamefont {Y.}~\bibnamefont {Cheng}}, \bibinfo {author} {\bibfnamefont
  {W.}~\bibnamefont {Jiang}}, \bibinfo {author} {\bibfnamefont
  {C.}~\bibnamefont {Lu}}, \bibinfo {author} {\bibfnamefont {T.}~\bibnamefont
  {Jiang}}, \bibinfo {author} {\bibfnamefont {D.}~\bibnamefont {Qian}},
  \bibinfo {author} {\bibfnamefont {Z.}~\bibnamefont {Wang}}, \bibinfo {author}
  {\bibfnamefont {W.}~\bibnamefont {Zhang}}, \bibinfo {author} {\bibfnamefont
  {P.}~\bibnamefont {Zhu}}, \bibinfo {author} {\bibfnamefont {X.}~\bibnamefont
  {Zou}}, \bibinfo {author} {\bibfnamefont {W.}~\bibnamefont {Wan}}, \bibinfo
  {author} {\bibfnamefont {D.}~\bibnamefont {Xiang}},\ and\ \bibinfo {author}
  {\bibfnamefont {J.}~\bibnamefont {Zhang}},\ }\bibfield  {title} {\bibinfo
  {title} {{Breaking 50 femtosecond resolution barrier in MeV ultrafast
  electron diffraction with a double bend achromat compressor}},\ }\href
  {https://doi.org/10.1103/PhysRevLett.124.134803} {\bibfield  {journal}
  {\bibinfo  {journal} {Phys. Rev. Lett.}\ }\textbf {\bibinfo {volume} {124}},\
  \bibinfo {pages} {134803} (\bibinfo {year} {2020})}\BibitemShut {NoStop}%
\bibitem [{\citenamefont {D{\"{u}}rr}\ \emph {et~al.}(2021)\citenamefont
  {D{\"{u}}rr}, \citenamefont {Ernstorfer},\ and\ \citenamefont
  {Siwick}}]{Durr2021}%
  \BibitemOpen
  \bibfield  {author} {\bibinfo {author} {\bibfnamefont {H.~A.}\ \bibnamefont
  {D{\"{u}}rr}}, \bibinfo {author} {\bibfnamefont {R.}~\bibnamefont
  {Ernstorfer}},\ and\ \bibinfo {author} {\bibfnamefont {B.~J.}\ \bibnamefont
  {Siwick}},\ }\bibfield  {title} {\bibinfo {title} {{Revealing
  momentum-dependent electron–phonon and phonon–phonon coupling in complex
  materials with ultrafast electron diffuse scattering}},\ }\href
  {https://doi.org/10.1557/s43577-021-00156-7} {\bibfield  {journal} {\bibinfo
  {journal} {MRS Bull.}\ }\textbf {\bibinfo {volume} {46}},\ \bibinfo {pages}
  {731} (\bibinfo {year} {2021})}\BibitemShut {NoStop}%
\end{thebibliography}
\end{document}